\documentclass{article}
\usepackage[preprint]{neurips_2021}

\usepackage[hyphens]{url}
\usepackage[colorlinks=false,hidelinks]{hyperref}
\usepackage{graphicx}
\usepackage{tabularx}
\usepackage{booktabs}
\usepackage{amssymb}
\usepackage{amsmath}
\usepackage{amsthm}
\usepackage{mathtools}
\usepackage{multirow}

\setcitestyle{numbers,square}

\title{Benchmarking JSON BinPack}

\author{
  Juan Cruz~Viotti\thanks{\url{https://www.jviotti.com}} \\
  Department of Computer Science \\
  University of Oxford \\
  Oxford, GB OX1 3QD \\
  \texttt{jv@jviotti.com} \\
  \and
  Mital~Kinderkhedia \\
  Department of Computer Science \\
  University of Oxford \\
  Oxford, GB OX1 3QD \\
  \texttt{mital.kinderkhedia@cs.ox.ac.uk} \\
}

\begin{document}
\maketitle

\begin{abstract}
In this paper, we present bechmark results for a pre-production implementation
of a novel serialization specification: JSON BinPack. JSON BinPack is a
schema-driven and schema-less sequential binary serialization specification
based on JSON Schema.  It is rich in diverse encodings, and is developed to
improve network performance and reduce the operational costs of Internet-based
software systems.  We present bechmark results for 27 JSON documents and for
each plot, we show the schema-driven and schema-less serialization
specifications that produce the smallest bit-strings.  Through extensive plots
and statistical comparisons, we show that JSON BinPack in schema-driven mode is
as space-efficient or more space-efficient than every other serialization
specification for the 27 documents under consideration.  In comparison to JSON,
JSON BinPack in schema-driven mode provides a median and average size
reductions of 86.7\% and 78.7\%, respectively.  We also show that the
schema-less mode of the JSON BinPack binary serialization specification is as
space-efficient or more space-efficient than every other schema-less
serialization specification for the 27 documents under consideration.  In
comparison to JSON, JSON BinPack in schema-less mode provides a median and
average size reductions of 30.6\% and 30.5\%, respectively.  Unlike other
considered schema-driven binary serialization specifications, JSON BinPack in
schema-driven mode is space-efficient in comparison to best-case compressed
JSON in terms of the median and average with size reductions of 76.1\% and
66.8\%, respectively.  We have made our benchmark results available at
\href{https://github.com/jviotti/binary-json-size-benchmark}{jviotti/binary-json-size-benchmark}
on GitHub.

\end{abstract}

\section{Introduction}
\label{sec:introduction}
For consumers of Internet-based software systems, substandard network
performance results in impaired user experience given that an increasing amount
of software is now accessed over the Internet. This type of software systems
are particularly sensitive to substandard network performance. Additionally,
software systems that operate over the Internet typically rely on
infrastructure that charges for inbound or outbound network communication.
Given the decentralized architecture and complex dynamics of the Internet,
network communications are unpredictable and often unreliable. Therefore,
transmitting data over the network directly translates to operational expenses.

To facilitate interoperability, Internet-based software systems transmit
information using data serialization specifications such as JSON
\cite{ECMA-404} and XML \cite{Paoli:06:EML}. \cite{viotti2022survey} identifies
JSON \cite{ECMA-404} as the dominant data interchange standard in the context
of cloud software systems. However, it concludes that JSON is neither a
space-efficient nor a runtime-efficient serialization specification.
\cite{krashinsky2003efficient} argues that the bottleneck of Internet network
communication is the time of transmission, making the runtime-efficiency aspect
of the choice of a serialization specification irrelevant in such context.
Therefore, the choice of a serialization specification and its space-efficiency
characteristics are crucial for improving network performance and reducing the
operational costs of Internet-based software systems.

In the pursuit of a solution to enhance network performance and reduce
operational costs, our previous works \cite{viotti2022survey}
\cite{viotti2022benchmark} explore in depth a set of 13 schema-driven and
schema-less JSON-compatible binary serialization specifications with different
space-efficiency, runtime-efficiency and architectural characteristics: ASN.1
\cite{asn1-per}, Apache Avro \cite{avro}, Microsoft Bond \cite{microsoft-bond},
Cap'n Proto \cite{capnproto}, FlatBuffers \cite{flatbuffers}, Protocol Buffers
\cite{protocolbuffers}, and Apache Thrift \cite{slee2007thrift}, BSON
\cite{bson}, CBOR \cite{RFC7049}, FlexBuffers \cite{flexbuffers}, MessagePack
\cite{messagepack}, Smile \cite{smile} and UBJSON \cite{ubjson}, and proposes
JSON BinPack - a novel open-source binary serialization specification that has
a strong focus on space-efficiency, is strictly JSON-compatible, is a hybrid
between schema-driven and schema-less and relies on the industry-standard
schema language for describing JSON documents: JSON Schema
\cite{jsonschema-core} (upcoming paper). Using the same set of JSON-compatible
binary serialization specifications and taking data compression into
consideration, we produced a comprehensive space-efficiency benchmark
\cite{viotti2022benchmark} that is based on a representative set of 27
real-world JSON documents and analyzed the results across different lens to
understand in which cases and why certain serialization specifications
outperform others.

\subsection{Paper Organization}
This paper is organized as follows. In \autoref{sec:introduction}, we discuss
related literature in the context of space-efficient binary serialization
specifications and space-efficiency benchmarking. We propose a set of
space-efficiency-oriented research questions for the novel JSON BinPack binary
serialization specification and we state our contributions.  In
\autoref{sec:methodology}, we review the benchmarking methodology introduced in
\cite{viotti2022benchmark} and apply it to evaluate the JSON BinPack binary
serialization specification. We then describe the alternative serialization
specifications we compare JSON BinPack to, linking them to the schema
definitions used as part of the benchmark study and expand on our strategy to
provide a fair benchmark.  In \autoref{sec:benchmark}, we present the results
of extending the benchmark study introduced in \cite{viotti2022benchmark} to
include the JSON BinPack binary serialization specification and state how our
experiments could be reproduced.  In \autoref{sec:conclusions}, we critically
evaluate the benchmark results in terms of each of the research questions
proposed in \autoref{sec:questions}.  In \autoref{sec:future-work}, we conclude
with an assessment of future work to be done for the JSON BinPack binary
serialization specification.

\subsection{Related Literature}
In \cite{viotti2022survey}, we introduce the problem of data serialization,
summarize the state-of-the-art in JSON-compatible data serialization
literature, discuss the history of JSON \cite{ECMA-404}, its relevance,
characteristics and shortcomings. Next, we study the history, the advantages,
the characteristics and encodings of 13 popular schema-driven (ASN.1, Apache
Avro, Microsoft Bond, Cap’n Proto, FlatBuffers, Protocol Buffers, and Apache
Thrift) and schema-less (BSON, CBOR, FlexBuffers, MessagePack, Smile, and
UBJSON) JSON-compatible binary serialization specifications and discuss the
problem of schema evolution in the context of JSON-compatible binary
schema-driven serialization specifications.

As a continuation of this work, \cite{viotti2022benchmark} makes a case for the
importance of space-efficiency in the context of data transmission as an
approach to improve network performance and reduce operational costs.  To
evaluate the space-efficiency characteristics of available JSON-compatible
serialization specifications, \cite{viotti2022benchmark} presents a
comprehensive and representative benchmark of JSON-compatible binary
serialization specifications. The benchmark involves the 13 schema-driven and
schema-less binary serialization specifications studied in
\cite{viotti2022survey}. The input data consists of a set of 27 real-world JSON
documents obtained from the SchemaStore open-source dataset. The pre-production
benchmark software is open-source and is designed to be continuously extended
with new JSON-compatible serialization specifications and input JSON documents.

\cite{viotti2022benchmark} concludes that ASN.1 PER Unaligned \cite{asn1-per}
and Apache Avro \cite{avro} are space-efficient in comparison to JSON and other
schema-driven and schema-less JSON-compatible binary serialization
specifications in most cases. Furthermore, space-efficient schema-driven
sequential binary serialization specifications tend to outperform JSON general
purpose data compression, especially on small JSON documents.  However, no
considered binary serialization specification is strictly superior than JSON
nor the other considered serialization specifications in every case. Out of the
selection of schema-driven binary serialization specifications, Apache Avro
\cite{avro} is the only serialization specification that was found to be
strictly JSON-compatible.

\begin{figure*}[ht!]
\frame{\includegraphics[width=\linewidth]{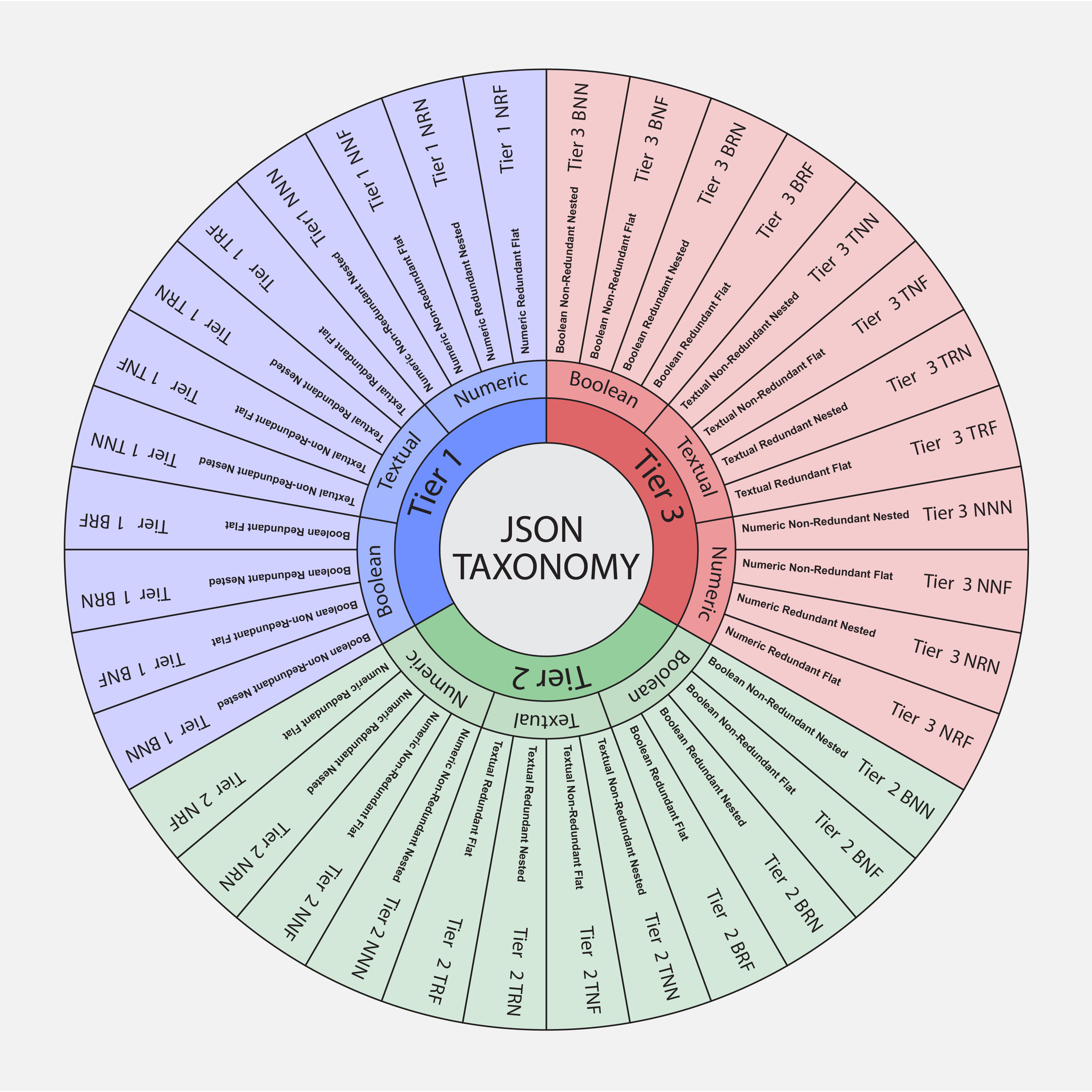}}
\caption{\cite{viotti2022benchmark} introduces a formal taxonomy to class JSON
documents that consists of 36 categories.} \label{fig:json-taxonomy}
\end{figure*}

To methodically solve the input data selection process and provide a fair
representative benchmark using the smallest possible set of input data,
\cite{https://doi.org/10.48550/arxiv.2211.11314} introduces a formal tiered
taxonomy for JSON \cite{ECMA-404} documents. This taxonomy consists of 36
categories classified as Tier 1, Tier 2 and Tier 3 as a common basis to class
JSON documents based on characteristics that are relevant in the context of
data serialization such as size, type of content, structural and redundancy
criteria.  The categories of the taxonomy are visually represented in
\autoref{fig:json-taxonomy}.

\subsection{Research Questions}
\label{sec:questions}
This paper aims to evaluate the space-efficiency characteristics of the JSON
BinPack pre-production implementation against the JSON-compatible schema-less
and schema-driven binary serialization specifications and general purpose data
compressors studied in \cite{viotti2022survey} and \cite{viotti2022benchmark}.

This benchmark aims to answer the following set of research questions:

\begin{itemize}

\item \textbf{Q1}: How does JSON BinPack in schema-driven mode compare to JSON
in terms of space-efficiency?

\item \textbf{Q2}: How does JSON BinPack in schema-less mode compare to JSON
in terms of space-efficiency?

\item \textbf{Q3}: How does JSON BinPack in schema-driven and schema-less mode
  compare to compressed JSON?

\end{itemize}

\subsection{Contributions}
We believe that the space-efficiency benchmark introduced in
\cite{viotti2022benchmark} is the first of its kind to produce a comprehensive,
reproducible, extensible and open-source study that considers a large and
representative input dataset of real-world JSON documents across industries and
takes data compression into consideration.

Our main contribution in this paper is the extension of the benchmark software
developed in \cite{viotti2022benchmark} to compare the novel JSON BinPack
binary serialization specification against the popular JSON-compatible
serialization specifications studied in \cite{viotti2022survey}.

\section{Methodology}
\label{sec:methodology}
To evaluate the space-efficiency characteristics of JSON BinPack in comparison
to the JSON-compatible binary serialization specifications and general-purpose
data compressors studied in \cite{viotti2022benchmark} and
\cite{viotti2022survey}, we benchmark the JSON BinPack pre-production
implementation using the open-source space-efficiency framework written in
\cite{viotti2022benchmark}. Our approach is the following:

\begin{enumerate}

  \item \textbf{JSON BinPack.} We extended the open-source automated benchmark
    software implemented in \cite{viotti2022benchmark} to recognize JSON
    BinPack as a JSON-compatible binary serialization specification.

  \item \textbf{Schema Definitions.} We wrote both \emph{strict} and
    \emph{loose} JSON Schema \cite{jsonschema-core-2020} definitions for the 27
    input JSON \cite{ECMA-404} documents selected in
    \cite{viotti2022benchmark}. In total, we wrote 54 schema definitions.

  \item \textbf{Benchmarking JSON BinPack against Alternatives.} We used the
    automated benchmark software to serialize and de-serialize the 27
    real-world JSON \cite{ECMA-404} documents selected in
    \cite{viotti2022benchmark} with JSON BinPack.  \cite{viotti2022survey}
    proposes the idea of schema-less as a subset of schema-driven. Under this
    concept, we considered serializing the input data using a loose schema
    definition that matches any instance, thus executing the JSON BinPack
    binary specification in \emph{schema-less} mode.  Conversely, we considered
    serializing the input data using the strict schema definition, thus
    executing the JSON BinPack binary specification in \emph{schema-driven}
    mode.

  \item \textbf{Analyzing the Results.} We analyzed the benchmark results of
    JSON BinPack in comparison to the 14 JSON-compatible serialization
    specifications and encodings listed in
    \autoref{table:benchmark-specifications-schema-driven} and
    \autoref{table:benchmark-specifications-schema-less} by visualizing and
    comparing them.

  \item \textbf{Conclusions.} We made deductions based on the benchmark results
    analyzing how JSON BinPack compared to the 14 JSON-compatible serialization
    specifications and encodings listed in
    \autoref{table:benchmark-specifications-schema-driven} and
    \autoref{table:benchmark-specifications-schema-less} in terms of
    space-efficiency and JSON-compatibility.

\end{enumerate}

\subsection{Serialization Specifications}
\label{sec:benchmark-specifications}

The following benchmark as referred to in \cite{viotti2022benchmark} compares
\textbf{9 schema-driven} JSON-compatible binary serialization specifications
and encodings listed in \autoref{table:benchmark-specifications-schema-driven}
and \textbf{7 schema-less} JSON-compatible binary serialization specifications
and encodings listed in \autoref{table:benchmark-specifications-schema-less},
with the exception of JSON BinPack.

\begin{table}[ht!]
\caption{The selection of schema-driven JSON-compatible binary serialization specifications based on our previous work \cite{viotti2022benchmark}.}
\label{table:benchmark-specifications-schema-driven}
\begin{tabularx}{\linewidth}{l|X|X|l}
\toprule
\textbf{Specification} & \textbf{Implementation} & \textbf{Encoding} & \textbf{License} \\
\midrule

ASN.1            & OSS ASN-1Step Version 10.0.2               & PER Unaligned \cite{asn1-per}                 & Proprietary \\ \hline
Apache Avro      & Python \texttt{avro} (pip) 1.10.0          & Binary Encoding \footnotemark with no framing & Apache-2.0 \\ \hline
Microsoft Bond   & C++ library 9.0.4                          & Compact Binary v1 \footnotemark               & MIT \\ \hline
Cap'n Proto      & \texttt{capnp} command-line tool 0.8.0     & Binary Encoding \footnotemark                 & MIT \\ \hline
Cap'n Proto      & \texttt{capnp} command-line tool 0.8.0     & Packed Encoding \footnotemark                 & MIT \\ \hline
FlatBuffers      & \texttt{flatc} command-line tool 1.12.0    & Binary Wire Format \footnotemark              & Apache-2.0 \\ \hline
JSON BinPack     & Pre-production implementation v1.1.2       & Binary Encoding & Apache-2.0 \\ \hline
Protocol Buffers & Python \texttt{protobuf} (pip) 3.15.3      & Binary Wire Format \footnotemark              & 3-Clause BSD \\ \hline
Apache Thrift    & Python \texttt{thrift} (pip) 0.13.0        & Compact Protocol \footnotemark                & Apache-2.0 \\

\bottomrule
\end{tabularx}
\end{table}

\footnotetext[\numexpr\thefootnote-6]{\url{https://avro.apache.org/docs/current/spec.html\#binary\_encoding}}
\footnotetext[\numexpr\thefootnote-5]{\url{https://microsoft.github.io/bond/reference/cpp/compact\_\_binary\_8h\_source.html}}
\footnotetext[\numexpr\thefootnote-4]{\url{https://capnproto.org/encoding.html\#packing}}
\footnotetext[\numexpr\thefootnote-3]{\url{https://capnproto.org/encoding.html}}
\footnotetext[\numexpr\thefootnote-2]{\url{https://google.github.io/flatbuffers/flatbuffers\_internals.html}}
\footnotetext[\numexpr\thefootnote-1]{\url{https://developers.google.com/protocol-buffers/docs/encoding}}
\footnotetext[\numexpr\thefootnote]{\url{https://github.com/apache/thrift/blob/master/doc/specs/thrift-compact-protocol.md}}

\begin{table}[ht!]
\caption{The selection of schema-less JSON-compatible binary serialization specifications based on our previous work \cite{viotti2022benchmark}.}
\label{table:benchmark-specifications-schema-less}
\begin{tabularx}{\linewidth}{l|X|l}
\toprule
\textbf{Specification} & \textbf{Implementation} & \textbf{License} \\
\midrule
BSON        & Node.js \texttt{bson} (npm) 4.2.2                                      & Apache-2.0 \\ \hline
CBOR        & Python \texttt{cbor2} (pip) 5.1.2                                      & MIT \\ \hline
FlexBuffers & \texttt{flatc} command-line tool 1.12.0                                & Apache-2.0 \\ \hline
JSON BinPack     & Pre-production implementation v1.1.2       & Apache-2.0 \\ \hline
MessagePack & \texttt{json2msgpack} command-line tool 0.6 with \texttt{MPack} 0.9dev & MIT \\ \hline
Smile       & Clojure \texttt{cheshire} 5.10.0                                       & MIT \\ \hline
UBJSON      & Python \texttt{py-ubjson} (pip) 0.16.1                                 & Apache-2.0 \\
\bottomrule
\end{tabularx}
\end{table}

\subsection{Schema Definitions}
\label{sec:schema-definitions}

For brevity, we do not include the schema definitions for each input JSON
document listed in \cite{viotti2022benchmark}. The schema definitions can be
found on the GitHub repository implemented as part of the benchmark study
\footnote{\url{https://github.com/jviotti/binary-json-size-benchmark}}.

\subsection{Fair Benchmarking}

We aim to provide a fair benchmark. Thus the resulting bit-strings are ensured
to be lossless encodings of the respective input JSON \cite{ECMA-404}
documents.

The implemented benchmark program validates that for each combination of
serialization specification listed in \autoref{sec:benchmark-specifications}
and input JSON document listed in \cite{viotti2022benchmark}, the produced
bit-strings encode the same information as the respective input JSON document.
The automated test consists in serializing the input JSON document using a
given binary serialization specification, deserializing the resulting
bit-string and asserting that the original JSON document is strictly equal to
the deserialized JSON document.

\clearpage
\section{Benchmark}
\label{sec:benchmark}
In this section, we present the bechmark results for the 27 JSON
\cite{ECMA-404} documents considered by \cite{viotti2022benchmark}.  For each
document, we present the plots and the related analysis. The schema-driven and
schema-less serialization specifications that produce the smallest bit-strings
for the corresponding documents are shown in full opacity. \emph{JSON BinPack
(Schema-driven)} corresponds to executing JSON BinPack with a strict schema
definition and \emph{JSON BinPack (Schema-less)} corresponds to executing JSON
BinPack with the wildcard schema definition that matches any instance.

\begin{table*}[ht!]
\caption{The 27 JSON documents considered by \cite{viotti2022benchmark} and their corresponding taxonomy categories.}

\label{table:benchmark-documents}
\begin{tabularx}{\linewidth}{X|l|l}
\toprule
\textbf{Description} & \textbf{Section} & \textbf{Category} \\
\midrule

JSON-e templating engine sort example                   & \autoref{sec:benchmark-jsonesort} & TNRF \\ \hline
JSON-e templating engine reverse sort example           & \autoref{sec:benchmark-jsonereversesort} & TNRN \\ \hline
CircleCI definition (blank)                             & \autoref{sec:benchmark-circleciblank} & TNNF \\ \hline
CircleCI matrix definition                              & \autoref{sec:benchmark-circlecimatrix} & TNNN \\ \hline
Grunt.js "clean" task definition                        & \autoref{sec:benchmark-gruntcontribclean} & TTRF \\ \hline
CommitLint configuration                                & \autoref{sec:benchmark-commitlint} & TTRN \\ \hline
TSLint linter definition (extends only)                 & \autoref{sec:benchmark-tslintextend} & TTNF \\ \hline
ImageOptimizer Azure Webjob configuration               & \autoref{sec:benchmark-imageoptimizerwebjob} & TTNN \\ \hline
SAP Cloud SDK Continuous Delivery Toolkit configuration & \autoref{sec:benchmark-sapcloudsdkpipeline} & TBRF \\ \hline
TSLint linter definition (multi-rule)                   & \autoref{sec:benchmark-tslintmulti} & TBRN \\ \hline
CommitLint configuration (basic)                        & \autoref{sec:benchmark-commitlintbasic} & TBNF \\ \hline
TSLint linter definition (basic)                        & \autoref{sec:benchmark-tslintbasic} & TBNN \\ \hline
GeoJSON example JSON document                           & \autoref{sec:benchmark-geojson} & SNRN \\ \hline
OpenWeatherMap API example JSON document                & \autoref{sec:benchmark-openweathermap} & SNNF \\ \hline
OpenWeather Road Risk API example                       & \autoref{sec:benchmark-openweatherroadrisk} & SNNN \\ \hline
TravisCI notifications configuration                    & \autoref{sec:benchmark-travisnotifications} & STRF \\ \hline
Entry Point Regulation manifest                         & \autoref{sec:benchmark-epr} & STRN \\ \hline
JSON Feed example document                              & \autoref{sec:benchmark-jsonfeed} & STNF \\ \hline
GitHub Workflow Definition                              & \autoref{sec:benchmark-githubworkflow} & STNN \\ \hline
GitHub FUNDING sponsorship definition (empty)           & \autoref{sec:benchmark-githubfundingblank} & SBRF \\ \hline
ECMAScript module loader definition                     & \autoref{sec:benchmark-esmrc} & SBNF \\ \hline
ESLint configuration document                           & \autoref{sec:benchmark-eslintrc} & LNRF \\ \hline
NPM Package.json Linter configuration manifest          & \autoref{sec:benchmark-packagejsonlintrc} & LTRF \\ \hline
.NET Core project.json                                  & \autoref{sec:benchmark-netcoreproject} & LTRN \\ \hline
NPM Package.json example manifest                       & \autoref{sec:benchmark-packagejson} & LTNF \\ \hline
JSON Resume                                             & \autoref{sec:benchmark-jsonresume} & LTNN \\ \hline
Nightwatch.js Test Framework Configuration              & \autoref{sec:benchmark-nightwatch} & LBRF \\

\bottomrule
\end{tabularx}
\end{table*}

\clearpage
\subsection{JSON-e Templating Engine Sort Example}
\label{sec:benchmark-jsonesort}

\begin{figure*}[ht!]
  \frame{\includegraphics[width=\linewidth]{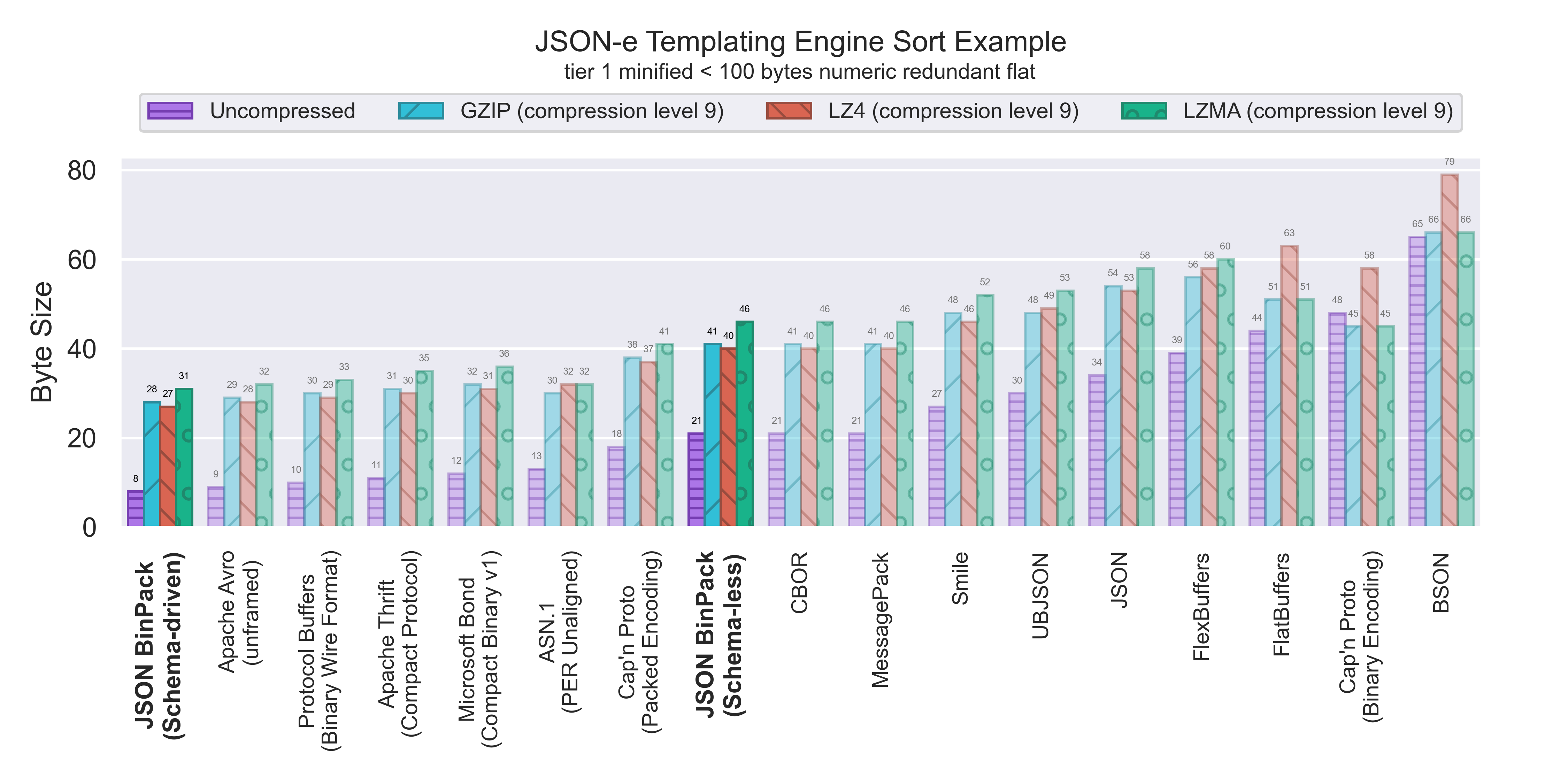}}
  \caption{The space-efficiency benchmark results for the JSON-e Templating Engine Sort Example test case selected from the SchemaStore open-source dataset test suite in \cite{viotti2022benchmark}.}
\label{fig:benchmark-jsonesort} \end{figure*}

JSON-e \footnote{\url{https://github.com/taskcluster/json-e}} is an open-source
JSON-based templating engine created by Mozilla as part of the TaskCluster
\footnote{\url{https://taskcluster.net}} project, the open-source task
execution framework that supports Mozilla's continuous integration and release
processes. In \autoref{fig:benchmark-jsonesort}, we demonstrate a \textbf{Tier
1 minified $<$ 100 bytes numeric redundant flat} (Tier 1 NRF from
\cite{viotti2022benchmark}) JSON document that consists of an example JSON-e
template definition to sort an array of numbers.

The smallest bit-string produced by JSON BinPack (Schema-driven) results in a
\textbf{11.1\%} (8 bytes) size reduction compared to the next best performing
specification: Apache Avro \cite{avro} (9 bytes).  JSON BinPack (Schema-driven)
achieves a \textbf{61.9\%} size reduction compared to the best performing
schema-less serialization specifications: JSON BinPack (Schema-less), CBOR
\cite{RFC7049} and MessagePack \cite{messagepack}.

Additionally, JSON BinPack (Schema-less) (21 bytes), along with CBOR
\cite{RFC7049} and MessagePack \cite{messagepack}, produce the smallest
bit-string for schema-less binary serialization specifications, resulting in a
\textbf{22.2\%} size reduction compared to the next best performing
specification: Smile \cite{smile} (27 bytes).


\textbf{Comparison to Uncompressed and Compressed JSON}. In comparison to JSON
\cite{ECMA-404} (34 bytes), JSON BinPack (Schema-driven) (8 bytes) and JSON
BinPack (Schema-less) (21 bytes) achieve a \textbf{76.4\%} and \textbf{38.2\%}
size reduction, respectively.  In comparison to best-case compressed JSON
\cite{ECMA-404} (53 bytes), JSON BinPack (Schema-driven) (8 bytes) and JSON
BinPack (Schema-less) (21 bytes) achieve a \textbf{84.9\%} and \textbf{60.3\%}
size reduction, respectively.

\clearpage
\subsection{JSON-e Templating Engine Reverse Sort Example}
\label{sec:benchmark-jsonereversesort}

\begin{figure*}[ht!]
  \frame{\includegraphics[width=\linewidth]{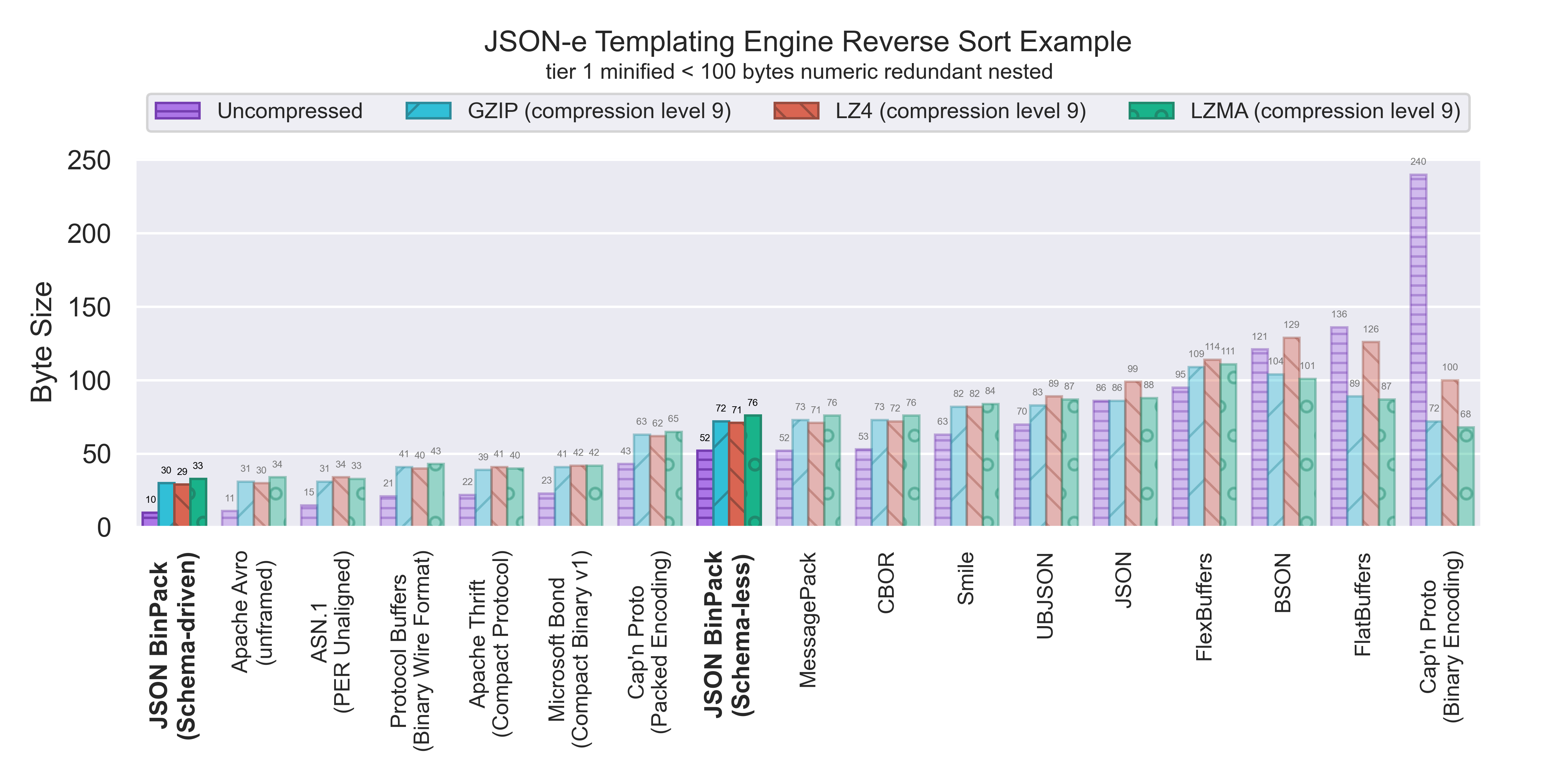}}
  \caption{The space-efficiency benchmark results for the JSON-e Templating Engine Reverse Sort Example test case selected from the SchemaStore open-source dataset test suite in \cite{viotti2022benchmark}.} \label{fig:benchmark-jsonereversesort}
\end{figure*}

JSON-e \footnote{\url{https://github.com/taskcluster/json-e}} is an open-source
JSON-based templating engine created by Mozilla as part of the TaskCluster
\footnote{\url{https://taskcluster.net}} project, the open-source task
execution framework that supports Mozilla's continuous integration and release
processes. In \autoref{fig:benchmark-jsonereversesort}, we demonstrate a
\textbf{Tier 1 minified $<$ 100 bytes numeric redundant nested} (Tier 1 NRN
from \cite{viotti2022benchmark}) JSON document that consists of an example
JSON-e template definition to sort and reverse an array of numbers.

The smallest bit-string produced by JSON BinPack (Schema-driven) (10 bytes)
results in a \textbf{9\%} size reduction compared to the next best performing
specification: Apache Avro \cite{avro} (11 bytes).  JSON BinPack
(Schema-driven) achieves a \textbf{80.7\%} size reduction compared to the best
performing schema-less serialization specifications: JSON BinPack (Schema-less)
and MessagePack \cite{messagepack}

Additionally, JSON BinPack (Schema-less) (52 bytes), along with MessagePack
\cite{messagepack}, produce the smallest bit-string for schema-less binary
serialization specifications, resulting in a \textbf{1.8\%} size reduction
compared to the next best performing specification: CBOR \cite{RFC7049} (53
bytes).


\textbf{Comparison to Uncompressed and Compressed JSON}. In comparison to both
JSON \cite{ECMA-404} (86 bytes) and best-case compressed JSON \cite{ECMA-404}
(86 bytes), JSON BinPack (Schema-driven) (10 bytes) and JSON BinPack
(Schema-less) (52 bytes) achieve a \textbf{88.3\%} and \textbf{39.5\%} size
reduction, respectively.

\clearpage
\subsection{CircleCI Definition (Blank)}
\label{sec:benchmark-circleciblank}

\begin{figure*}[ht!]
  \frame{\includegraphics[width=\linewidth]{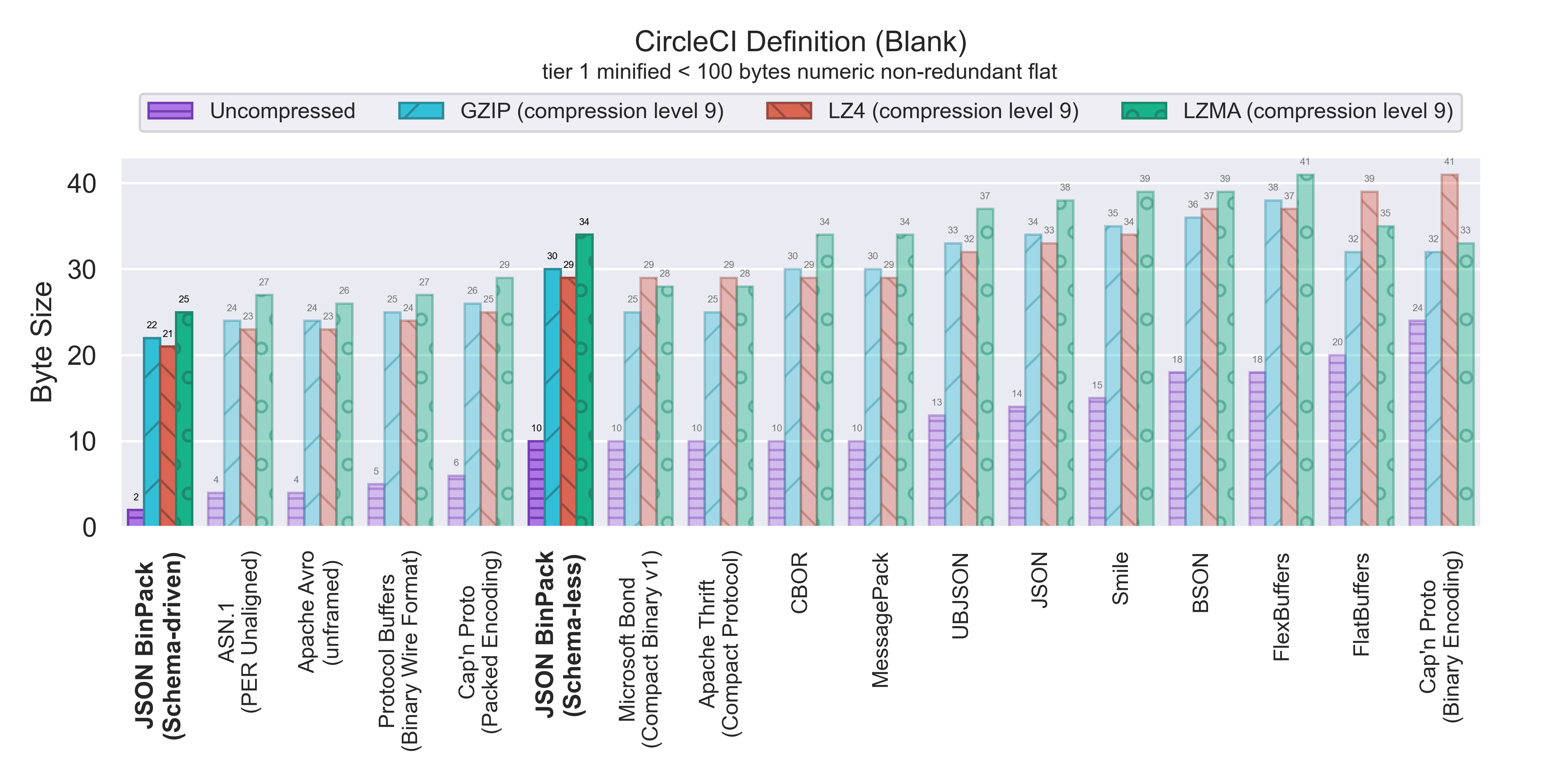}}
  \caption{The space-efficiency benchmark results for the CircleCI Definition (Blank) test case selected from the SchemaStore open-source dataset test suite in \cite{viotti2022benchmark}.}
\label{fig:benchmark-circleciblank} \end{figure*}

CircleCI \footnote{\url{https://circleci.com}} is a commercial cloud-provider
of continuous integration and deployment pipelines used by a wide range of
companies in the software development industry such as Facebook, Spotify, and
Heroku \footnote{\url{https://circleci.com/customers/}}. In
\autoref{fig:benchmark-circleciblank}, we demonstrate a \textbf{Tier 1
minified $<$ 100 bytes numeric non-redundant flat} (Tier 1 NNF from
\cite{viotti2022benchmark}) JSON document that represents a simple pipeline
configuration file for CircleCI that declares the desired CircleCI version
without defining any workflows.

The smallest bit-string produced by JSON BinPack (Schema-driven) (2 bytes)
results in a \textbf{50\%} size reduction compared to the next best performing
specifications: ASN.1 PER Unaligned \cite{asn1-per} and Apache Avro \cite{avro}
(4 bytes).  JSON BinPack (Schema-driven) achieves a \textbf{80\%} size
reduction compared to the best performing schema-less serialization
specifications: JSON BinPack (Schema-less) and CBOR \cite{RFC7049}.

Additionally, JSON BinPack (Schema-less) (10 bytes), along with CBOR
\cite{RFC7049} and MessagePack \cite{messagepack}, produce the smallest
bit-string for schema-less binary serialization specifications, resulting in a
\textbf{23\%} size reduction compared to the next best performing
specification: UBJSON \cite{ubjson} (13 bytes).


\textbf{Comparison to Uncompressed and Compressed JSON}. In comparison to JSON
\cite{ECMA-404} (14 bytes), JSON BinPack (Schema-driven) (2 bytes) and JSON
BinPack (Schema-less) (10 bytes) achieve a \textbf{85.7\%} and \textbf{28.5\%}
size reduction, respectively.  In comparison to best-case compressed JSON
\cite{ECMA-404} (33 bytes), JSON BinPack (Schema-driven) (2 bytes) and JSON
BinPack (Schema-less) (10 bytes) achieve a \textbf{93.9\%} and \textbf{69.6\%}
size reduction, respectively.

\clearpage
\subsection{CircleCI Matrix Definition}
\label{sec:benchmark-circlecimatrix}

\begin{figure*}[ht!]
  \frame{\includegraphics[width=\linewidth]{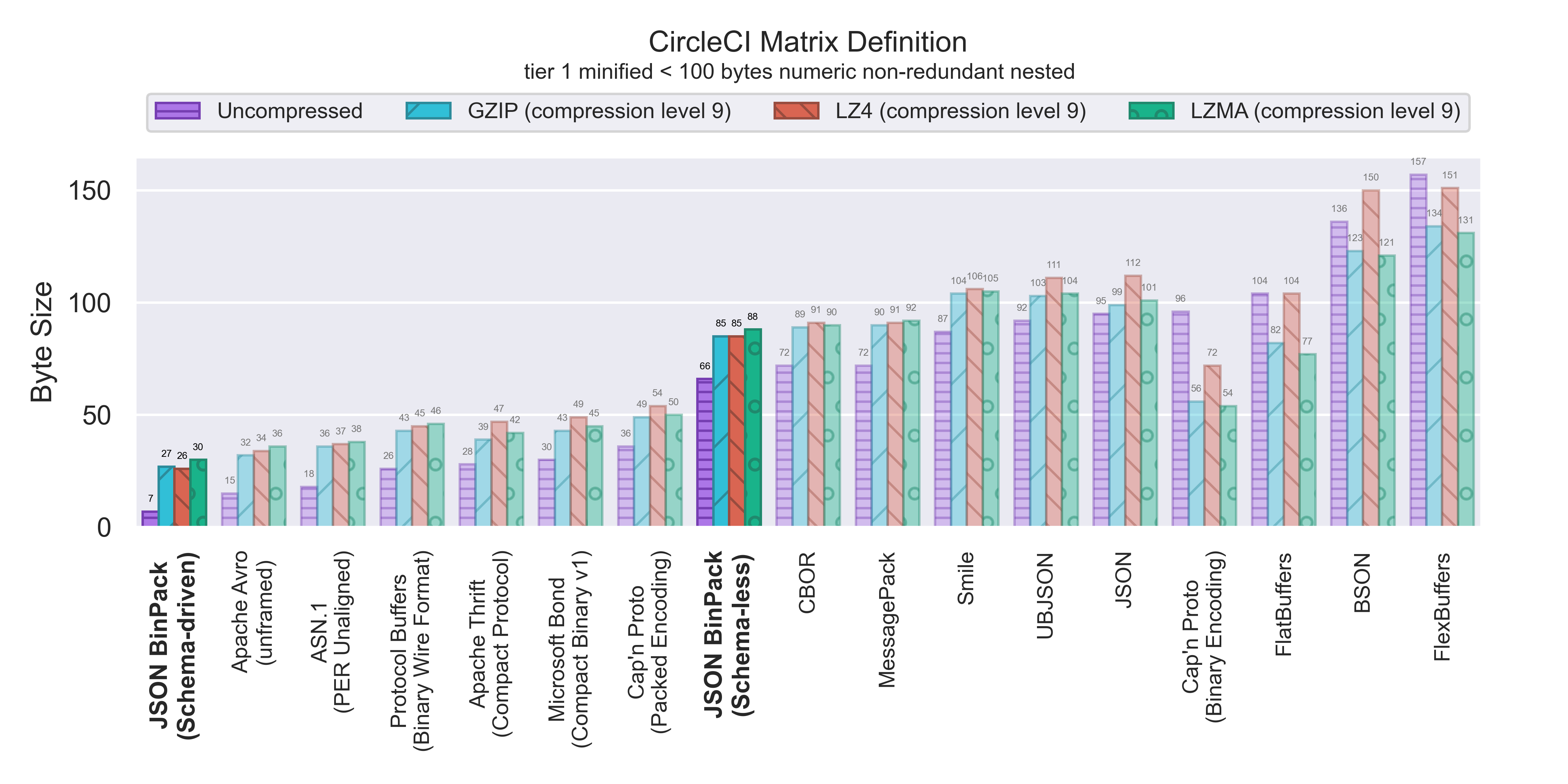}}
  \caption{The space-efficiency benchmark results for the CircleCI Matrix Definition test case selected from the SchemaStore open-source dataset test suite in \cite{viotti2022benchmark}.}
\label{fig:benchmark-circlecimatrix} \end{figure*}

CircleCI \footnote{\url{https://circleci.com}} is a commercial cloud-provider
of continuous integration and deployment pipelines used by a wide range of
companies in the software development industry such as Facebook, Spotify, and
Heroku \footnote{\url{https://circleci.com/customers/}}. In
\autoref{fig:benchmark-circlecimatrix}, we demonstrate a \textbf{Tier 1
minified $<$ 100 bytes numeric non-redundant nested} (Tier 1 NNN from
\cite{viotti2022benchmark}) JSON document that represents a pipeline
configuration file for CircleCI that declares the desired CircleCI version and
defines a workflow that contains a single blank matrix-based job.

The smallest bit-string produced by JSON BinPack (Schema-driven) (7 bytes)
results in a \textbf{53.3\%} size reduction compared to the next best
performing specification: Apache Avro \cite{avro} (15 bytes).  JSON BinPack
(Schema-driven) achieves a \textbf{89.3\%} size reduction compared to the best
performing schema-less serialization specification: JSON BinPack (Schema-less).

Additionally, JSON BinPack (Schema-less) (66 bytes) produces the smallest
bit-string for schema-less binary serialization specifications,
resulting in a \textbf{8.3\%} size reduction compared to the next best
performing specifications: CBOR \cite{RFC7049} and MessagePack
\cite{messagepack} (72 bytes).


\textbf{Comparison to Uncompressed and Compressed JSON}. In comparison to JSON
\cite{ECMA-404} (95 bytes), JSON BinPack (Schema-driven) (7 bytes) and JSON
BinPack (Schema-less) (66 bytes) achieve a \textbf{92.6\%} and \textbf{30.5\%}
size reduction, respectively.  In comparison to best-case compressed JSON
\cite{ECMA-404} (99 bytes) , JSON BinPack (Schema-driven) (7 bytes) and JSON
BinPack (Schema-less) (66 bytes) achieve a \textbf{92.9\%} and \textbf{33.3\%}
size reduction, respectively.

\clearpage
\subsection{Grunt.js Clean Task Definition}
\label{sec:benchmark-gruntcontribclean}

\begin{figure*}[ht!]
  \frame{\includegraphics[width=\linewidth]{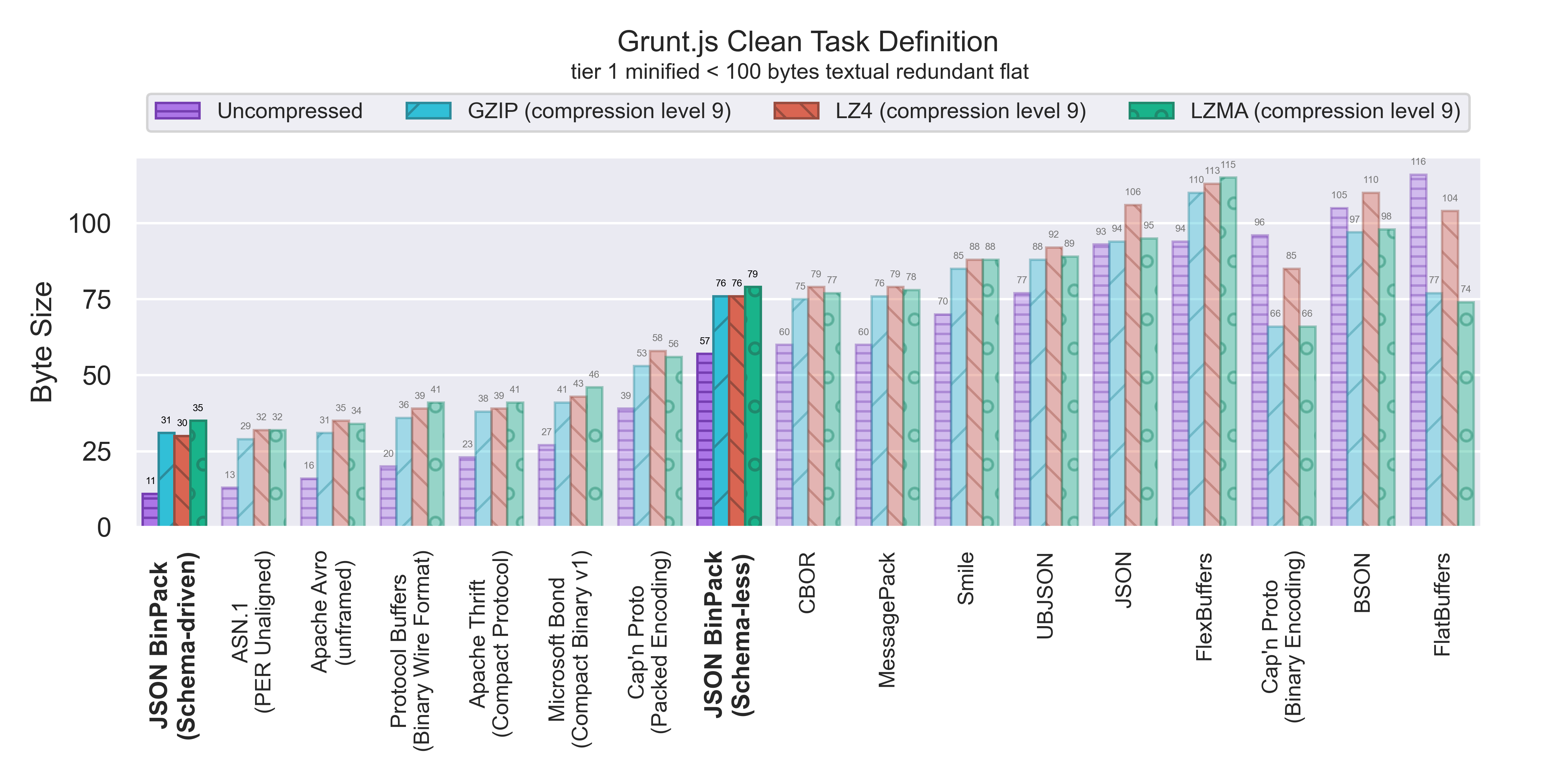}}
  \caption{The space-efficiency benchmark results for the Grunt.js Clean Task Definition test case selected from the SchemaStore open-source dataset test suite in \cite{viotti2022benchmark}.}
\label{fig:benchmark-gruntcontribclean} \end{figure*}

Grunt.js \footnote{\url{https://gruntjs.com}} is an open-source task runner for
the JavaScript \cite{ECMA-262} programming language used by a wide range of
companies in the software development industry such as Twitter, Adobe, and
Mozilla \footnote{\url{https://gruntjs.com/who-uses-grunt}}. In
\autoref{fig:benchmark-gruntcontribclean}, we demonstrate a \textbf{Tier 1
minified $<$ 100 bytes textual redundant flat} (Tier 1 TRF from
\cite{viotti2022benchmark}) JSON document that consists of an example
configuration for a built-in plugin to clear files and folders called
\texttt{grunt-contrib-clean}
\footnote{\url{https://github.com/gruntjs/grunt-contrib-clean}}.

The smallest bit-string produced by JSON BinPack (Schema-driven) (11 bytes)
results in a \textbf{15.3\%} size reduction compared to the next best
performing specification: ASN.1 PER Unaligned \cite{asn1-per} (13 bytes).  JSON
BinPack (Schema-driven) achieves a \textbf{80.7\%} size reduction compared to
the best performing schema-less serialization specification: JSON BinPack
(Schema-less).

Additionally, JSON BinPack (Schema-less) (57 bytes) produces the smallest
bit-string for schema-less binary serialization specifications, resulting in a
\textbf{5\%} size reduction compared to the next best performing specification:
CBOR \cite{RFC7049} and MessagePack \cite{messagepack} (60 bytes).


\textbf{Comparison to Uncompressed and Compressed JSON}. In comparison to JSON
\cite{ECMA-404} (93 bytes), JSON BinPack (Schema-driven) (11 bytes) and JSON
BinPack (Schema-less) (57 bytes) achieve a \textbf{88.1\%} and \textbf{38.7\%}
size reduction, respectively.  In comparison to best-case compressed JSON
\cite{ECMA-404} (94 bytes), JSON BinPack (Schema-driven) (11 bytes) and JSON
BinPack (Schema-less) (57 bytes) achieve a \textbf{88.2\%} and \textbf{39.3\%}
size reduction, respectively.

\clearpage
\subsection{CommitLint Configuration}
\label{sec:benchmark-commitlint}

\begin{figure*}[ht!]
  \frame{\includegraphics[width=\linewidth]{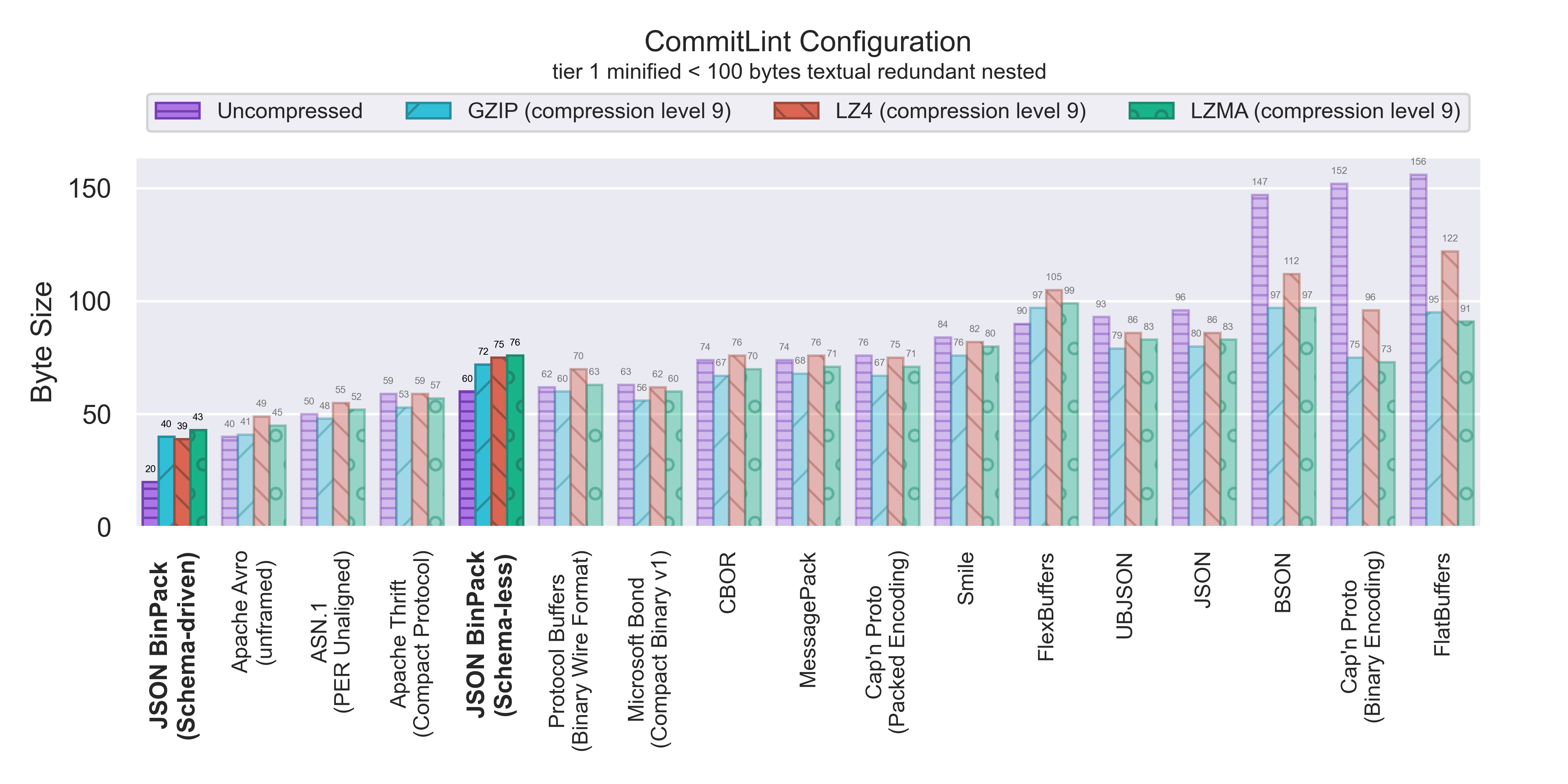}}
  \caption{The space-efficiency benchmark results for the CommitLint Configuration test case selected from the SchemaStore open-source dataset test suite in \cite{viotti2022benchmark}.}
\label{fig:benchmark-commitlint} \end{figure*}

CommitLint \footnote{\url{https://commitlint.js.org/\#/}} is an open-source
command-line tool to enforce version-control commit conventions in software
engineering projects. CommitLint is a community effort under the Conventional
Changelog \footnote{\url{https://github.com/conventional-changelog}}
organization formed by employees from companies including GitHub
\footnote{\url{https://github.com/zeke}} and Google
\footnote{\url{https://github.com/bcoe}}. In
\autoref{fig:benchmark-commitlint}, we demonstrate a \textbf{Tier 1 minified
$<$ 100 bytes textual redundant nested} (Tier 1 TRN from
\cite{viotti2022benchmark}) JSON document that represents a CommitLint
configuration file which declares that the subject and the scope of any commit
must be written in lower-case form.

The smallest bit-string produced by JSON BinPack (Schema-driven) (20 bytes)
results in a \textbf{50\%} size reduction compared to the next best performing
specification: Apache Avro \cite{avro} (40 bytes).  JSON BinPack
(Schema-driven) achieves a \textbf{66.6\%} size reduction compared to the best
performing schema-less serialization specification: JSON BinPack (Schema-less).

Additionally, JSON BinPack (Schema-less) (60 bytes) produces the smallest
bit-string for schema-less binary serialization specifications, resulting in a
\textbf{18.9\%} size reduction compared to the next best performing
specification: CBOR \cite{RFC7049} and MessagePack \cite{messagepack} (74
bytes).


\textbf{Comparison to Uncompressed and Compressed JSON}. In comparison to JSON
\cite{ECMA-404} (96 bytes), JSON BinPack (Schema-driven) (20 bytes) and JSON
BinPack (Schema-less) (60 bytes) achieve a \textbf{79.1\%} and \textbf{37.5\%}
size reduction, respectively.  In comparison to best-case compressed JSON
\cite{ECMA-404} (80 bytes), JSON BinPack (Schema-driven) (20 bytes) and JSON
BinPack (Schema-less) (60 bytes) achieve a \textbf{75\%} and \textbf{25\%} size
reduction, respectively.

\clearpage
\subsection{TSLint Linter Definition (Extends Only)}
\label{sec:benchmark-tslintextend}

\begin{figure*}[ht!]
  \frame{\includegraphics[width=\linewidth]{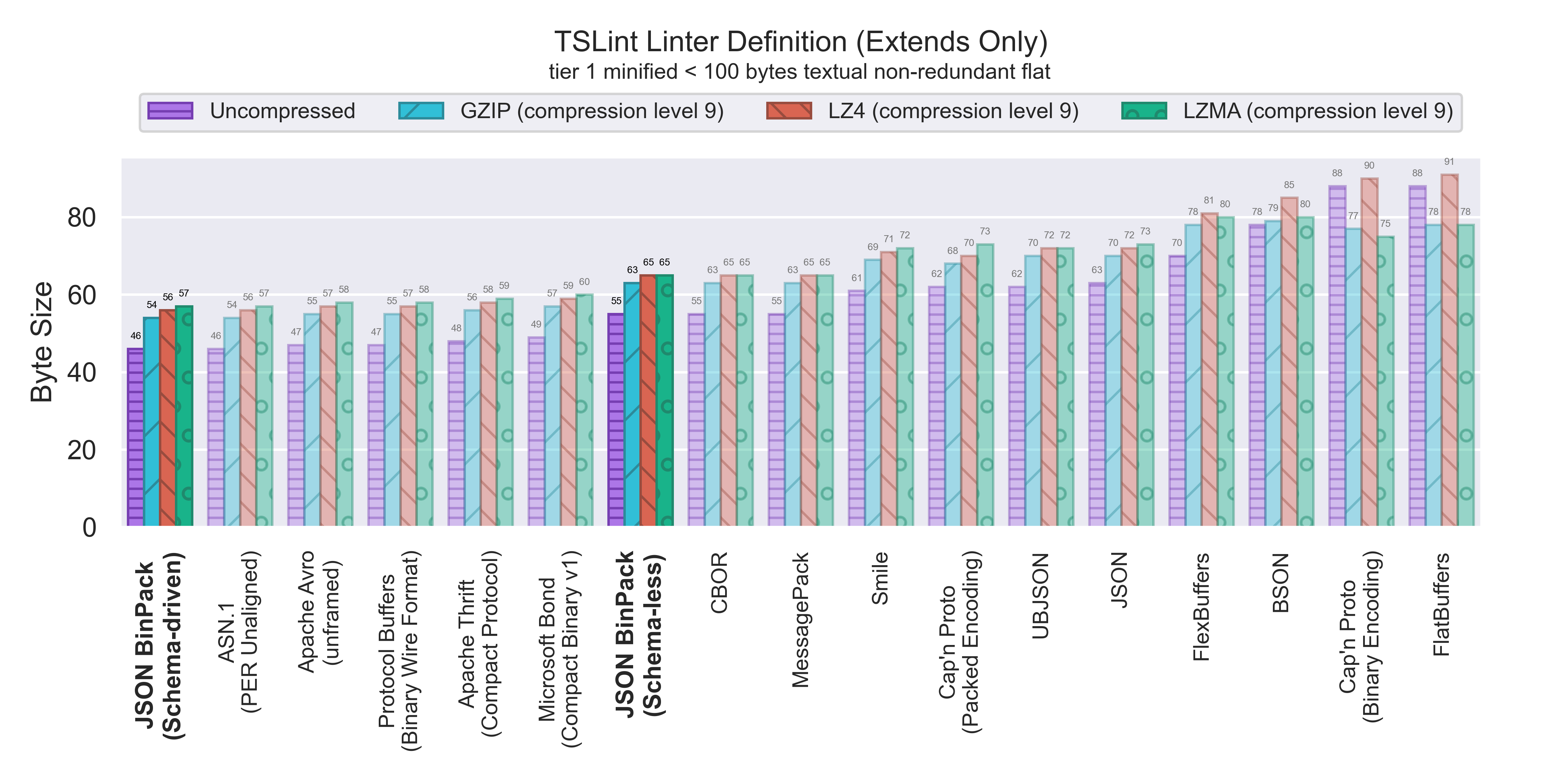}}
  \caption{The space-efficiency benchmark results for the TSLint Linter Definition (Extends Only) test case selected from the SchemaStore open-source dataset test suite in \cite{viotti2022benchmark}.}
\label{fig:benchmark-tslintextend} \end{figure*}

TSLint \footnote{\url{https://palantir.github.io/tslint}} is now an obsolete
open-source linter for the TypeScript
\footnote{\url{https://www.typescriptlang.org}} programming language. TSLint
was created by the Big Data analytics company Palantir
\footnote{\url{https://www.palantir.com}} and was merged with the ESLint
open-source JavaScript linter in 2019
\footnote{\url{https://github.com/palantir/tslint/issues/4534}}. In
\autoref{fig:benchmark-tslintextend}, we demonstrate a \textbf{Tier 1 minified
$<$ 100 bytes textual non-redundant flat} (Tier 1 TNF from
\cite{viotti2022benchmark}) JSON document that consists of a basic TSLint
configuration that only extends a set of existing TSLint configurations.

The smallest bit-string produced by both JSON BinPack (Schema-driven) and ASN.1
PER Unaligned \cite{asn1-per} (46 bytes) results in a \textbf{2.1\%} size
reduction compared to the next best performing specifications: Apache Avro
\cite{avro} and Protocol Buffers \cite{protocolbuffers} (47 bytes).  JSON
BinPack (Schema-driven) achieves a \textbf{16.3\%} size reduction compared to
the best performing schema-less serialization specifications: JSON BinPack
(Schema-less), CBOR \cite{RFC7049} and MessagePack \cite{messagepack}.

Additionally, JSON BinPack (Schema-less) (55 bytes), along with CBOR
\cite{RFC7049} and MessagePack \cite{messagepack}, produce the smallest
bit-string for schema-less binary serialization specifications, resulting in a
\textbf{9.8\%} size reduction compared to the next best performing
specification: Smile \cite{smile} (61 bytes).


\textbf{Comparison to Uncompressed and Compressed JSON}. In comparison to JSON
\cite{ECMA-404} (63 bytes), JSON BinPack (Schema-driven) (46 bytes) and JSON
BinPack (Schema-less) (55 bytes) achieve a \textbf{26.9\%} and \textbf{12.6\%}
size reduction, respectively.  In comparison to best-case compressed JSON
\cite{ECMA-404} (70 bytes), JSON BinPack (Schema-driven) (46 bytes) and JSON
BinPack (Schema-less) (55 bytes) achieve a \textbf{34.2\%} and \textbf{21.4\%}
size reduction, respectively.

\clearpage
\subsection{ImageOptimizer Azure Webjob Configuration}
\label{sec:benchmark-imageoptimizerwebjob}

\begin{figure*}[ht!]
  \frame{\includegraphics[width=\linewidth]{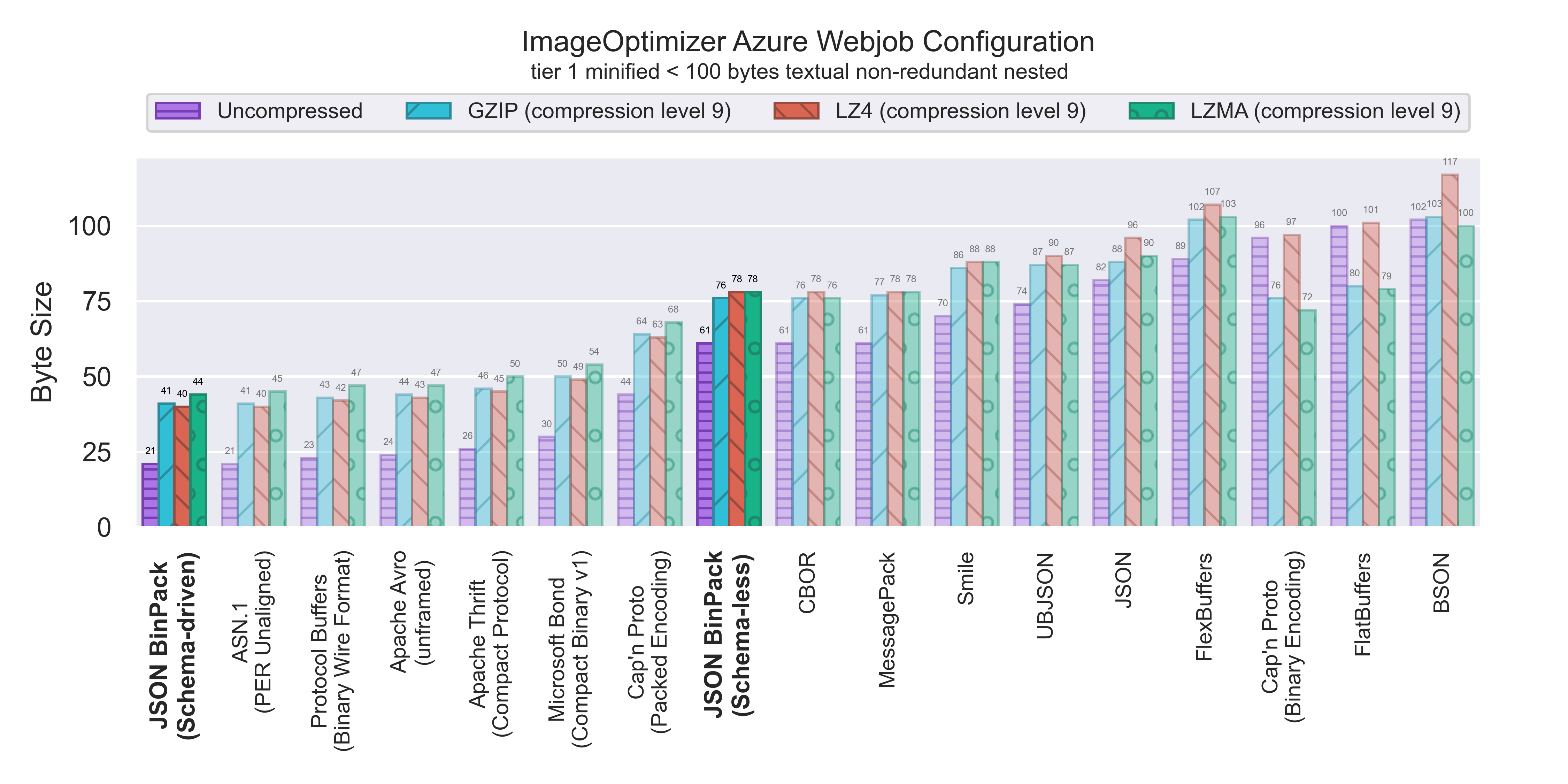}}
  \caption{The space-efficiency benchmark results for the ImageOptimizer Azure Webjob Configuration test case selected from the SchemaStore open-source dataset test suite in \cite{viotti2022benchmark}.}
\label{fig:benchmark-imageoptimizerwebjob} \end{figure*}

Image Optimizer
\footnote{\url{https://github.com/madskristensen/ImageOptimizerWebJob}} is an
Azure App Services WebJob
\footnote{\url{https://docs.microsoft.com/en-us/azure/app-service/webjobs-create}}
to compress website images used in the web development industry. In
\autoref{fig:benchmark-imageoptimizerwebjob}, we demonstrate a \textbf{Tier 1
minified $<$ 100 bytes textual non-redundant nested} (Tier 1 TNN from
\cite{viotti2022benchmark}) JSON document that consists of an Image
Optimizer configuration to perform lossy compression on images inside a
particular folder.

The smallest bit-string produced by both JSON BinPack (Schema-driven) and ASN.1
PER Unaligned \cite{asn1-per} (21 bytes) results in a \textbf{8.6\%} size
reduction compared to the next best performing specification: Protocol Buffers
\cite{protocolbuffers} (23 bytes).  JSON BinPack (Schema-driven) achieves a
\textbf{65.5\%} size reduction compared to the best performing schema-less
serialization specifications: JSON BinPack (Schema-less), CBOR \cite{RFC7049}
and MessagePack \cite{messagepack}.

Additionally, JSON BinPack (Schema-less) (61 bytes), along with CBOR
\cite{RFC7049} and MessagePack \cite{messagepack}, produce the smallest
bit-string for schema-less binary serialization specifications, resulting in a
\textbf{12.8\%} size reduction compared to the next best performing
specification: Smile \cite{smile} (70 bytes).


\textbf{Comparison to Uncompressed and Compressed JSON}. In comparison to JSON
\cite{ECMA-404} (82 bytes), JSON BinPack (Schema-driven) (21 bytes) and JSON
BinPack (Schema-less) (61 bytes) achieve a \textbf{74.3\%} and \textbf{25.6\%}
size reduction, respectively.  In comparison to best-case compressed JSON
\cite{ECMA-404} (88 bytes), JSON BinPack (Schema-driven) (21 bytes) and JSON
BinPack (Schema-less) (61 bytes) achieve a \textbf{76.1\%} and \textbf{30.6\%}
size reduction, respectively.

\clearpage
\subsection{SAP Cloud SDK Continuous Delivery Toolkit Configuration}
\label{sec:benchmark-sapcloudsdkpipeline}

\begin{figure*}[ht!]
  \frame{\includegraphics[width=\linewidth]{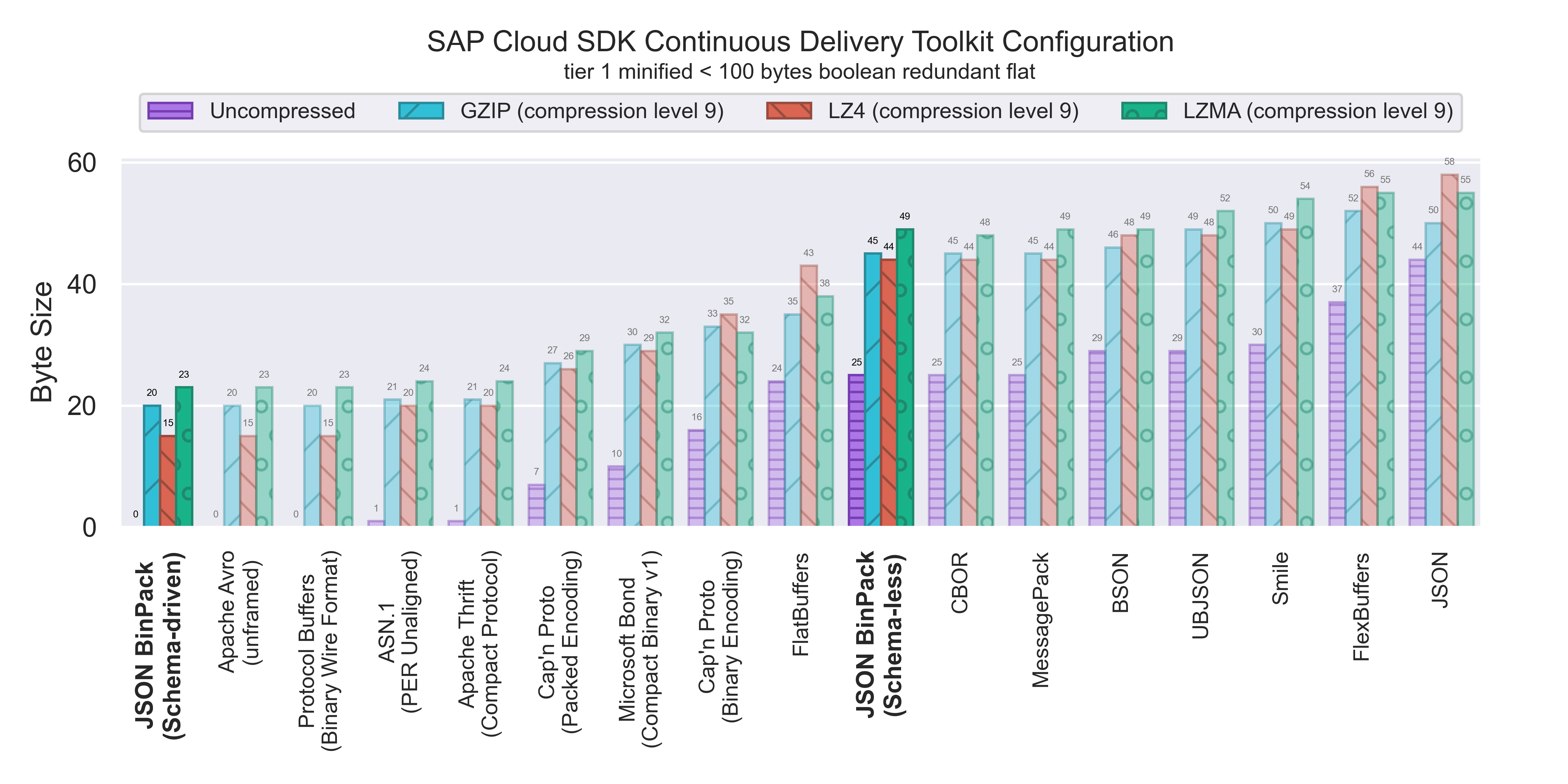}}
  \caption{The space-efficiency benchmark results for the SAP Cloud SDK Continuous Delivery Toolkit Configuration test case selected from the SchemaStore open-source dataset test suite in \cite{viotti2022benchmark}.}
\label{fig:benchmark-sapcloudsdkpipeline} \end{figure*}

SAP Cloud SDK \footnote{\url{https://sap.github.io/cloud-sdk/}} is a framework
that includes support for continuous integration and delivery pipelines to
develop applications for the SAP
\footnote{\url{https://www.sap.com/index.html}} enterprise resource planning
platform used by industries such as finance, healthcare and retail. In
\autoref{fig:benchmark-sapcloudsdkpipeline}, we demonstrate a \textbf{Tier 1
minified $<$ 100 bytes boolean redundant flat} (Tier 1 BRF from
\cite{viotti2022benchmark}) JSON document that defines a blank pipeline with
no declared steps.

The smallest bit-string produced by JSON BinPack (Schema-driven), Apache Avro
\cite{avro} and Protocol Buffers \cite{protocolbuffers} (0 bytes) results in a
\textbf{100\%} size reduction compared to the next best performing
specifications: ASN.1 PER Unaligned \cite{asn1-per} and Apache Thrift
\cite{slee2007thrift} (1 byte).

Additionally, JSON BinPack (Schema-less) (25 bytes), along with CBOR
\cite{RFC7049} and MessagePack \cite{messagepack}, produce the smallest
bit-string for schema-less binary serialization specifications, resulting in a
\textbf{13.7\%} size reduction compared to the next best performing
specifications: BSON \cite{bsonspec} and UBJSON \cite{ubjson} (29 bytes).


\textbf{Comparison to Uncompressed and Compressed JSON}. In comparison to JSON
\cite{ECMA-404} (44 bytes), JSON BinPack (Schema-driven) (0 bytes) and JSON
BinPack (Schema-less) (25 bytes) achieve a \textbf{100\%} and \textbf{43.1\%}
size reduction, respectively.  In comparison to best-case compressed JSON
\cite{ECMA-404} (50 bytes), JSON BinPack (Schema-driven) (0 bytes) and JSON
BinPack (Schema-less) (25 bytes) achieve a \textbf{100\%} and \textbf{50\%}
size reduction, respectively.

\clearpage
\subsection{TSLint Linter Definition (Multi-rule)}
\label{sec:benchmark-tslintmulti}

\begin{figure*}[ht!]
  \frame{\includegraphics[width=\linewidth]{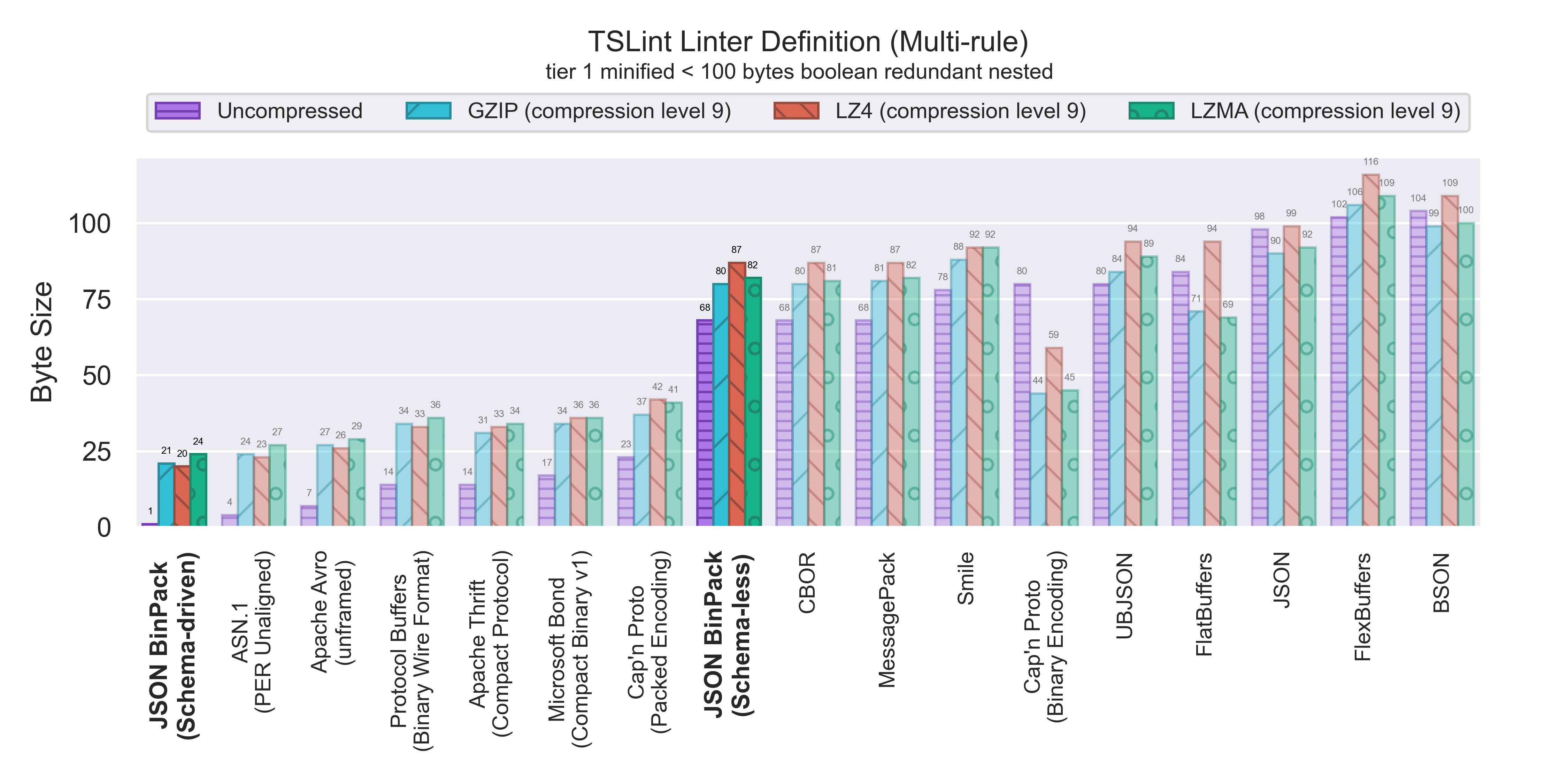}}
  \caption{The space-efficiency benchmark results for the TSLint Linter Definition (Multi-rule) test case selected from the SchemaStore open-source dataset test suite in \cite{viotti2022benchmark}.}
\label{fig:benchmark-tslintmulti} \end{figure*}

TSLint \footnote{\url{https://palantir.github.io/tslint}} is now an obsolete
open-source linter for the TypeScript
\footnote{\url{https://www.typescriptlang.org}} programming language. TSLint
was created by the Big Data analytics company Palantir
\footnote{\url{https://www.palantir.com}} and was merged with the ESLint
open-source JavaScript linter in 2019
\footnote{\url{https://github.com/palantir/tslint/issues/4534}}. In
\autoref{fig:benchmark-tslintmulti}, we demonstrate a \textbf{Tier 1 minified
$<$ 100 bytes boolean redundant nested} (Tier 1 BRN from
\cite{viotti2022benchmark}) JSON document that consists of a TSLint
configuration that enables and configures a set of built-in rules.

The smallest bit-string produced by JSON BinPack (Schema-driven) (1 byte)
results in a \textbf{75\%} size reduction compared to the next best performing
specification: ASN.1 PER Unaligned \cite{asn1-per} (4 bytes).  JSON BinPack
(Schema-driven) achieves a \textbf{98.5\%} size reduction compared to the best
performing schema-less serialization specifications: JSON BinPack (Schema-less),
CBOR \cite{RFC7049} and MessagePack \cite{messagepack}.

Additionally, JSON BinPack (Schema-less) (68 bytes), along with CBOR
\cite{RFC7049} and MessagePack \cite{messagepack}, produce the smallest
bit-string for schema-less binary serialization specifications, resulting in a
\textbf{12.8\%} size reduction compared to the next best performing
specification: Smile \cite{smile} (78 bytes).


\textbf{Comparison to Uncompressed and Compressed JSON}. In comparison to JSON
\cite{ECMA-404} (98 bytes), JSON BinPack (Schema-driven) (1 byte) and JSON
BinPack (Schema-less) (68 bytes) achieve a \textbf{98.9\%} and \textbf{30.6\%}
size reduction, respectively.  In comparison to best-case compressed JSON
\cite{ECMA-404} (90 bytes), JSON BinPack (Schema-driven) (1 byte) and JSON
BinPack (Schema-less) (68 bytes) achieve a \textbf{98.8\%} and \textbf{24.4\%}
size reduction, respectively.

\clearpage
\subsection{CommitLint Configuration (Basic)}
\label{sec:benchmark-commitlintbasic}

\begin{figure*}[ht!]
  \frame{\includegraphics[width=\linewidth]{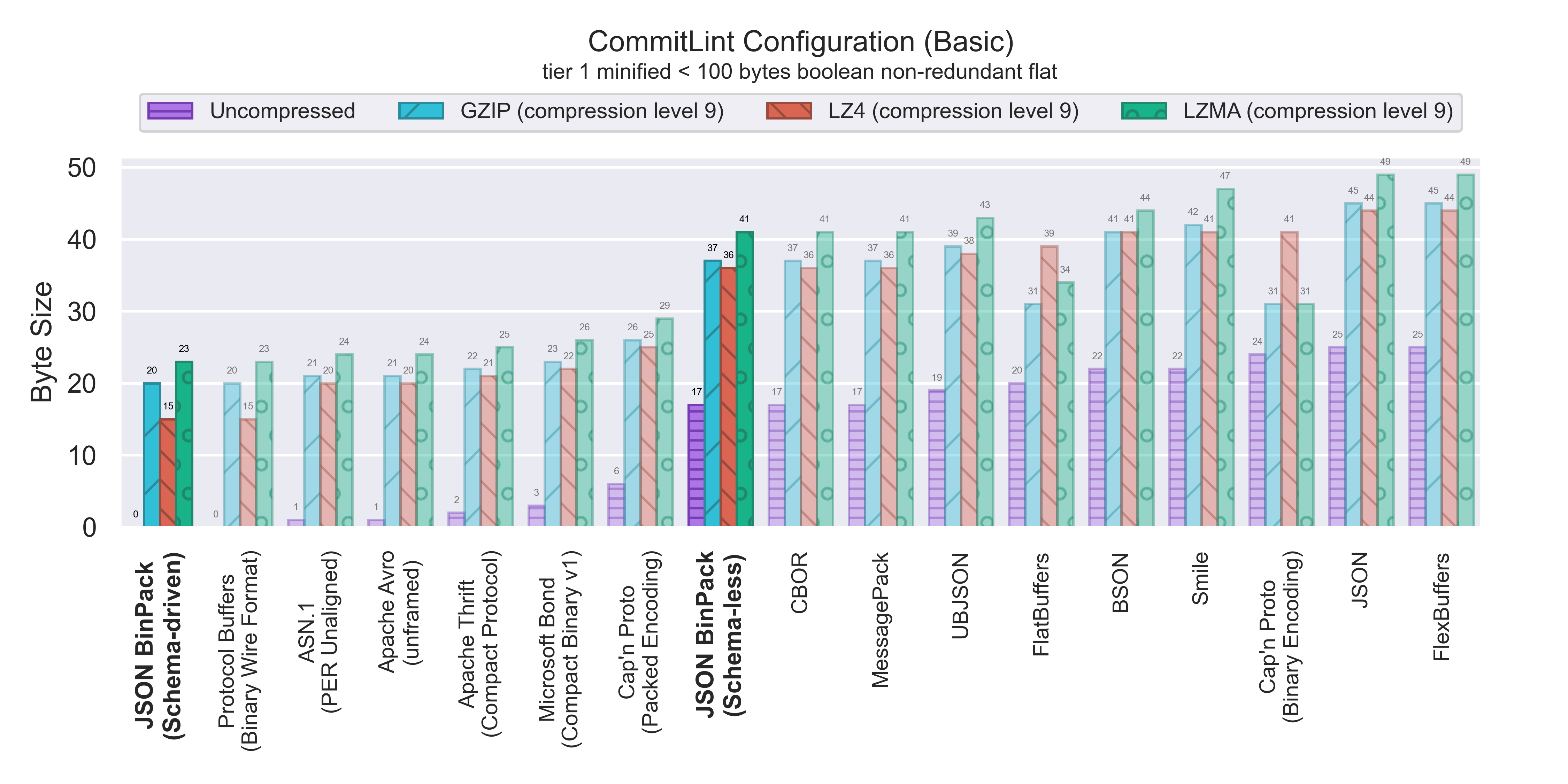}}
  \caption{The space-efficiency benchmark results for the CommitLint Configuration (Basic) test case selected from the SchemaStore open-source dataset test suite in \cite{viotti2022benchmark}.}
\label{fig:benchmark-commitlintbasic} \end{figure*}

CommitLint \footnote{\url{https://commitlint.js.org/\#/}} is an open-source
command-line tool to enforce version-control commit conventions in software
engineering projects. CommitLint is a community effort under the Conventional
Changelog \footnote{\url{https://github.com/conventional-changelog}}
organization formed by employees from companies including GitHub
\footnote{\url{https://github.com/zeke}} and Google
\footnote{\url{https://github.com/bcoe}}. In
\autoref{fig:benchmark-commitlintbasic}, we demonstrate a \textbf{Tier 1
minified $<$ 100 bytes boolean non-redundant flat} (Tier 1 BNF from
\cite{viotti2022benchmark}) JSON document that represents a CommitLint
configuration file which declares that CommitLint must not use its default
commit ignore rules.

The smallest bit-string produced by both JSON BinPack (Schema-driven) and
Protocol Buffers \cite{protocolbuffers} (0 bytes) results in a \textbf{100\%}
size reduction compared to the next best performing specifications: ASN.1 PER
Unaligned \cite{asn1-per} and Apache Avro \cite{avro} (1 byte).

Additionally, JSON BinPack (Schema-less) (17 bytes), along with CBOR
\cite{RFC7049} and MessagePack \cite{messagepack}, produce the smallest
bit-string for schema-less binary serialization specifications, resulting in a
\textbf{10.5\%} size reduction compared to the next best performing
specification: UBJSON \cite{ubjson} (19 bytes).


\textbf{Comparison to Uncompressed and Compressed JSON}. In comparison to JSON
\cite{ECMA-404} (25 bytes), JSON BinPack (Schema-driven) (0 bytes) and JSON
BinPack (Schema-less) (17 bytes) achieve a \textbf{100\%} and \textbf{32\%}
size reduction, respectively.  In comparison to best-case compressed JSON
\cite{ECMA-404} (44 bytes), JSON BinPack (Schema-driven) (0 bytes) and JSON
BinPack (Schema-less) (17 bytes) achieve a \textbf{100\%} and \textbf{61.3\%}
size reduction, respectively.

\clearpage
\subsection{TSLint Linter Definition (Basic)}
\label{sec:benchmark-tslintbasic}

\begin{figure*}[ht!]
  \frame{\includegraphics[width=\linewidth]{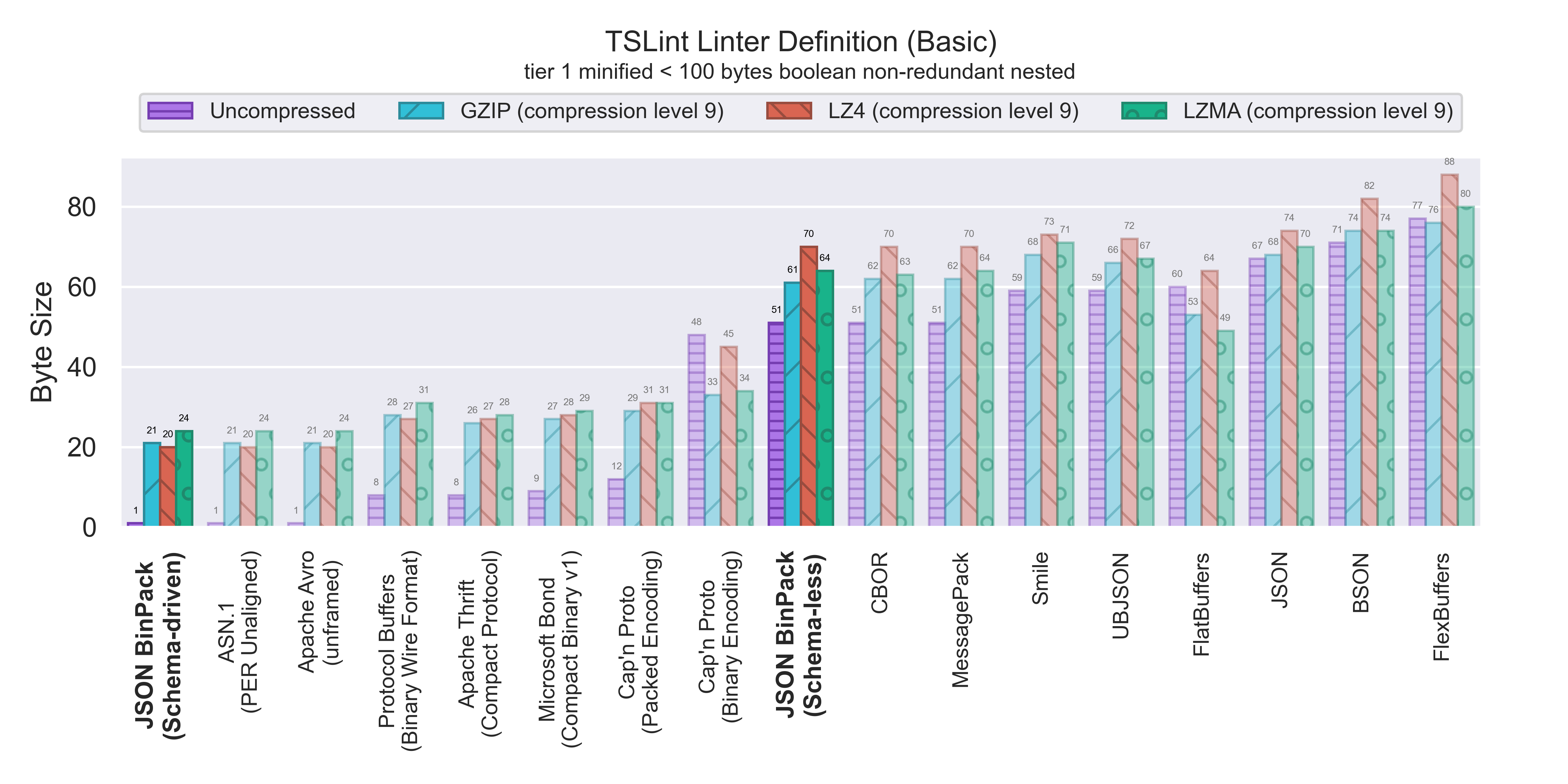}}
  \caption{The space-efficiency benchmark results for the TSLint Linter Definition (Basic) test case selected from the SchemaStore open-source dataset test suite in \cite{viotti2022benchmark}.}
\label{fig:benchmark-tslintbasic} \end{figure*}

TSLint \footnote{\url{https://palantir.github.io/tslint}} is now an obsolete
open-source linter for the TypeScript
\footnote{\url{https://www.typescriptlang.org}} programming language. TSLint
was created by the Big Data analytics company Palantir
\footnote{\url{https://www.palantir.com}} and was merged with the ESLint
open-source JavaScript linter in 2019
\footnote{\url{https://github.com/palantir/tslint/issues/4534}}. In
\autoref{fig:benchmark-tslintbasic}, we demonstrate a \textbf{Tier 1 minified
$<$ 100 bytes boolean non-redundant nested} (Tier 1 BNN from
\cite{viotti2022benchmark}) JSON document that consists of a basic TSLint
configuration that enforces grouped alphabetized imports.

The smallest bit-string produced by JSON BinPack (Schema-driven), ASN.1 PER
Unaligned \cite{asn1-per} and Apache Avro \cite{avro} (1 byte) results in a
\textbf{87.5\%} size reduction compared to the next best performing
specifications: Protocol Buffers \cite{protocolbuffers} and Apache Thrift
\cite{slee2007thrift} (8 bytes).  JSON BinPack (Schema-driven) achieves a
\textbf{98\%} size reduction compared to the best performing schema-less
serialization specifications: JSON BinPack (Schema-less), CBOR \cite{RFC7049}
and MessagePack \cite{messagepack}.

Additionally, JSON BinPack (Schema-less) (51 bytes), along with CBOR
\cite{RFC7049} and MessagePack \cite{messagepack}, produce the smallest
bit-string for schema-less binary serialization specifications, resulting in a
\textbf{13.5\%} size reduction compared to the next best performing
specification: Smile \cite{smile} and UBJSON \cite{ubjson} (59 bytes).


\textbf{Comparison to Uncompressed and Compressed JSON}. In comparison to JSON
\cite{ECMA-404} (67 bytes), JSON BinPack (Schema-driven) (1 byte) and JSON
BinPack (Schema-less) (51 bytes) achieve a \textbf{98.5\%} and \textbf{23.8\%}
size reduction, respectively.  In comparison to best-case compressed JSON
\cite{ECMA-404} (68 bytes), JSON BinPack (Schema-driven) (1 byte) and JSON
BinPack (Schema-less) (51 bytes) achieve a \textbf{98.5\%} and \textbf{25\%}
size reduction, respectively.

\clearpage
\subsection{GeoJSON Example Document}
\label{sec:benchmark-geojson}

\begin{figure*}[ht!]
  \frame{\includegraphics[width=\linewidth]{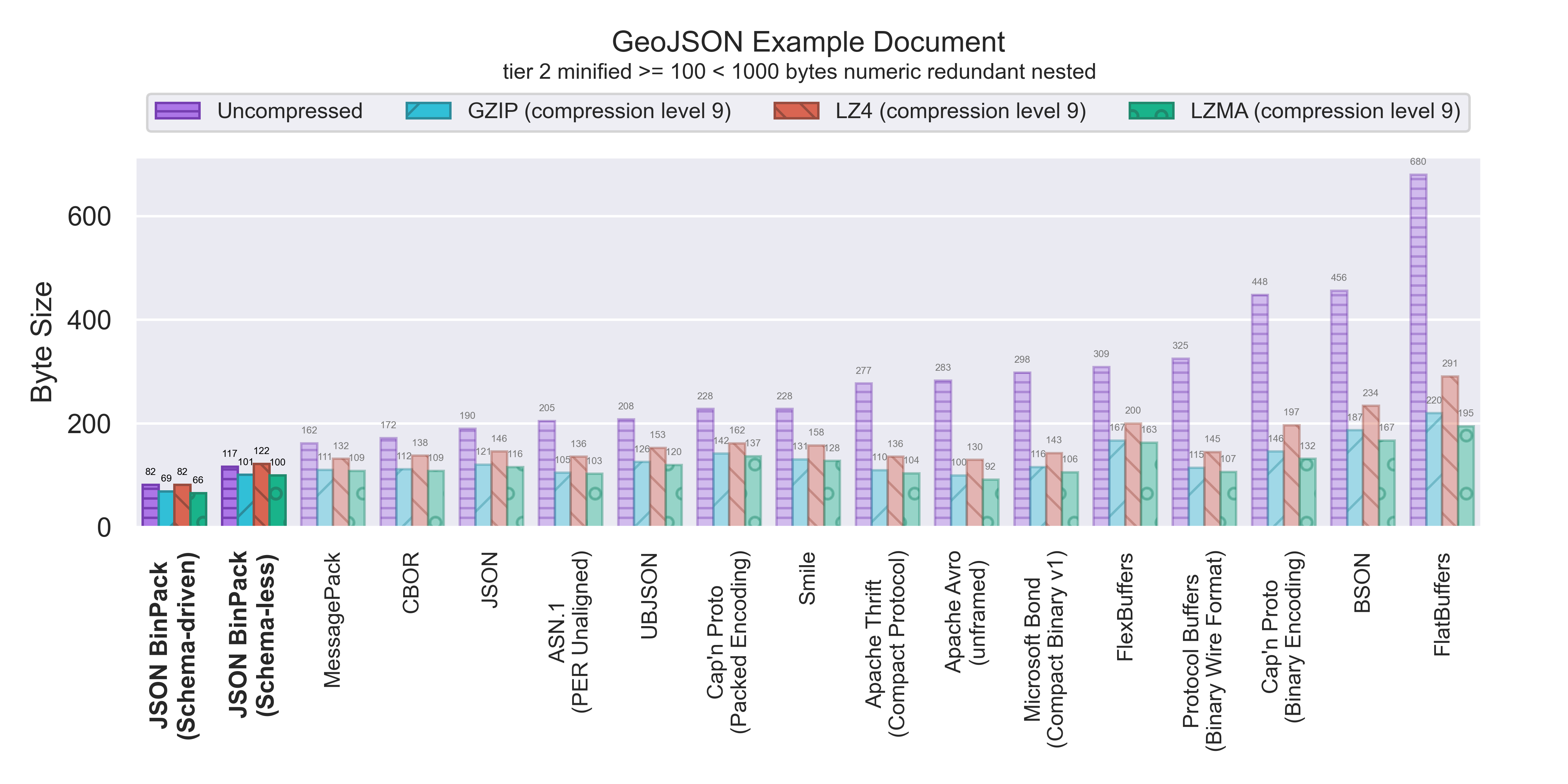}}
  \caption{The space-efficiency benchmark results for the GeoJSON Example Document test case selected from the SchemaStore open-source dataset test suite in \cite{viotti2022benchmark}.}
\label{fig:benchmark-geojson} \end{figure*}

GeoJSON \cite{RFC7946} is a standard to encode geospatial information using
JSON. GeoJSON is used in industries that have geographical and geospatial use
cases such as engineering, logistics and telecommunications. In
\autoref{fig:benchmark-geojson}, we demonstrate a \textbf{Tier 2 minified
$\geq$ 100 $<$ 1000 bytes numeric redundant nested} (Tier 2 NRN from
\cite{viotti2022benchmark}) JSON document that defines an example polygon
using the GeoJSON format.

The smallest bit-string produced by JSON BinPack (Schema-driven) (82 bytes)
results in a \textbf{29.9\%} size reduction compared to the next best
performing specification: JSON BinPack (Schema-less) (117 bytes).  JSON BinPack
(Schema-driven) achieves a \textbf{29.9\%} size reduction compared to the best
performing schema-less serialization specification: JSON BinPack (Schema-less).

Additionally, JSON BinPack (Schema-less) (117 bytes) produces the smallest
bit-string for schema-less binary serialization specifications, resulting in a
\textbf{27.7\%} size reduction compared to the next best performing
specification: MessagePack \cite{messagepack} (162 bytes).


\textbf{Comparison to Uncompressed and Compressed JSON}. In comparison to JSON
\cite{ECMA-404} (190 bytes), JSON BinPack (Schema-driven) (82 bytes) and JSON
BinPack (Schema-less) (117 bytes) achieve a \textbf{56.8\%} and \textbf{38.4\%}
size reduction, respectively.  In comparison to best-case compressed JSON
\cite{ECMA-404} (116 bytes), JSON BinPack (Schema-driven) (82 bytes) and JSON
BinPack (Schema-less) (117 bytes) achieve a \textbf{29.3\%} and negative
\textbf{0.8\%} size reduction, respectively.

\clearpage
\subsection{OpenWeatherMap API Example Document}
\label{sec:benchmark-openweathermap}

\begin{figure*}[ht!]
  \frame{\includegraphics[width=\linewidth]{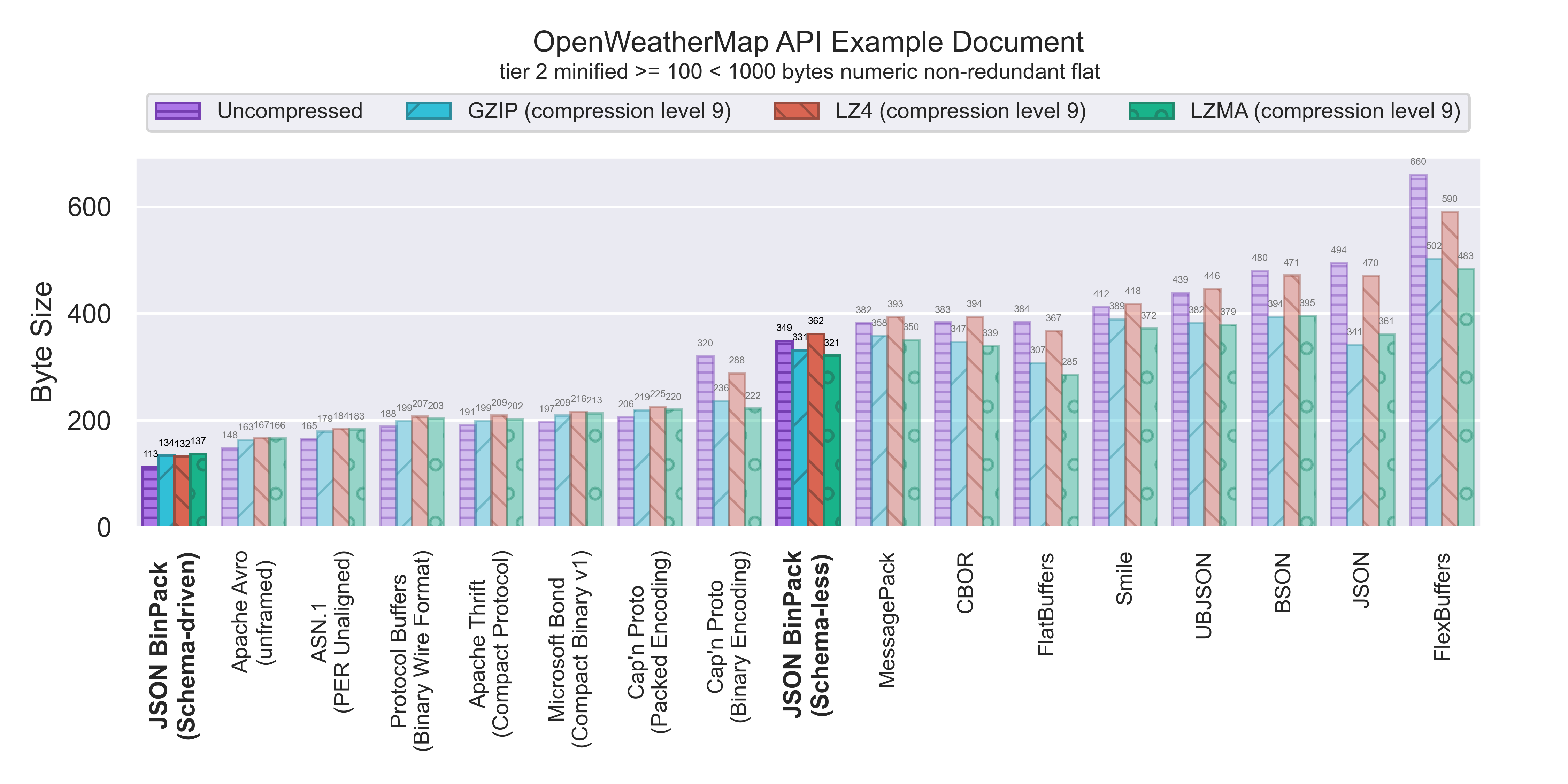}}
  \caption{The space-efficiency benchmark results for the OpenWeatherMap API Example Document test case selected from the SchemaStore open-source dataset test suite in \cite{viotti2022benchmark}.}
\label{fig:benchmark-openweathermap} \end{figure*}

OpenWeatherMap \footnote{\url{https://openweathermap.org}} is a weather data
and forecast API provider used in industries such as energy, agriculture,
transportation and construction. In \autoref{fig:benchmark-openweathermap}, we
demonstrate a \textbf{Tier 2 minified $\geq$ 100 $<$ 1000 bytes numeric
non-redundant flat} (Tier 2 NNF from \cite{viotti2022benchmark}) JSON
document that consists of an HTTP/1.1 \cite{RFC7231} response of the weather
information in Mountain View, California on June 12, 2019 at 2:44:05 PM GMT.

The smallest bit-string produced by JSON BinPack (Schema-driven) (113 bytes)
results in a \textbf{23.6\%} size reduction compared to the next best
performing specification: Apache Avro \cite{avro} (148 bytes).  JSON BinPack
(Schema-driven) achieves a \textbf{67.6\%} size reduction compared to the best
performing schema-less serialization specification: JSON BinPack (Schema-less).

Additionally, JSON BinPack (Schema-less) (349 bytes) produces the smallest
bit-string for schema-less binary serialization specifications, resulting in a
\textbf{8.6\%} size reduction compared to the next best performing
specification: MessagePack \cite{messagepack} (382 bytes).


\textbf{Comparison to Uncompressed and Compressed JSON}. In comparison to JSON
\cite{ECMA-404} (494 bytes), JSON BinPack (Schema-driven) (113 bytes) and JSON
BinPack (Schema-less) (349 bytes) achieve a \textbf{77.1\%} and \textbf{29.3\%}
size reduction, respectively.  In comparison to best-case compressed JSON
\cite{ECMA-404} (341 bytes), JSON BinPack (Schema-driven) (113 bytes) and JSON
BinPack (Schema-less) (349 bytes) achieve a \textbf{66.8\%} and negative
\textbf{2.3\%} size reduction, respectively.

\clearpage
\subsection{OpenWeather Road Risk API Example}
\label{sec:benchmark-openweatherroadrisk}

\begin{figure*}[ht!]
  \frame{\includegraphics[width=\linewidth]{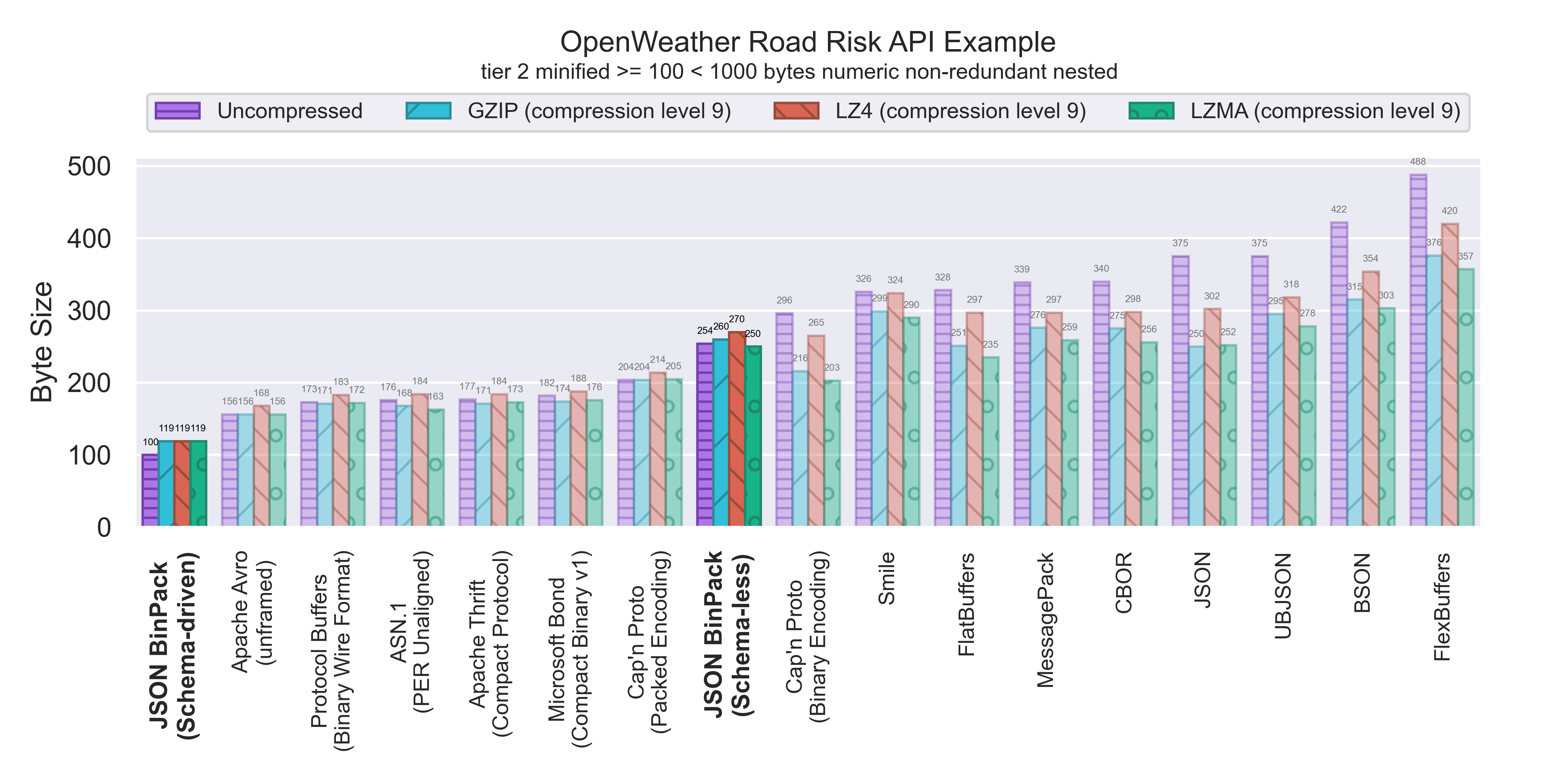}}
  \caption{The space-efficiency benchmark results for the OpenWeatherMap Road Risk API Example test case selected from the SchemaStore open-source dataset test suite in \cite{viotti2022benchmark}.}
\label{fig:benchmark-openweatherroadrisk} \end{figure*}

OpenWeatherMap \footnote{\url{https://openweathermap.org}} is a weather data
and forecast API provider used in industries such as energy, agriculture,
transportation and construction. In
\autoref{fig:benchmark-openweatherroadrisk}, we demonstrate a \textbf{Tier 2
minified $\geq$ 100 $<$ 1000 bytes numeric non-redundant nested} (Tier 2 NNN
from \cite{viotti2022benchmark}) JSON document that consists of an example
HTTP/1.1 \cite{RFC7231} Road Risk API response from the official API
documentation that provides weather data and national alerts along a specific
route.

The smallest bit-string produced by JSON BinPack (Schema-driven) (100 bytes)
results in a \textbf{35.8\%} size reduction compared to the next best
performing specification: Apache Avro \cite{avro} (156 bytes).  JSON BinPack
(Schema-driven) achieves a \textbf{60.6\%} size reduction compared to the best
performing schema-less serialization specification: JSON BinPack (Schema-less).

Additionally, JSON BinPack (Schema-less) (254 bytes) produce the smallest
bit-string for schema-less binary serialization specifications, resulting in a
\textbf{22\%} size reduction compared to the next best performing
specification: Smile \cite{smile} (326 bytes).


\textbf{Comparison to Uncompressed and Compressed JSON}. In comparison to JSON
\cite{ECMA-404} (375 bytes), JSON BinPack (Schema-driven) (100 bytes) and JSON
BinPack (Schema-less) (254 bytes) achieve a \textbf{73.3\%} and \textbf{32.2\%}
size reduction, respectively.  In comparison to best-case compressed JSON
\cite{ECMA-404} (250 bytes), JSON BinPack (Schema-driven) (100 bytes) and JSON
BinPack (Schema-less) (254 bytes) achieve a \textbf{60\%} and negative
\textbf{1.6\%} size reduction, respectively.

\clearpage
\subsection{TravisCI Notifications Configuration}
\label{sec:benchmark-travisnotifications}

\begin{figure*}[ht!]
  \frame{\includegraphics[width=\linewidth]{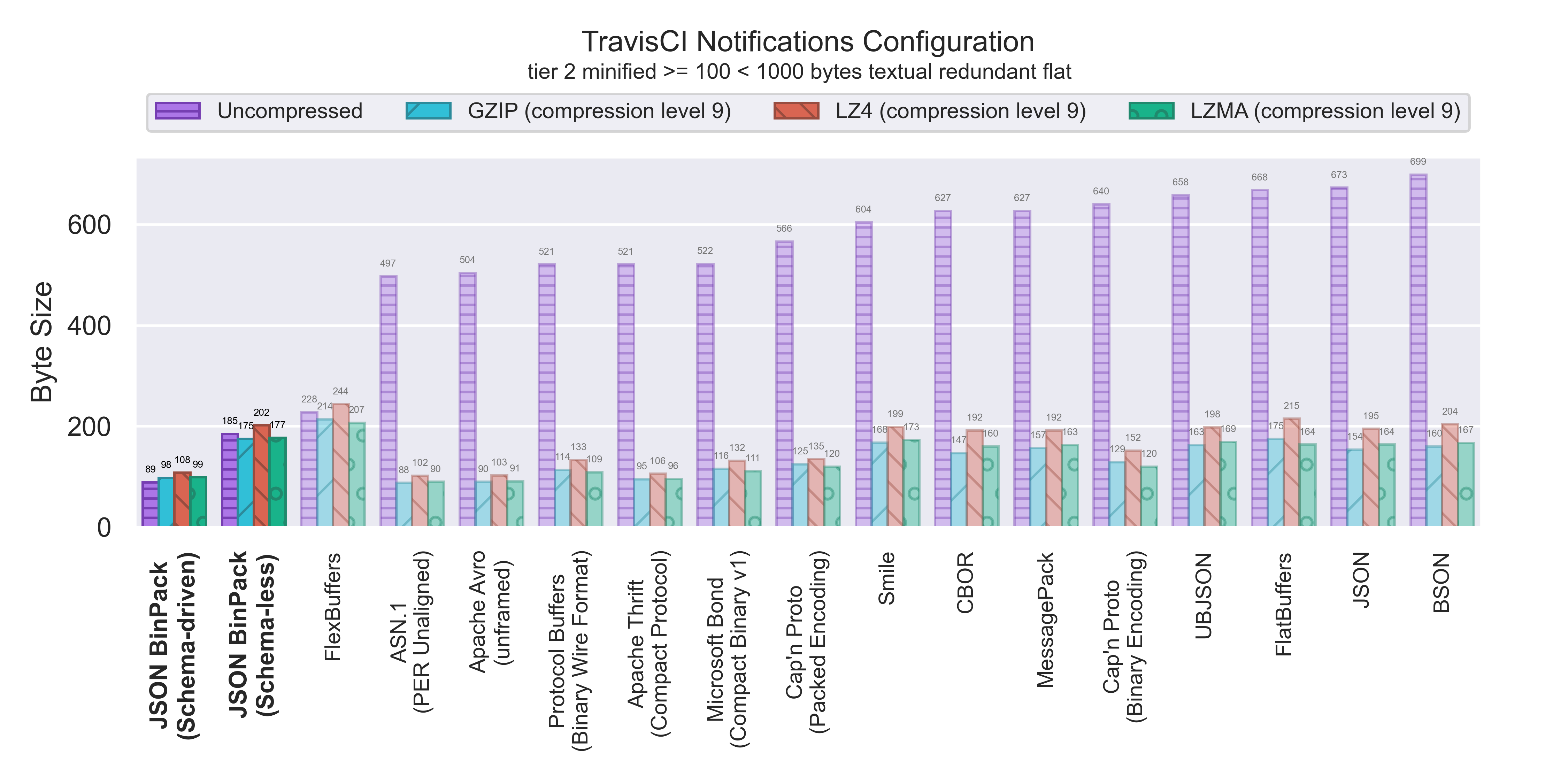}}
  \caption{The space-efficiency benchmark results for the TravisCI Notifications Configuration test case selected from the SchemaStore open-source dataset test suite in \cite{viotti2022benchmark}.}
\label{fig:benchmark-travisnotifications} \end{figure*}

TravisCI \footnote{\url{https://travis-ci.com}} is a commercial cloud-provider
of continuous integration and deployment pipelines  used by a wide range of
companies in the software development industry such as ZenDesk, BitTorrent, and
Engine Yard. In \autoref{fig:benchmark-travisnotifications}, we demonstrate a
\textbf{Tier 2 minified $\geq$ 100 $<$ 1000 bytes textual redundant flat} (Tier
2 TRF from \cite{viotti2022benchmark}) JSON document that consists of an
example pipeline configuration for TravisCI that declares a set of credentials
to post build notifications to various external services.

The smallest bit-string produced by JSON BinPack (Schema-driven) (89 bytes)
results in a \textbf{51.8\%} size reduction compared to the next best
performing specification: JSON BinPack (Schema-less) (185 bytes).  JSON BinPack
(Schema-driven) achieves a \textbf{51.8\%} size reduction compared to the best
performing schema-less serialization specification: JSON BinPack (Schema-less).

Additionally, JSON BinPack (Schema-less) (185 bytes) produces the smallest
bit-string for schema-less binary serialization specifications, resulting in a
\textbf{18.8\%} size reduction compared to the next best performing
specification: FlexBuffers \cite{flexbuffers} (228 bytes).


\textbf{Comparison to Uncompressed and Compressed JSON}. In comparison to JSON
\cite{ECMA-404} (673 bytes), JSON BinPack (Schema-driven) (89 bytes) and JSON
BinPack (Schema-less) (185 bytes) achieve a \textbf{86.7\%} and \textbf{72.5\%}
size reduction, respectively.  In comparison to best-case compressed JSON
\cite{ECMA-404} (154 bytes), JSON BinPack (Schema-driven) (89 bytes) and JSON
BinPack (Schema-less) (185 bytes) achieve a \textbf{42.2\%} and negative
\textbf{20.1\%} size reduction, respectively.

\clearpage
\subsection{Entry Point Regulation Manifest}
\label{sec:benchmark-epr}

\begin{figure*}[ht!]
  \frame{\includegraphics[width=\linewidth]{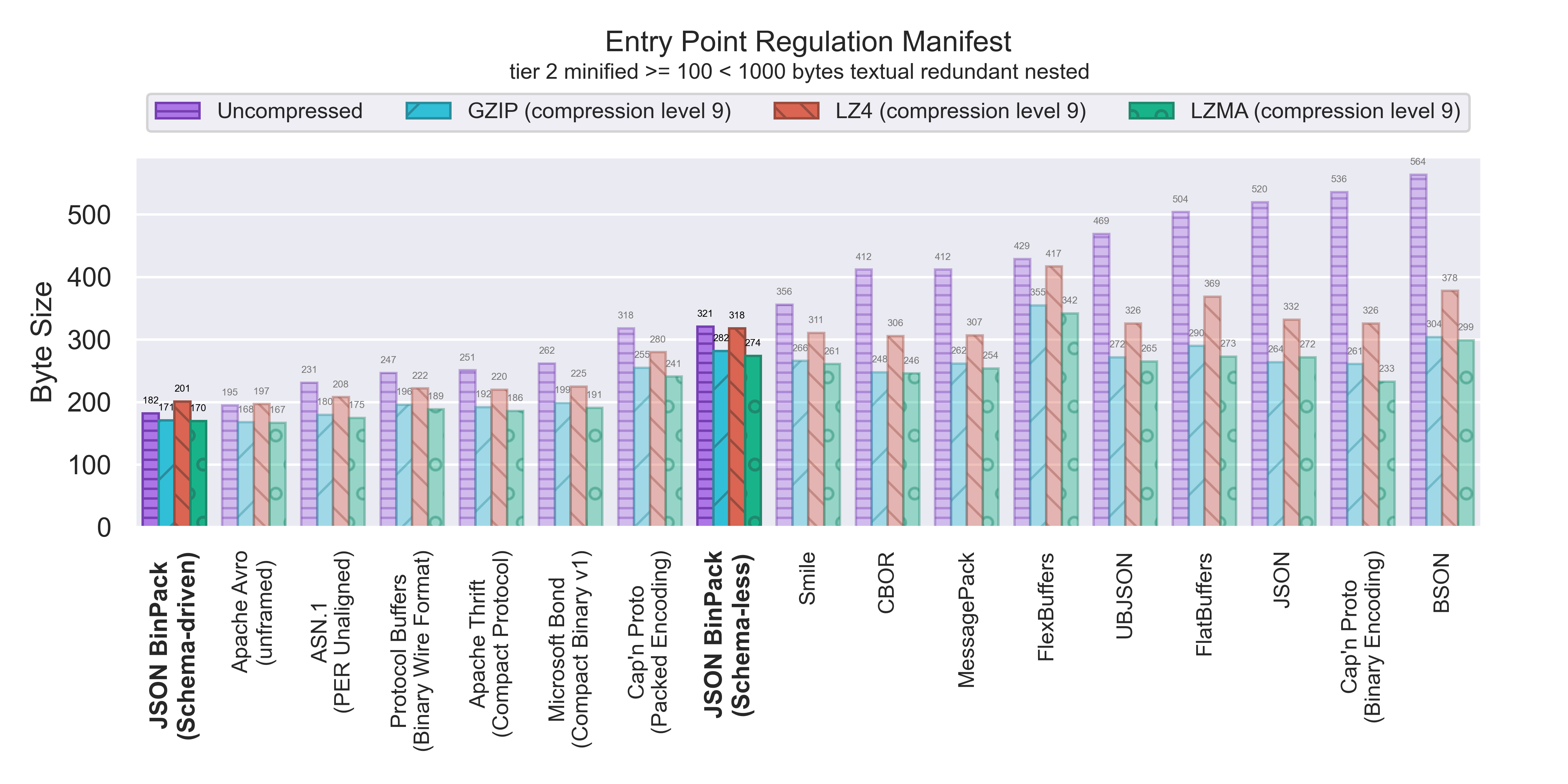}}
  \caption{The space-efficiency benchmark results for the Entry Point Regulation Manifest test case selected from the SchemaStore open-source dataset test suite in \cite{viotti2022benchmark}.}
\label{fig:benchmark-epr} \end{figure*}

Entry Point Regulation (EPR) \cite{EPR} is a W3C proposal led by Google that
defines a manifest that protects websites against cross-site scripting attacks
by allowing the developer to mark the areas of the application that can be
externally referenced. EPR manifests are used in the web industry. In
\autoref{fig:benchmark-epr}, we demonstrate a \textbf{Tier 2 minified $\geq$
100 $<$ 1000 bytes textual redundant nested} (Tier 2 TRN from
\cite{viotti2022benchmark}) JSON document that defines an example EPR policy
for a fictitious website.

The smallest bit-string produced by JSON BinPack (Schema-driven) (182 bytes)
results in a \textbf{6.6\%} size reduction compared to the next best performing
specification: Apache Avro \cite{avro} (195 bytes).  JSON BinPack
(Schema-driven) achieves a \textbf{43.3\%} size reduction compared to the best
performing schema-less serialization specification: JSON BinPack (Schema-less).

Additionally, JSON BinPack (Schema-less) (321 bytes) produces the smallest
bit-string for schema-less binary serialization specifications, resulting in a
\textbf{9.8\%} size reduction compared to the next best performing
specification: Smile \cite{smile} (356 bytes).


\textbf{Comparison to Uncompressed and Compressed JSON}. In comparison to JSON
\cite{ECMA-404} (520 bytes), JSON BinPack (Schema-driven) (182 bytes) and JSON
BinPack (Schema-less) (321 bytes) achieve a \textbf{65\%} and \textbf{38.2\%}
size reduction, respectively.  In comparison to best-case compressed JSON
\cite{ECMA-404} (264 bytes), JSON BinPack (Schema-driven) (182 bytes) and JSON
BinPack (Schema-less) (321 bytes) achieve a \textbf{31\%} and negative
\textbf{21.5\%} size reduction, respectively.

\clearpage
\subsection{JSON Feed Example Document}
\label{sec:benchmark-jsonfeed}

\begin{figure*}[ht!]
  \frame{\includegraphics[width=\linewidth]{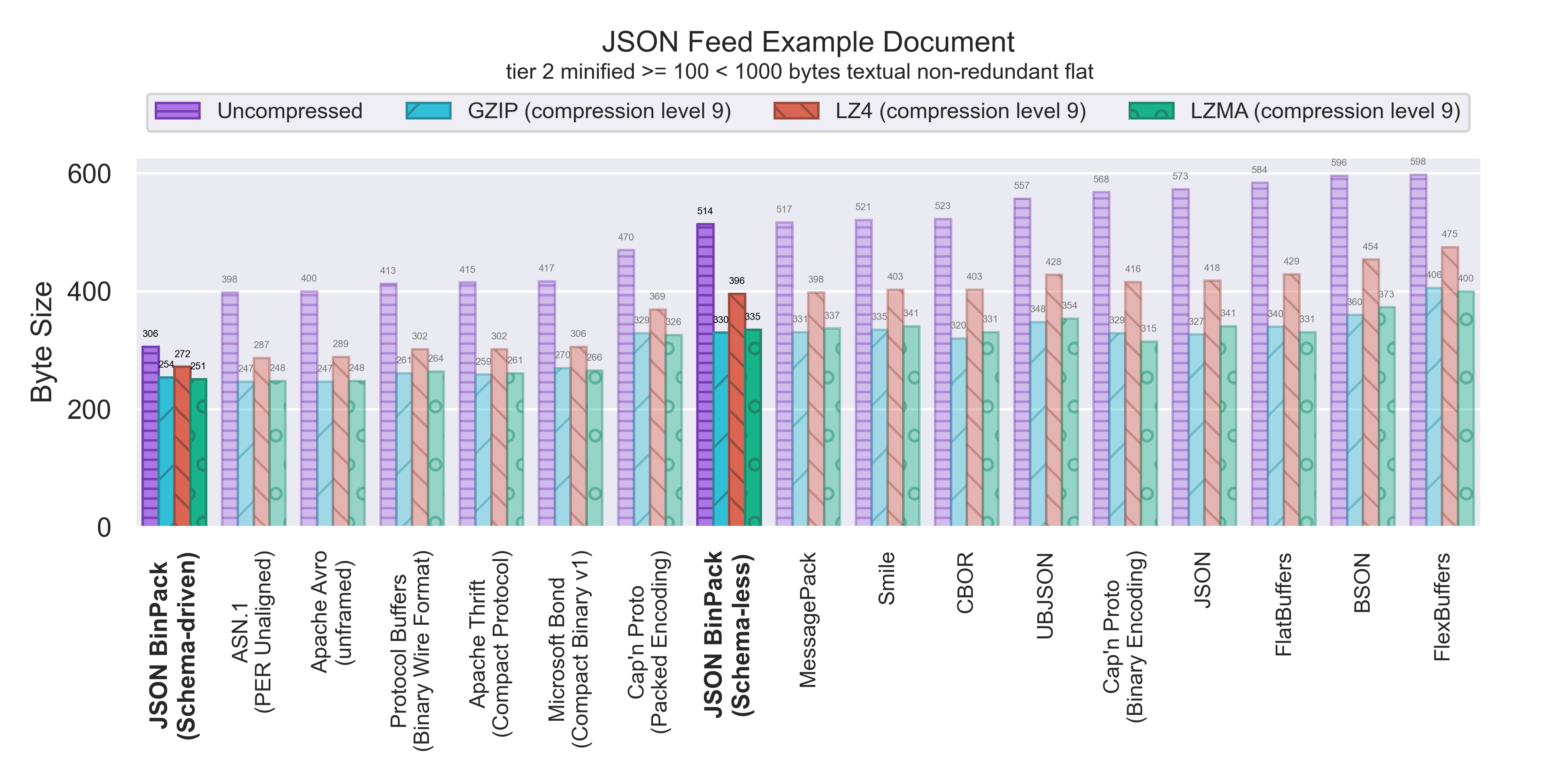}}
  \caption{The space-efficiency benchmark results for the JSON Feed Example Document test case selected from the SchemaStore open-source dataset test suite in \cite{viotti2022benchmark}.}
\label{fig:benchmark-jsonfeed} \end{figure*}

JSON Feed \cite{jsonfeed} is a specification for a syndication JSON format
similar to RSS \cite{rss} and Atom \cite{RFC4287} used in the publishing
\footnote{\url{https://micro.blog}} and media
\footnote{\url{https://npr.codes/npr-now-supports-json-feed-1c8af29d0ce7}}
industries. In \autoref{fig:benchmark-jsonfeed}, we demonstrate a \textbf{Tier
2 minified $\geq$ 100 $<$ 1000 bytes textual non-redundant flat} (Tier 2 TNF
from \cite{viotti2022benchmark}) JSON document that consists of a JSON Feed
manifest for an example website that contains a single blog entry.

The smallest bit-string produced by JSON BinPack (Schema-driven) (306 bytes)
results in a \textbf{23.1\%} size reduction compared to the next best
performing specification: ASN.1 PER Unaligned \cite{asn1-per} (398 bytes).
JSON BinPack (Schema-driven) achieves a \textbf{40.4\%} size reduction compared
to the best performing schema-less serialization specification: JSON BinPack
(Schema-less).

Additionally, JSON BinPack (Schema-less) (514 bytes) produces the smallest
bit-string for schema-less binary serialization specifications, resulting in a
\textbf{0.5\%} size reduction compared to the next best performing
specification: MessagePack \cite{messagepack} (517 bytes).


\textbf{Comparison to Uncompressed and Compressed JSON}. In comparison to JSON
\cite{ECMA-404} (573 bytes), JSON BinPack (Schema-driven) (306 bytes) and JSON
BinPack (Schema-less) (514 bytes) achieve a \textbf{46.5\%} and \textbf{10.2\%}
size reduction, respectively.  In comparison to best-case compressed JSON
\cite{ECMA-404} (327 bytes), JSON BinPack (Schema-driven) (306 bytes) and JSON
BinPack (Schema-less) (514 bytes) achieve a \textbf{6.4\%} and negative
\textbf{57.1\%} size reduction, respectively.

\clearpage
\subsection{GitHub Workflow Definition}
\label{sec:benchmark-githubworkflow}

\begin{figure*}[ht!]
  \frame{\includegraphics[width=\linewidth]{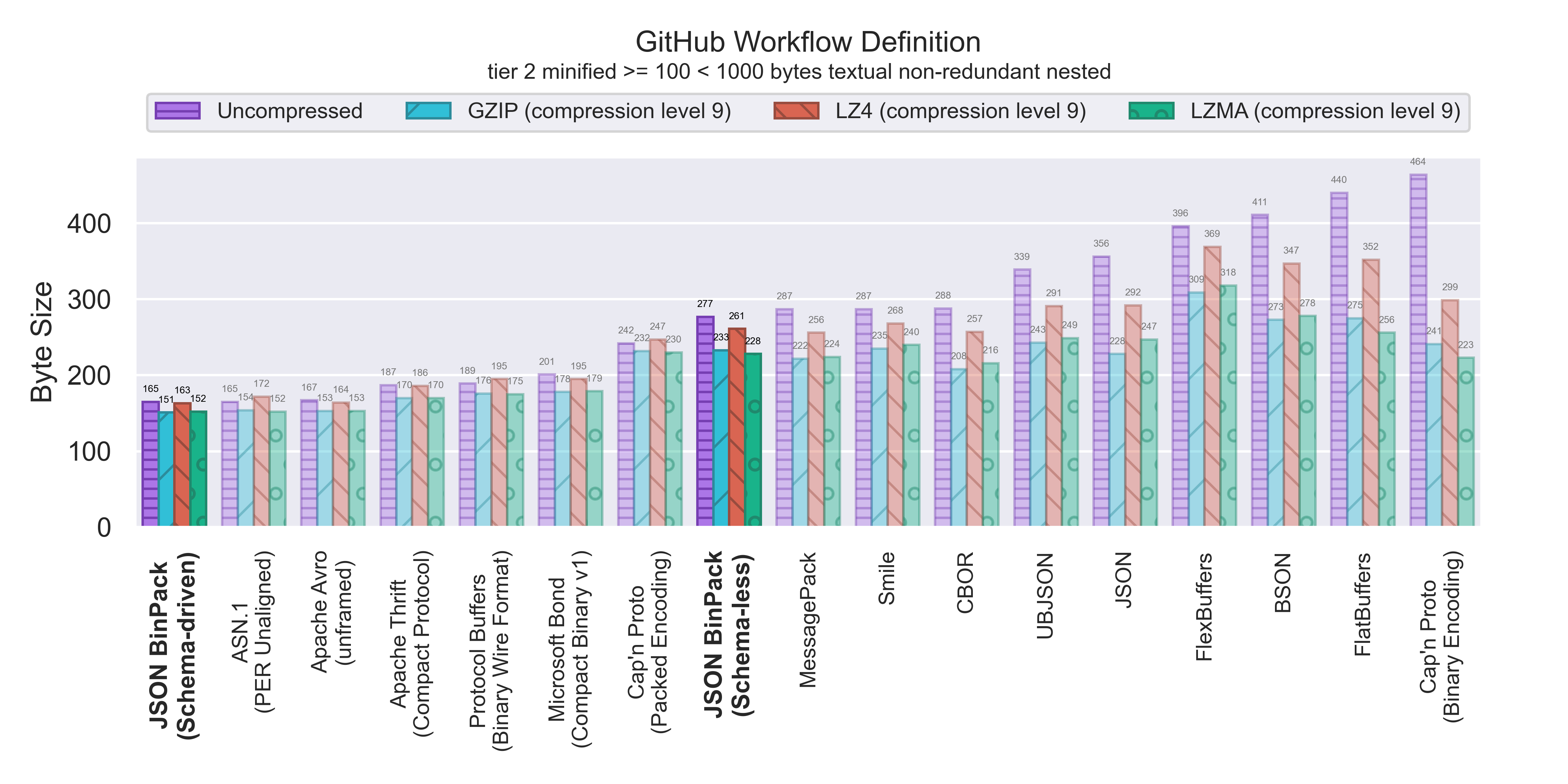}}
  \caption{The space-efficiency benchmark results for the GitHub Workflow Definition test case selected from the SchemaStore open-source dataset test suite in \cite{viotti2022benchmark}.}
\label{fig:benchmark-githubworkflow} \end{figure*}

The GitHub \footnote{\url{https://github.com}} software hosting provider has an
automation service called GitHub Actions
\footnote{\url{https://github.com/features/actions}} for projects to define
custom workflows. GitHub Actions is used primarily by the open-source software
industry. In \autoref{fig:benchmark-githubworkflow}, we demonstrate a
\textbf{Tier 2 minified $\geq$ 100 $<$ 1000 bytes textual non-redundant nested}
(Tier 2 TNN from \cite{viotti2022benchmark}) JSON document that consists of
a simple example workflow definition.

The smallest bit-string produced by both JSON BinPack (Schema-driven) and ASN.1
PER Unaligned \cite{asn1-per} (165 bytes) results in a \textbf{1.1\%} size
reduction compared to the next best performing specification: Apache Avro
\cite{avro} (167 bytes).  JSON BinPack (Schema-driven) achieves a
\textbf{40.4\%} size reduction compared to the best performing schema-less
serialization specification: JSON BinPack (Schema-less).

Additionally, JSON BinPack (Schema-less) (227 bytes) produces the smallest
bit-string for schema-less binary serialization specifications, resulting in a
\textbf{3.4\%} size reduction compared to the next best performing
specifications: MessagePack \cite{messagepack} and Smile \cite{smile} (287
bytes).


\textbf{Comparison to Uncompressed and Compressed JSON}. In comparison to JSON
\cite{ECMA-404} (356 bytes), JSON BinPack (Schema-driven) (165 bytes) and JSON
BinPack (Schema-less) (227 bytes) achieve a \textbf{53.6\%} and \textbf{22.1\%}
size reduction, respectively.  In comparison to best-case compressed JSON
\cite{ECMA-404} (228 bytes), JSON BinPack (Schema-driven) (165 bytes) and JSON
BinPack (Schema-less) (227 bytes) achieve a \textbf{27.6\%} and negative
\textbf{21.4\%} size reduction, respectively.

\clearpage
\subsection{GitHub FUNDING Sponsorship Definition (Empty)}
\label{sec:benchmark-githubfundingblank}

\begin{figure*}[ht!]
  \frame{\includegraphics[width=\linewidth]{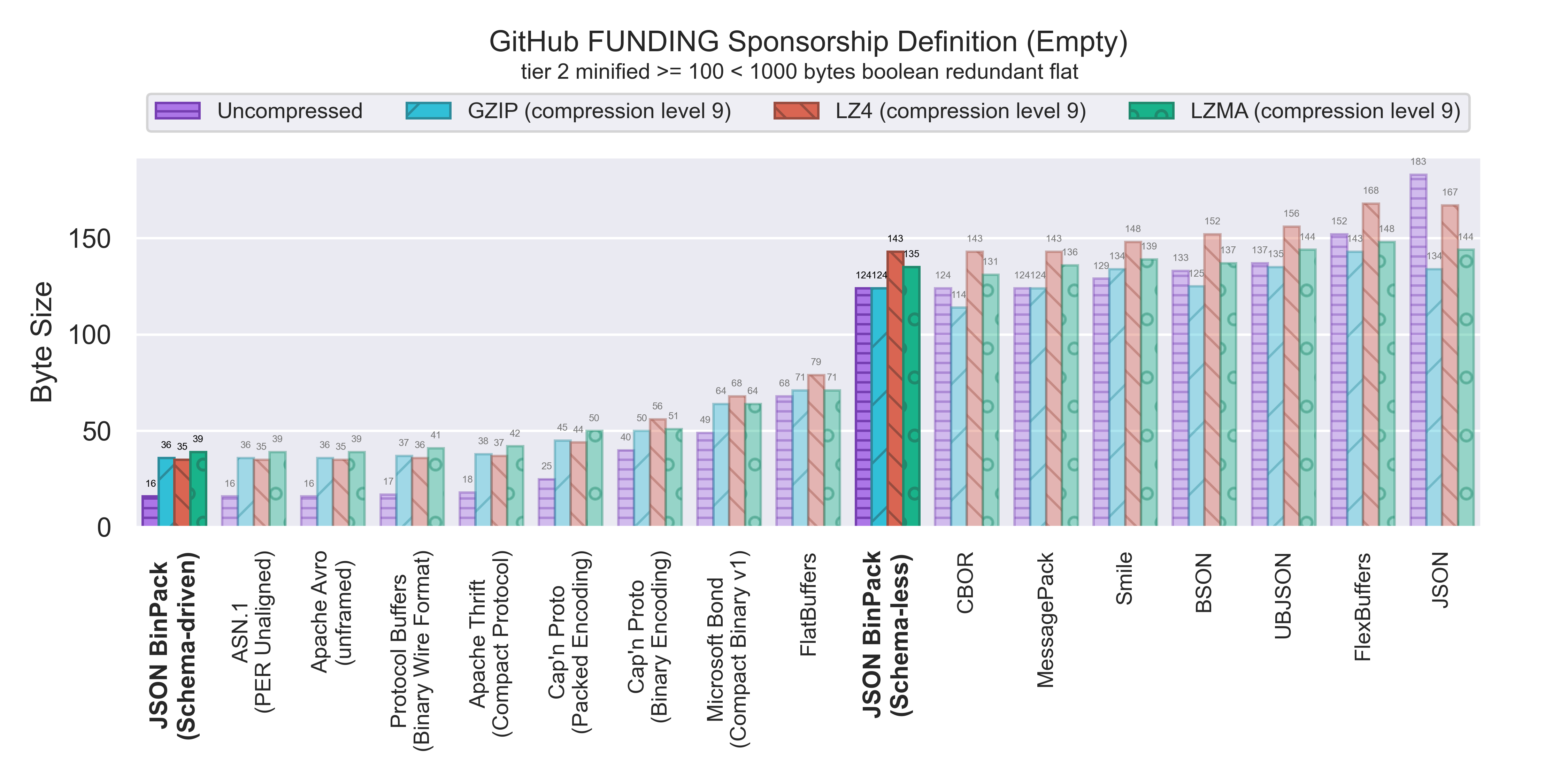}}
  \caption{The space-efficiency benchmark results for the GitHub FUNDING Sponsorship Definition (Empty) test case selected from the SchemaStore open-source dataset test suite in \cite{viotti2022benchmark}.}
\label{fig:benchmark-githubfundingblank} \end{figure*}

The GitHub \footnote{\url{https://github.com}} software hosting provider
defines a \texttt{FUNDING}
\footnote{\url{https://docs.github.com/en/github/administering-a-repository/managing-repository-settings/displaying-a-sponsor-button-in-your-repository}}
file format to declare the funding platforms that an open-source project
supports. The \texttt{FUNDING} file format is used by the open-source software
industry. In \autoref{fig:benchmark-githubfundingblank}, we demonstrate a
\textbf{Tier 2 minified $\geq$ 100 $<$ 1000 bytes boolean redundant flat} (Tier
2 BRF from \cite{viotti2022benchmark}) JSON document that consists of a
definition that does not declare any supported funding platforms.

The smallest bit-string produced by JSON BinPack (Schema-driven), ASN.1 PER
Unaligned \cite{asn1-per} and Apache Avro \cite{avro} (16 bytes) results in a
\textbf{5.8\%} size reduction compared to the next best performing
specification: Protocol Buffers \cite{protocolbuffers} (17 bytes).  JSON
BinPack (Schema-driven) achieves a \textbf{87\%} size reduction compared to the
best performing schema-less serialization specifications: JSON BinPack
(Schema-less), CBOR \cite{RFC7049} and MessagePack \cite{messagepack}.

Additionally, JSON BinPack (Schema-less) (124 bytes), along with CBOR
\cite{RFC7049} and MessagePack \cite{messagepack}, produce the smallest
bit-string for schema-less binary serialization specifications, resulting in a
\textbf{3.8\%} size reduction compared to the next best performing
specifications: Smile \cite{smile} (129 bytes).


\textbf{Comparison to Uncompressed and Compressed JSON}. In comparison to JSON
\cite{ECMA-404} (183 bytes), JSON BinPack (Schema-driven) (16 bytes) and JSON
BinPack (Schema-less) (124 bytes) achieve a \textbf{91.2\%} and \textbf{32.2\%}
size reduction, respectively.  In comparison to best-case compressed JSON
\cite{ECMA-404} (134 bytes), JSON BinPack (Schema-driven) (16 bytes) and JSON
BinPack (Schema-less) (124 bytes) achieve a \textbf{88\%} and \textbf{7.4\%}
size reduction, respectively.

\clearpage
\subsection{ECMAScript Module Loader Definition}
\label{sec:benchmark-esmrc}

\begin{figure*}[ht!]
  \frame{\includegraphics[width=\linewidth]{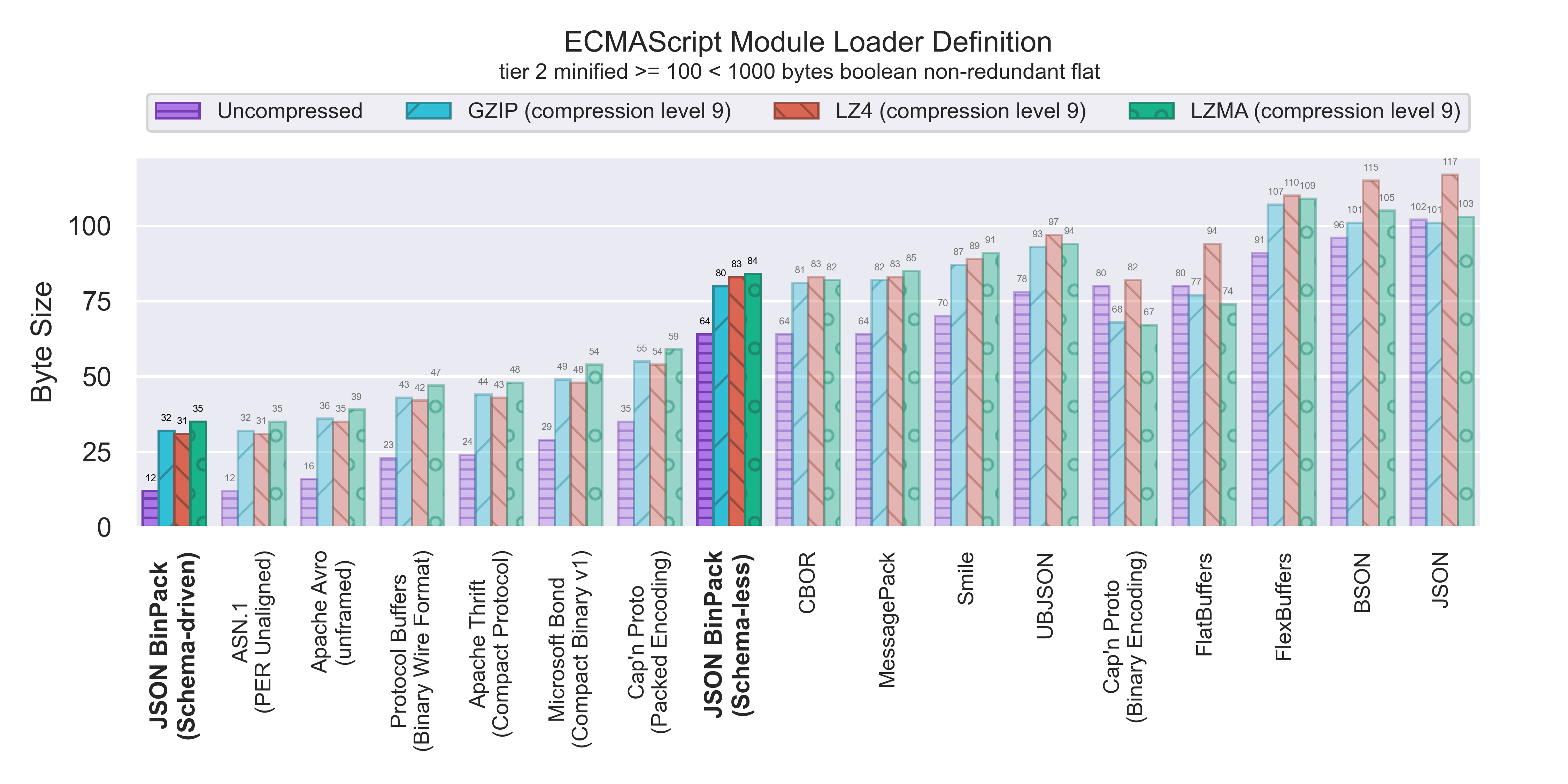}}
  \caption{The space-efficiency benchmark results for the ECMAScript Module Loader Definition test case selected from the SchemaStore open-source dataset test suite in \cite{viotti2022benchmark}.}
\label{fig:benchmark-esmrc} \end{figure*}

\texttt{esm} \footnote{\url{https://github.com/standard-things/esm}} is an
open-source ECMAScript \cite{ECMA-262} module loader for the Node.js
\footnote{\url{https://nodejs.org}} JavaScript runtime that allows developers
to use the modern \texttt{import} module syntax on older runtime versions.
\texttt{esm} is used in the web industry. In \autoref{fig:benchmark-esmrc}, we
demonstrate a \textbf{Tier 2 minified $\geq$ 100 $<$ 1000 bytes boolean
non-redundant flat} (Tier 2 BNF from \cite{viotti2022benchmark}) JSON
document that defines an example \texttt{esm} configuration.

The smallest bit-string produced by both JSON BinPack (Schema-driven) and ASN.1
PER Unaligned \cite{asn1-per} (12 bytes) results in a \textbf{25\%} size
reduction compared to the next best performing specification: Apache Avro
\cite{avro} (16 bytes).  JSON BinPack (Schema-driven) achieves a
\textbf{81.2\%} size reduction compared to the best performing schema-less
serialization specifications: JSON BinPack (Schema-less), CBOR \cite{RFC7049}
and MessagePack \cite{messagepack}.

Additionally, JSON BinPack (Schema-less) (64 bytes), along with CBOR
\cite{RFC7049} and MessagePack \cite{messagepack}, produce the smallest
bit-string for schema-less binary serialization specifications, resulting in a
\textbf{8.5\%} size reduction compared to the next best performing
specifications: Smile \cite{smile} (70 bytes).


\textbf{Comparison to Uncompressed and Compressed JSON}. In comparison to JSON
\cite{ECMA-404} (102 bytes), JSON BinPack (Schema-driven) (12 bytes) and JSON
BinPack (Schema-less) (64 bytes) achieve a \textbf{88.2\%} and \textbf{37.2\%}
size reduction, respectively.  In comparison to best-case compressed JSON
\cite{ECMA-404} (101 bytes), JSON BinPack (Schema-driven) (12 bytes) and JSON
BinPack (Schema-less) (64 bytes) achieve a \textbf{88.1\%} and \textbf{36.6\%}
size reduction, respectively.

\clearpage
\subsection{ESLint Configuration Document}
\label{sec:benchmark-eslintrc}

\begin{figure*}[ht!]
  \frame{\includegraphics[width=\linewidth]{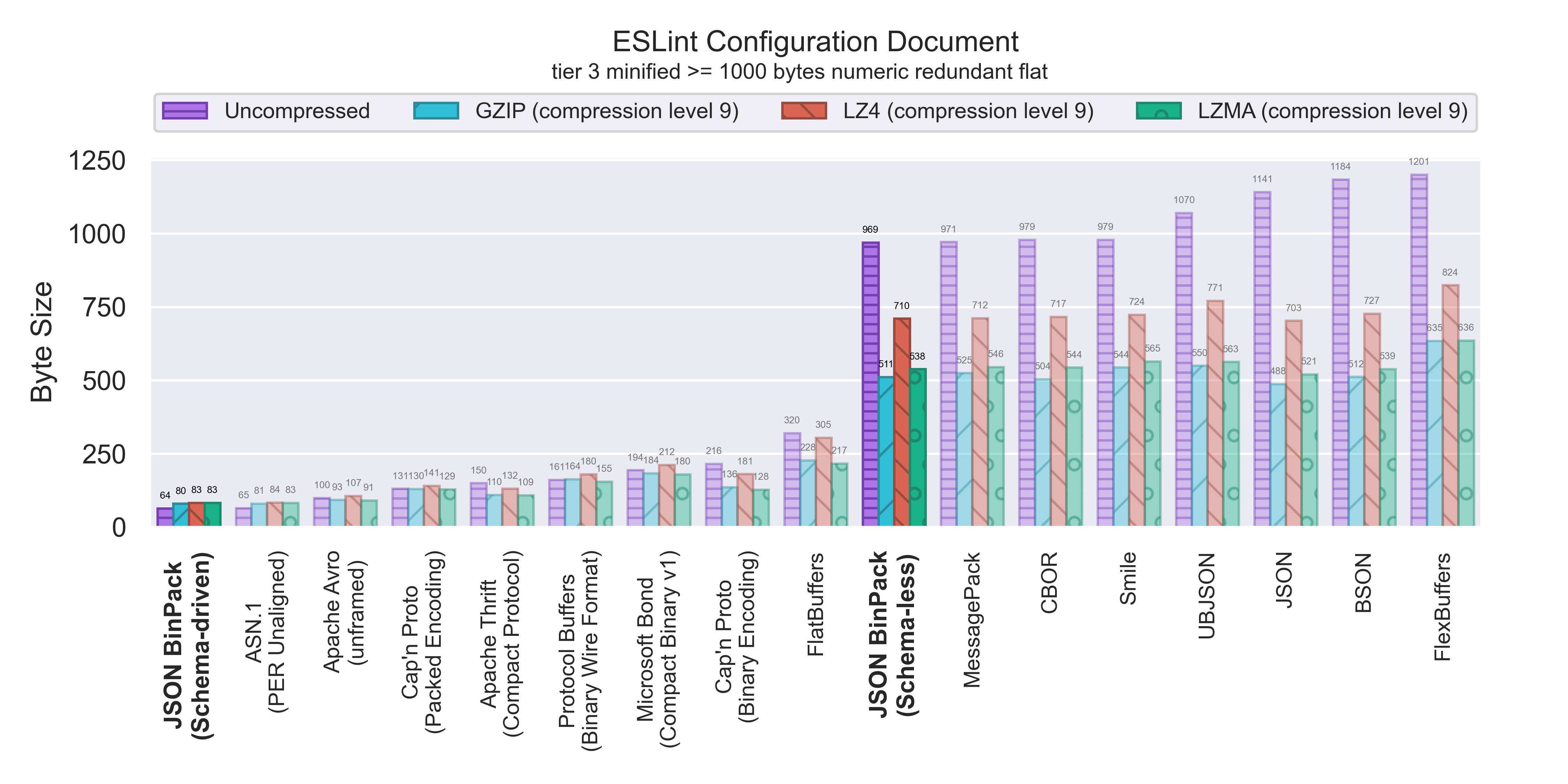}}
  \caption{The space-efficiency benchmark results for the ESLint Configuration Document test case selected from the SchemaStore open-source dataset test suite in \cite{viotti2022benchmark}.}
\label{fig:benchmark-eslintrc} \end{figure*}

ESLint \footnote{\url{https://eslint.org}} is a popular open-source extensible
linter for the JavaScript \cite{ECMA-262} programming language used by a wide
range of companies in the software development industry such as Google,
Salesforce, and Airbnb. In \autoref{fig:benchmark-eslintrc}, we demonstrate a
\textbf{Tier 3 minified $\geq$ 1000 bytes numeric redundant flat} (Tier 3 NRF
from \cite{viotti2022benchmark}) JSON document that defines a browser and
Node.js linter configuration that defines general-purposes and
\emph{React.js}-specific \footnote{\url{https://reactjs.org}} linting rules.

The smallest bit-string produced by JSON BinPack (Schema-driven) (64 bytes)
results in a \textbf{1.5\%} size reduction compared to the next best performing
specification: ASN.1 PER Unaligned \cite{asn1-per} (65 bytes).  JSON BinPack
(Schema-driven) achieves a \textbf{93.3\%} size reduction compared to the best
performing schema-less serialization specification: JSON BinPack (Schema-less).

Additionally, JSON BinPack (Schema-less) (969 bytes) produces the smallest
bit-string for schema-less binary serialization specifications, resulting in a
\textbf{0.2\%} size reduction compared to the next best performing
specifications: MessagePack \cite{messagepack} (971 bytes).


\textbf{Comparison to Uncompressed and Compressed JSON}. In comparison to JSON
\cite{ECMA-404} (1141 bytes), JSON BinPack (Schema-driven) (64 bytes) and JSON
BinPack (Schema-less) (969 bytes) achieve a \textbf{94.3\%} and \textbf{15\%}
size reduction, respectively.  In comparison to best-case compressed JSON
\cite{ECMA-404} (488 bytes), JSON BinPack (Schema-driven) (64 bytes) and JSON
BinPack (Schema-less) (969 bytes) achieve a \textbf{86.8\%} and negative
\textbf{98.5\%} size reduction, respectively.

\clearpage
\subsection{NPM Package.json Linter Configuration Manifest}
\label{sec:benchmark-packagejsonlintrc}

\begin{figure*}[ht!]
  \frame{\includegraphics[width=\linewidth]{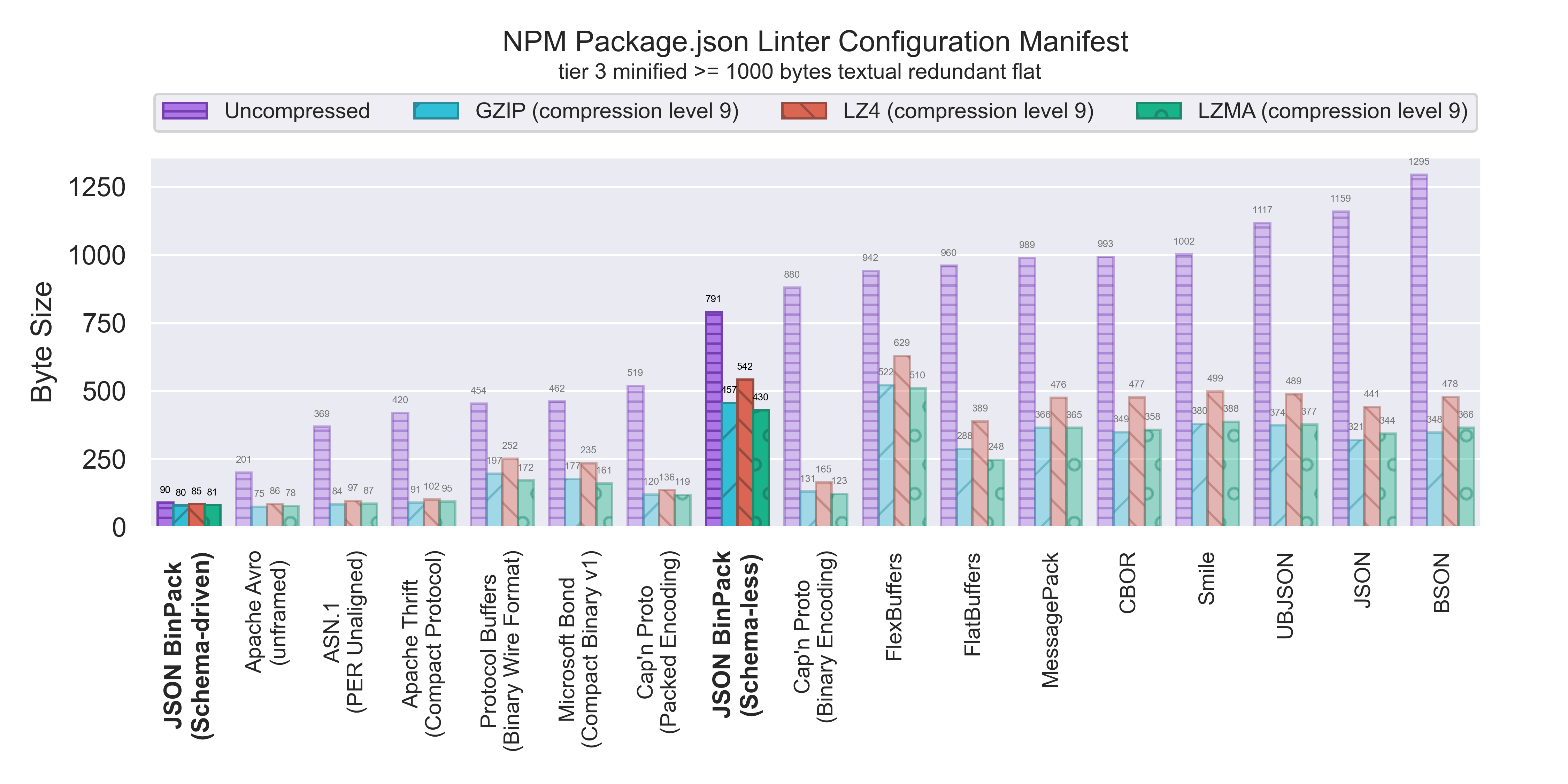}}
  \caption{The space-efficiency benchmark results for the NPM Package.json Linter Configuration Manifest test case selected from the SchemaStore open-source dataset test suite in \cite{viotti2022benchmark}.}
\label{fig:benchmark-packagejsonlintrc} \end{figure*}

Node.js Package Manager (NPM) \footnote{\url{https://www.npmjs.com}} is an
open-source package manager for Node.js \footnote{\url{https://nodejs.org}}, a
JavaScript \cite{ECMA-262} runtime targetted at the web development industry.
\texttt{npm-package-json-lint}
\footnote{\url{https://npmpackagejsonlint.org/en/}} is an open-source tool to
enforce a set of configurable rules for a Node.js Package Manager (NPM)
\footnote{\url{https://www.npmjs.com}} configuration manifest. In
\autoref{fig:benchmark-packagejsonlintrc}, we demonstrate a \textbf{Tier 3
minified $\geq$ 1000 bytes textual redundant flat} (Tier 3 TRF from
\cite{viotti2022benchmark}) JSON document that consists of an example
\texttt{npm-package-json-lint} configuration.

The smallest bit-string produced by JSON BinPack (Schema-driven) (90 bytes)
results in a \textbf{55.2\%} size reduction compared to the next best
performing specification: Apache Avro \cite{avro} (201 bytes).  JSON BinPack
(Schema-driven) achieves a \textbf{88.6\%} size reduction compared to the best
performing schema-less serialization specification: JSON BinPack (Schema-less).

Additionally, JSON BinPack (Schema-less) (791 bytes) produces the smallest
bit-string for schema-less binary serialization specifications, resulting in a
\textbf{16\%} size reduction compared to the next best performing
specifications: FlexBuffers \cite{flexbuffers} (942 bytes).


\textbf{Comparison to Uncompressed and Compressed JSON}. In comparison to JSON
\cite{ECMA-404} (1159 bytes), JSON BinPack (Schema-driven) (90 bytes) and JSON
BinPack (Schema-less) (791 bytes) achieve a \textbf{92.2\%} and \textbf{31.7\%}
size reduction, respectively.  In comparison to best-case compressed JSON
\cite{ECMA-404} (321 bytes), JSON BinPack (Schema-driven) (90 bytes) and JSON
BinPack (Schema-less) (791 bytes) achieve a \textbf{71.9\%} and negative
\textbf{146.4\%} size reduction, respectively.

\clearpage
\subsection{.NET Core Project}
\label{sec:benchmark-netcoreproject}

\begin{figure*}[ht!]
  \frame{\includegraphics[width=\linewidth]{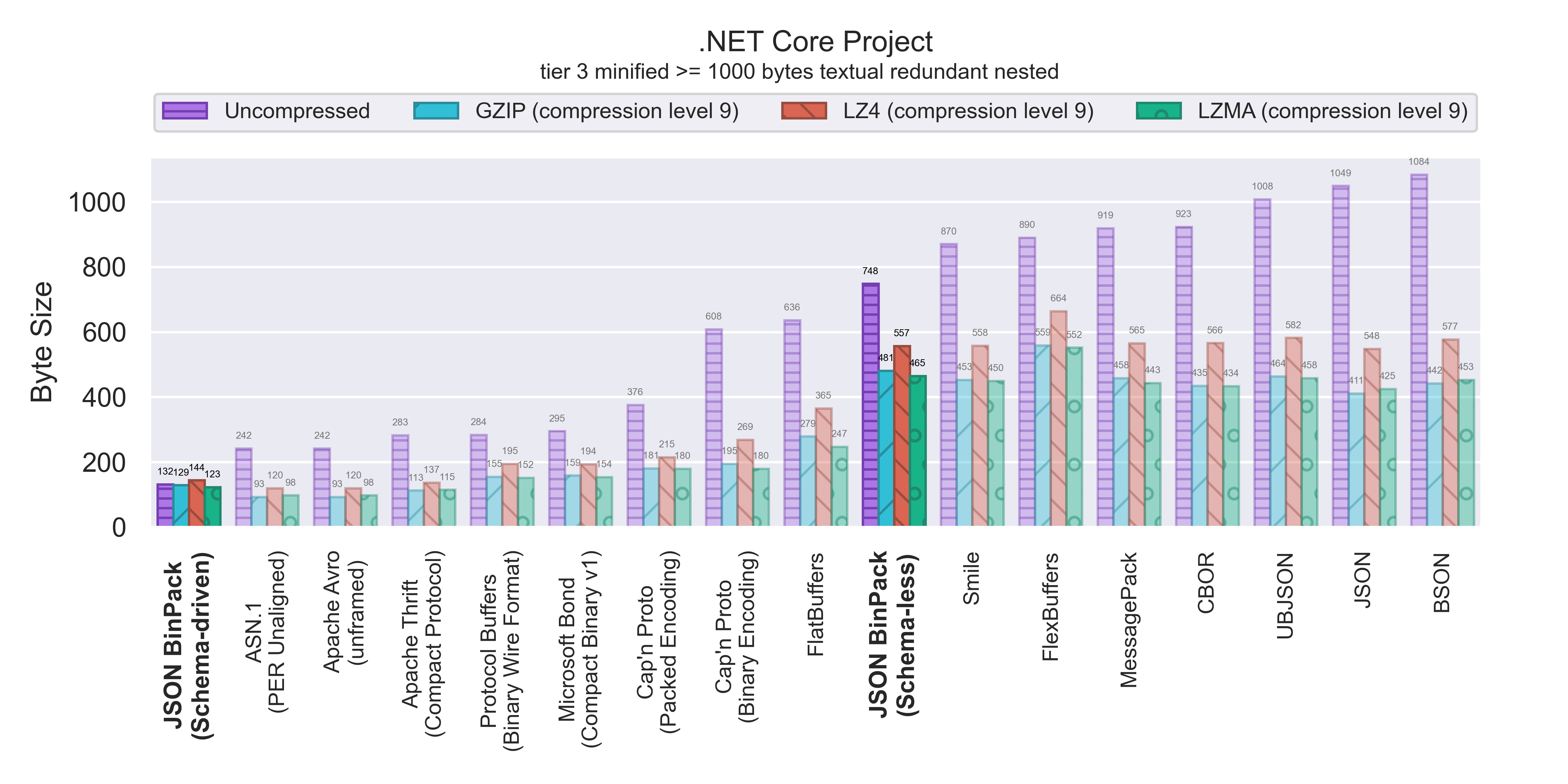}}
  \caption{The space-efficiency benchmark results for the .NET Core Project test case selected from the SchemaStore open-source dataset test suite in \cite{viotti2022benchmark}.}
\label{fig:benchmark-netcoreproject} \end{figure*}

The ASP.NET \footnote{\url{https://dotnet.microsoft.com/apps/aspnet}} Microsoft
web-application framework defined a now-obsolete JSON-based project manifest
called \texttt{project.json}
\footnote{\url{http://web.archive.org/web/20150322033428/https://github.com/aspnet/Home/wiki/Project.json-file}}
used in the web industry. In \autoref{fig:benchmark-netcoreproject}, we
demonstrate a \textbf{Tier 3 minified $\geq$ 1000 bytes textual redundant
nested} (Tier 3 TRN from \cite{viotti2022benchmark}) JSON document that
consists of a detailed an example \texttt{project.json} manifest that lists
several dependencies.

The smallest bit-string produced by JSON BinPack (Schema-driven) (132 bytes)
results in a \textbf{45.4\%} size reduction compared to the next best
performing specification: ASN.1 PER Unaligned \cite{asn1-per} and Apache Avro
\cite{avro} (242 bytes).  JSON BinPack (Schema-driven) achieves a
\textbf{82.3\%} size reduction compared to the best performing schema-less
serialization specification: JSON BinPack (Schema-less).

Additionally, JSON BinPack (Schema-less) (748 bytes) produces the smallest
bit-string for schema-less binary serialization specifications, resulting in a
\textbf{14\%} size reduction compared to the next best performing
specifications: Smile \cite{smile} (870 bytes).


\textbf{Comparison to Uncompressed and Compressed JSON}. In comparison to JSON
\cite{ECMA-404} (1049 bytes), JSON BinPack (Schema-driven) (132 bytes) and JSON
BinPack (Schema-less) (748 bytes) achieve a \textbf{87.4\%} and \textbf{28.6\%}
size reduction, respectively.  In comparison to best-case compressed JSON
\cite{ECMA-404} (411 bytes), JSON BinPack (Schema-driven) (132 bytes) and JSON
BinPack (Schema-less) (748 bytes) achieve a \textbf{67.8\%} and negative
\textbf{81.9\%} size reduction, respectively.

\clearpage
\subsection{NPM Package.json Example Manifest}
\label{sec:benchmark-packagejson}

\begin{figure*}[ht!]
  \frame{\includegraphics[width=\linewidth]{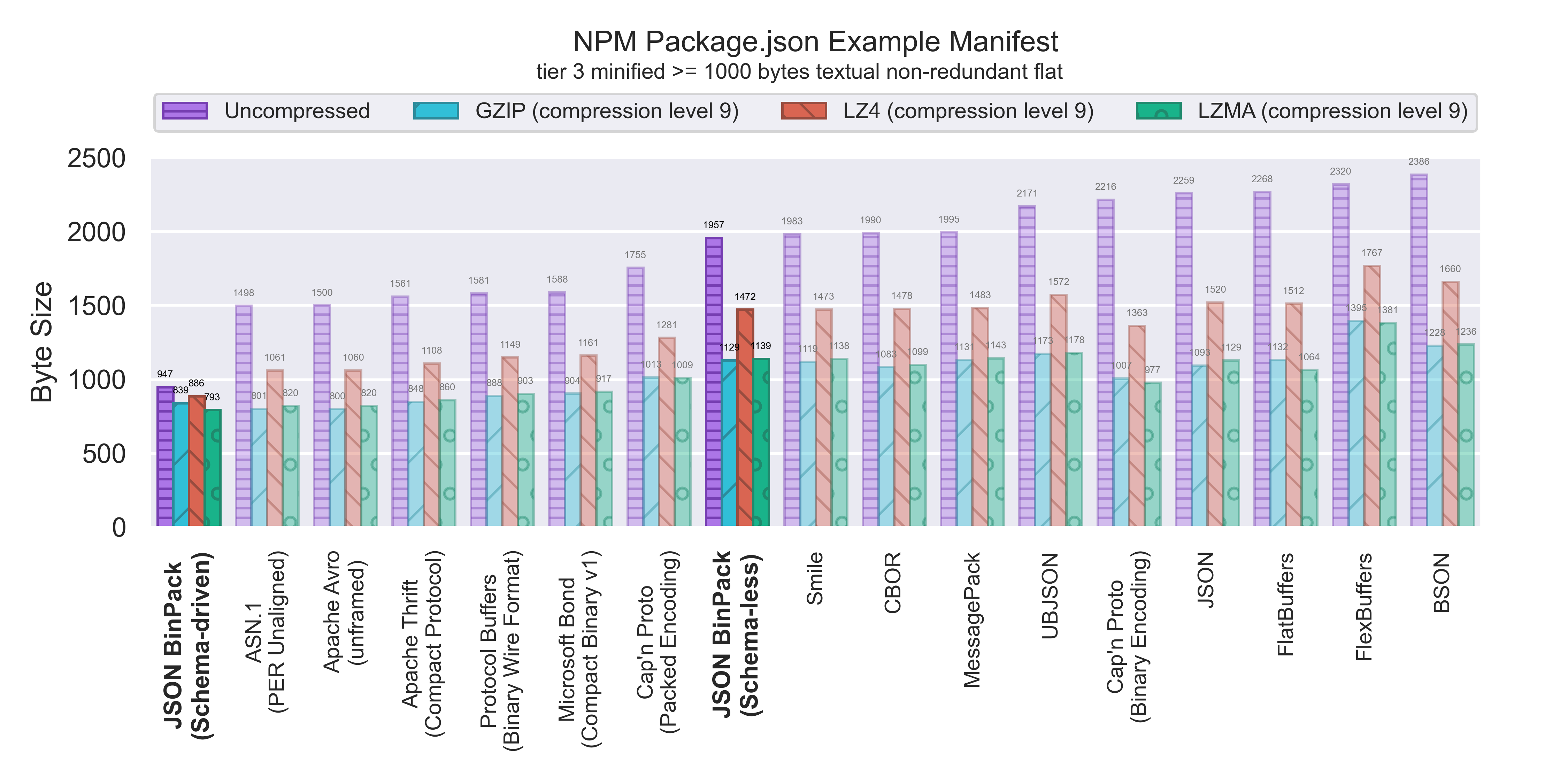}}
  \caption{The space-efficiency benchmark results for the NPM Package.json Example Manifest test case selected from the SchemaStore open-source dataset test suite in \cite{viotti2022benchmark}.}
\label{fig:benchmark-packagejson} \end{figure*}

Node.js Package Manager (NPM) \footnote{\url{https://www.npmjs.com}} is an
open-source package manager for Node.js \footnote{\url{https://nodejs.org}}, a
JavaScript \cite{ECMA-262} runtime targetted at the web development industry. A
package that is published to NPM is declared using a JSON file called
\texttt{package.json}
\footnote{\url{https://docs.npmjs.com/cli/v6/configuring-npm/package-json}}.
In \autoref{fig:benchmark-packagejson}, we demonstrate a \textbf{Tier 3
minified $\geq$ 1000 bytes textual non-redundant flat} (Tier 3 TNF from
\cite{viotti2022benchmark}) JSON document that consists of a
\texttt{package.json} manifest that declares a particular version of the
Grunt.js \footnote{\url{https://gruntjs.com}} task runner.

The smallest bit-string produced by JSON BinPack (Schema-driven) (947 bytes)
results in a \textbf{36.7\%} size reduction compared to the next best
performing specification: ASN.1 PER Unaligned \cite{asn1-per} (1498 bytes).
JSON BinPack (Schema-driven) achieves a \textbf{51.6\%} size reduction compared
to the best performing schema-less serialization specification: JSON BinPack
(Schema-less).

Additionally, JSON BinPack (Schema-less) (1957 bytes) produces the smallest
bit-string for schema-less binary serialization specifications, resulting in a
\textbf{1.3\%} size reduction compared to the next best performing
specifications: Smile \cite{smile} (1983 bytes).


\textbf{Comparison to Uncompressed and Compressed JSON}. In comparison to JSON
\cite{ECMA-404} (2259 bytes), JSON BinPack (Schema-driven) (947 bytes) and JSON
BinPack (Schema-less) (1957 bytes) achieve a \textbf{58\%} and \textbf{13.3\%}
size reduction, respectively.  In comparison to best-case compressed JSON
\cite{ECMA-404} (1093 bytes), JSON BinPack (Schema-driven) (947 bytes) and JSON
BinPack (Schema-less) (1957 bytes) achieve a \textbf{13.3\%} and negative
\textbf{79\%} size reduction, respectively.

\clearpage
\subsection{JSON Resume Example}
\label{sec:benchmark-jsonresume}

\begin{figure*}[ht!]
  \frame{\includegraphics[width=\linewidth]{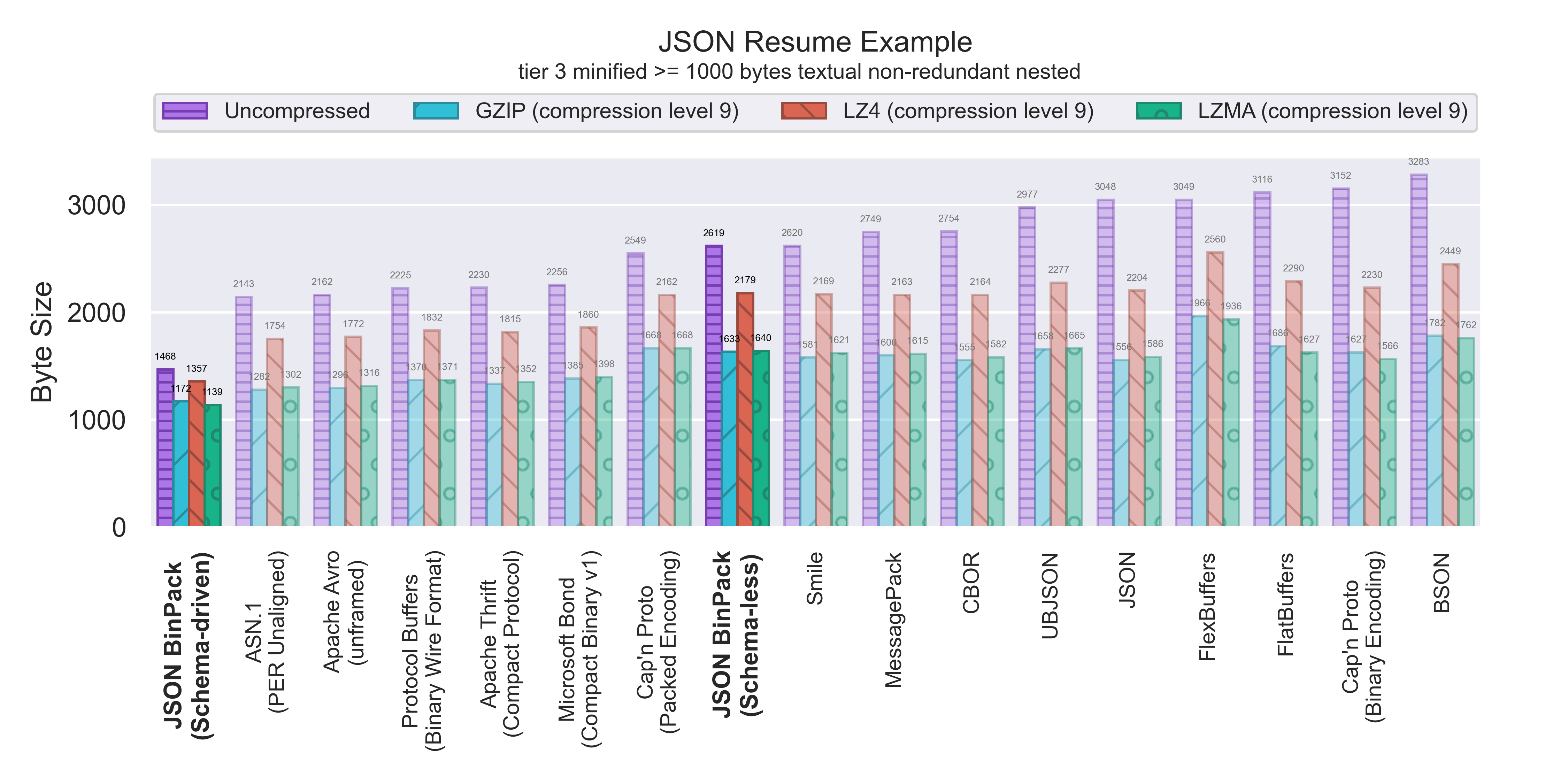}}
  \caption{The space-efficiency benchmark results for the JSON Resume Example test case selected from the SchemaStore open-source dataset test suite in \cite{viotti2022benchmark}.}
\label{fig:benchmark-jsonresume} \end{figure*}

JSON Resume \footnote{\url{https://jsonresume.org}} is a community-driven
proposal for a JSON-based file format that declares and renders themable
resumes used in the recruitment industry. In
\autoref{fig:benchmark-jsonresume}, we demonstrate a \textbf{Tier 3 minified
$\geq$ 1000 bytes textual non-redundant nested} (Tier 3 TNN from
\cite{viotti2022benchmark}) JSON document that consists of a detailed
example resume for a fictitious software programmer.

The smallest bit-string produced by JSON BinPack (Schema-driven) (1468 bytes)
results in a \textbf{31.4\%} size reduction compared to the next best
performing specification: ASN.1 PER Unaligned \cite{asn1-per} (2143 bytes).
JSON BinPack (Schema-driven) achieves a \textbf{43.9\%} size reduction compared
to the best performing schema-less serialization specification: JSON BinPack
(Schema-less).

Additionally, JSON BinPack (Schema-less) (2619 bytes) produces the smallest
bit-string for schema-less binary serialization specifications, resulting in a
\textbf{0.03\%} size reduction compared to the next best performing
specifications: Smile \cite{smile} (2620 bytes).


\textbf{Comparison to Uncompressed and Compressed JSON}. In comparison to JSON
\cite{ECMA-404} (3048 bytes), JSON BinPack (Schema-driven) (1468 bytes) and
JSON BinPack (Schema-less) (2619 bytes) achieve a \textbf{51.8\%} and
\textbf{14\%} size reduction, respectively.  In comparison to best-case
compressed JSON \cite{ECMA-404} (1556 bytes), JSON BinPack (Schema-driven)
(1468 bytes) and JSON BinPack (Schema-less) (2619 bytes) achieve a
\textbf{5.6\%} and negative \textbf{68.3\%} size reduction, respectively.

\clearpage
\subsection{Nightwatch.js Test Framework Configuration}
\label{sec:benchmark-nightwatch}

\begin{figure*}[ht!]
  \frame{\includegraphics[width=\linewidth]{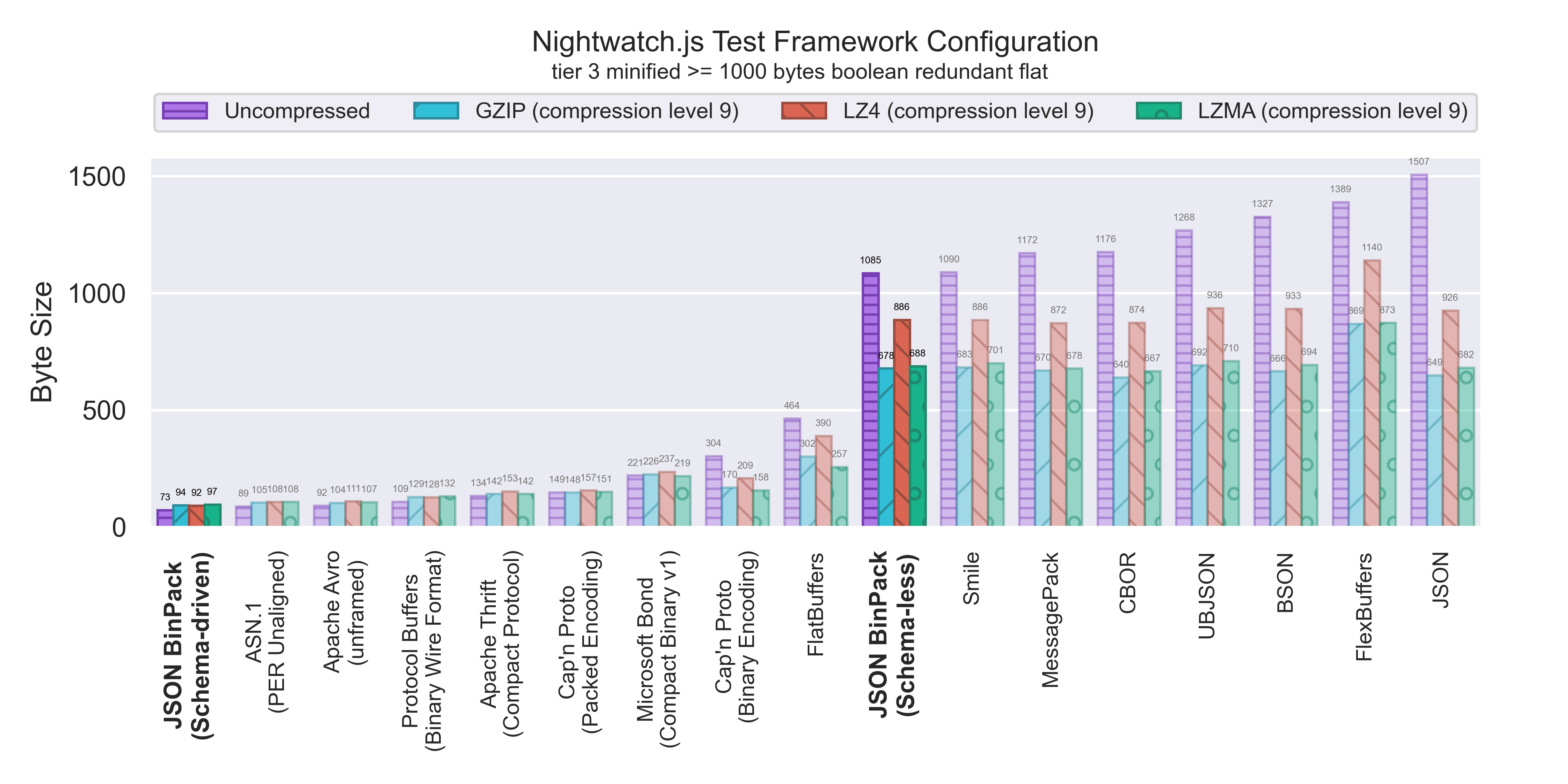}}
  \caption{The space-efficiency benchmark results for the Nightwatch.js Test Framework Configuration test case selected from the SchemaStore open-source dataset test suite in \cite{viotti2022benchmark}.}
\label{fig:benchmark-nightwatch} \end{figure*}

Nightwatch.js \footnote{\url{https://nightwatchjs.org}} is an open-source
browser automation solution used in the software testing industry. In
\autoref{fig:benchmark-nightwatch}, we demonstrate a \textbf{Tier 3 minified
$\geq$ 1000 bytes boolean redundant flat} (Tier 3 BRF from
\cite{viotti2022benchmark}) JSON document that consists of a Nightwatch.js
configuration file that defines a set of general-purpose WebDriver
\cite{webdriver} and Selenium \footnote{\url{https://www.selenium.dev}}
options.

The smallest bit-string produced by JSON BinPack (Schema-driven) (73 bytes)
results in a \textbf{17.9\%} size reduction compared to the next best
performing specification: ASN.1 PER Unaligned \cite{asn1-per} (89 bytes).  JSON
BinPack (Schema-driven) achieves a \textbf{93.2\%} size reduction compared to
the best performing schema-less serialization specification: JSON BinPack
(Schema-less).

Additionally, JSON BinPack (Schema-less) (1085 bytes) produces the smallest
bit-string for schema-less binary serialization specifications, resulting in a
\textbf{0.4\%} size reduction compared to the next best performing
specifications: Smile \cite{smile} (1090 bytes).


\textbf{Comparison to Uncompressed and Compressed JSON}. In comparison to JSON
\cite{ECMA-404} (1507 bytes), JSON BinPack (Schema-driven) (73 bytes) and JSON
BinPack (Schema-less) (1085 bytes) achieve a \textbf{95.1\%} and \textbf{28\%}
size reduction, respectively.  In comparison to best-case compressed JSON
\cite{ECMA-404} (649 bytes), JSON BinPack (Schema-driven) (73 bytes) and JSON
BinPack (Schema-less) (1085 bytes) achieve a \textbf{88.7\%} and negative
\textbf{67.1\%} size reduction, respectively.

\subsection{Reproducibility}
To make our pre-production benchmark reproducible, we followed the
reproducibility levels introduced by \cite{sam_harrison_2021_4761867}.  We
aimed for Level 3, the highest-level of reproducibility, as justified in the
following sections.

The pre-production implementation of JSON BinPack is written using the
TypeScript programming language. It contains valid encoding rules, mapping
rules and canonicalization rules but it is slow and memory-consuming compared
to a production-ready implementation.

\begin{itemize}

\item \textbf{Supported Environments for the Pre-production JSON BinPack Implementation}:

The pre-production JSON BinPack implementation and benchmark software are
    written to work on the macOS (Intel processors) and GNU/Linux operating
    systems. We do not make any effort to support the Microsoft Windows
    operating system, but we do expect the benchmark software to run on an
    \emph{msys2} \footnote{\url{https://www.msys2.org}} or \emph{Windows
    Subsystem for Linux}
    \footnote{\url{https://docs.microsoft.com/en-us/windows/wsl/}} environment
    with minor changes at most. The benchmark is exclusively concerned with the
    byte-size of the bit-strings produced by the binary serialization
    specifications.  Therefore, the CPU, memory, and network bandwidth
    characteristics of the test machine would not affect the results of the
    benchmark. No further conditions apart from the exact software versions of
    the dependencies required by the project are necessary to replicate the
    results.

\item \textbf{Automation.} The benchmark software, from the generation of the
serialized bit-strings to the generation of the plots using Matplotlib
\footnote{\url{https://matplotlib.org}}, is automated through a GNU Make
\footnote{\url{https://www.gnu.org/software/make/}} declarative and
parallelizable build definition.

\item \textbf{Testing.} The POSIX shell and Python scripts distributed with the
  benchmark are automatically linted using the \emph{shellcheck}
    \footnote{\url{https://www.shellcheck.net}} and \emph{flake8}
    \footnote{\url{https://flake8.pycqa.org/}} open-source tools, respectively.
    The serialization and deserialization procedures of the benchmark are
    automatically tested as explained in \cite{viotti2022benchmark}.

\item \textbf{Documentation and Readability.} The \texttt{README} file
  \footnote{\url{https://github.com/jviotti/binary-json-size-benchmark\#running-locally}}
    in the repository contains precise instructions for running the benchmark
    locally and generate the data files and plots.  The project documentation
    includes a detailed list of the system dependencies that are required to
    succesfully execute every part of the benchmark and a detailed list of the
    required binary serialization specifications, implementations, versions,
    and encodings. The benchmark source code is compact and easy to understand
    and navigate due to the declarative rule definition nature of GNU Make.

\item \textbf{DOI.} The version of the benchmark software described in this
  study is archived with a DOI \cite{juan_cruz_viotti_2022_7076479}. The DOI
    includes the source code for reproducing the benchmark and the associated
    results.

\item \textbf{Dependencies.} The benchmark software is implemented using
  established open-source software with the exception of the ASN-1Step
    \footnote{\url{https://www.oss.com/asn1/products/asn-1step/asn-1step.html}}
    command-line tool, which is a proprietary implementation of ASN.1
    \cite{asn1} distributed by OSS Nokalva with a 30 days free trial. Every
    binary serialization specification implementation used in the benchmark
    with the exception of ASN-1Step is pinned to its specific version to ensure
    reproducibility.  As explained in the online documentation, the benchmark
    software expects the ASN-1Step command-line tool version 10.0.2 to be
    installed and globally-accessible in the system in order to benchmark the
    ASN.1 PER Unaligned \cite{asn1-per} binary serialization specification.

\item \textbf{Version Control.} The benchmark repository utilises the
  \emph{git} \footnote{\url{https://git-scm.com}} version control system and
    its publicly hosted on GitHub
    \footnote{\url{https://github.com/jviotti/binary-json-size-benchmark}} as
    recommended by \cite{peng2011reproducible}.

\item \textbf{Continuous Integration.} The GitHub repository hosting the
  benchmark software is setup with the GitHub Actions
  \footnote{\url{https://github.com/features/actions}} continuous integration
  provider to re-run the benchmark automatically on new commits using a
  GNU/Linux Ubuntu 20.04 LTS cloud worker. This process prevents changes to the
  benchmark software from introducing regressions and new software errors. We
  make use of this process to validate GitHub internal and external pull
  requests before merging them into the trunk.

\item \textbf{Availability.} The benchmark software and results are publicly
  available and governed by the \emph{Apache License 2.0}
  \footnote{\url{https://www.apache.org/licenses/LICENSE-2.0.html}} open-source
  software license. The results of the benchmark are also published as a
  website hosted at \url{https://www.jviotti.com/binary-json-size-benchmark/}
  using the GitHub Pages free static hosting provider. The website provides
  direct links to the JSON \cite{ECMA-404} documents being encoded by the
  benchmark and direct links to the schema definitions used in every case. Both
  the JSON documents and the schema definitions are hosted in the benchmark
  GitHub repository to ensure their availability even if the original sources
  do not exist anymore.

\item \textbf{Continuity.} We plan to continue extending the benchmark software
  in the future to test new versions of the current selection of binary
  serialization specifications and to include new JSON-compatible binary
  serialization specifications. We hope for this project to become a
  collaborative effort to measure the space-efficiency of every new
  JSON-compatible serialization specifications and we are comitted to accepting
  open-source contributions.

\end{itemize}

\section{Conclusions}
\label{sec:conclusions}
In this section, we critically evaluate the JSON BinPack pre-production
implementation against schema-driven and schema-less alternative binary
serialization specifications, motivating our discussion starting with the
summary in \autoref{table:jsonbinpack-summary-results-uncompressed}.


\begin{table}[ht!] \caption{A summary of the size reduction provided by the
  JSON BinPack binary serialization specification proof-of-concept
  implementation in both schema-driven and schema-less mode in comparison to
  JSON \cite{ECMA-404} given the input data listed in
  \cite{viotti2022benchmark}.}

\label{table:jsonbinpack-summary-results-uncompressed}
\begin{tabularx}{\linewidth}{X|l|l|l|l|l|l}
\toprule
\multirow{2}{*}{\textbf{Mode}} & \multicolumn{5}{c|}{\textbf{Size Reductions in Comparison To JSON}} & \multirow{2}{*}{\textbf{Negative Cases}} \\ \cline{2-6}
& \textbf{Maximum} & \textbf{Minimum} & \textbf{Range} & \textbf{Median} & \textbf{Average} & \\
\midrule
JSON BinPack (Schema-driven) & 100\% & 26.9\% & 73 & 86.7\% & 78.7\% & 0 / 27 (0\%) \\ \hline
JSON BinPack (Schema-less)   & 72.5\% &    10.2\% &  62.2 & 30.6\% &  30.5\% & 0 / 27 (0\%) \\ \hline
\textbf{Averages} & \textbf{86.2\%} & \textbf{18.6\%} & \textbf{67.6} & \textbf{58.6\%} & \textbf{54.6\%} & \textbf{0\%} \\
\bottomrule
\end{tabularx}
\end{table}

\subsection{Q1: How does JSON BinPack in schema-driven mode compare to JSON in terms of space-efficiency?}
\label{sec:jsonbinpack-conclusions-schemadriven}

As demonstrated in \autoref{sec:benchmark} and
\autoref{table:jsonbinpack-summary-schema-driven}, the JSON BinPack
schema-driven binary serialization specification, denoted as \emph{JSON BinPack
(Schema-driven)}, is comparatively as or more space-efficient than every other
serialization specification for the \textbf{27} proposed input data considered
by \cite{viotti2022benchmark}. Unlike any other considered binary serialization
specification, JSON BinPack is strictly space-efficient in comparison to JSON
\cite{ECMA-404} given the input data.

In \cite{viotti2022benchmark}, we found that the most space-efficient
JSON-compatible binary serialization specifications are ASN.1 PER Unaligned
\cite{asn1-per} and Apache Avro \cite{avro}. ASN.1 PER Unaligned
\cite{asn1-per} results in \textbf{71.4\%} and \textbf{65.7\%} median and
average size reductions with a maximum of \textbf{98.5\%} and a minimum of
negative \textbf{7.9\%}. Apache Avro \cite{avro} results in \textbf{73.5\%} and
\textbf{65.7\%} median and average size reductions with a maximum of
\textbf{100\%} and a minimum of negative \textbf{48.9\%}. In comparison, JSON
BinPack produces strictly space-efficient results with a smaller range and no
negative cases: \textbf{86.7\%} and \textbf{78.7\%} median and average size
reductions with a maximum of \textbf{100\%} and a minimum of \textbf{26.9\%} as
shown in \autoref{table:jsonbinpack-summary-results-uncompressed}.
Additionally, JSON BinPack provides improvements in terms of space-efficiency
in comparison to the best-performing schema-driven binary serialization
specifications for documents that are highly redundant or nested according to
the JSON \cite{ECMA-404} taxonomy introduced in \cite{viotti2022benchmark}. For
example, JSON BinPack produces bit-strings that are \textbf{82\%},
\textbf{75\%} and \textbf{60\%} smaller than the best-performing schema-driven
alternatives for \emph{TravisCI Notifications Configuration} (see
\autoref{sec:benchmark-travisnotifications}), \emph{TSLint Linter Definition
(Multi-rule)} (see \autoref{sec:benchmark-tslintmulti}) and \emph{GeoJSON
Example Document} (see \autoref{sec:benchmark-geojson}), respectively.

JSON BinPack matches but not increases the size reduction characteristics
provided by the most space-efficient serialization specifications in \textbf{8}
out of the \textbf{27} cases. In terms of the taxonomy for JSON \cite{ECMA-404}
documents introduced in \cite{viotti2022benchmark}, this list includes
\textbf{5} out of the \textbf{7} documents considered boolean and \textbf{3}
out of the \textbf{6} documents considered textual and non-redundant. These
\textbf{8} documents represent cases where ASN.1 PER Unaligned \cite{asn1-per},
Apache Avro \cite{avro} and Protocol Buffers \cite{protocolbuffers} perform at
or close to the optimal level in terms of space-efficiency. For example, JSON
BinPack and Protocol Buffers \cite{protocolbuffers}, and JSON BinPack, Apache
Avro \cite{avro} and Protocol Buffers \cite{protocolbuffers}, serialize the
\emph{CommitLint Configuration (Basic)} \textbf{Tier 1 Minified $<$ 100 bytes,
boolean, non-redundant and flat} (see \autoref{sec:benchmark-commitlintbasic})
and the \emph{SAP Cloud SDK Continuous Delivery Toolkit Configuration}
\textbf{Tier 1 Minified $<$ 100 bytes, boolean, redundant and flat} (see
\autoref{sec:benchmark-sapcloudsdkpipeline}) JSON \cite{ECMA-404} documents,
respectively, into \textbf{0-byte} bit-strings.

\begin{table}[ht!] \caption{A summary of the top-performing and second
  top-performing schema-driven serialization specifications for every document
  considered in \cite{viotti2022benchmark}. Percentages represent the size
  reduction achieved in comparison to JSON. More is better.}

\label{table:jsonbinpack-summary-schema-driven}
\begin{tabularx}{\linewidth}{X|X|X}
\toprule
\textbf{Document Name} & \textbf{Best Performing Specifications} & \textbf{Second Best Performing Specifications} \\
\midrule

{ \small JSON-e Templating Engine Sort Example                   } & { \small JSON BinPack (76.4\%) }  & { \small Apache Avro (73.5\%) } \\ \hline
{ \small JSON-e Templating Engine Reverse Sort Example           } & { \small JSON BinPack (88.3\%) }  & { \small Apache Avro (87.2\%) } \\ \hline
{ \small CircleCI Definition (Blank)                             } & { \small JSON BinPack (85.7\%) }  & { \small ASN.1 and Apache Avro (71.4\%) } \\ \hline
{ \small CircleCI Matrix Definition                              } & { \small JSON BinPack (92.6\%) }  & { \small Apache Avro (84.2\%) } \\ \hline
{ \small Grunt.js Clean Task Definition                          } & { \small JSON BinPack (88.1\%) }  & { \small ASN.1 (86\%) } \\ \hline
{ \small CommitLint Configuration                                } & { \small JSON BinPack (79.1\%) }  & { \small Apache Avro (58.3\%) } \\ \hline
{ \small TSLint Linter Definition (Extends Only)                 } & { \small JSON BinPack and ASN.1 (26.9\%) } & { \small Apache Avro and Protocol Buffers (25.3\%) } \\ \hline
{ \small ImageOptimizer Azure Webjob Configuration               } & { \small JSON BinPack and ASN.1 (74.3\%) } & { \small Protocol Buffers (71.9\%) } \\ \hline
{ \small SAP Cloud SDK Continuous Delivery Toolkit Configuration } & { \small JSON BinPack, Apache Avro and Protocol Buffers (100\%) } & { \small ASN.1 and Apache Thrift (97.7\%) } \\ \hline
{ \small TSLint Linter Definition (Multi-rule)                   } & { \small JSON BinPack (98.9\%) }   & { \small ASN.1 (95.9\%) } \\ \hline
{ \small CommitLint Configuration (Basic)                        } & { \small JSON BinPack and Protocol Buffers (100\%) } & { \small ASN.1 and Apache Avro (96\%) } \\ \hline
{ \small TSLint Linter Definition (Basic)                        } & { \small JSON BinPack, ASN.1 and Apache Avro (98.5\%) } & { \small Protocol Buffers and Apache Thrift (88\%) } \\ \hline
{ \small GeoJSON Example Document                                } & { \small JSON BinPack (56.8\%) }  & { \small ASN.1 (-7.8\%) } \\ \hline
{ \small OpenWeatherMap API Example Document                     } & { \small JSON BinPack (77.1\%) }  & { \small Apache Avro (70\%) } \\ \hline
{ \small OpenWeather Road Risk API Example                       } & { \small JSON BinPack (73.3\%) }  & { \small Apache Avro (58.4\%) } \\ \hline
{ \small TravisCI Notifications Configuration                    } & { \small JSON BinPack (86.7\%) }  & { \small ASN.1 (26.1\%) } \\ \hline
{ \small Entry Point Regulation Manifest                         } & { \small JSON BinPack (65\%) }    & { \small Apache Avro (62.5\%) } \\ \hline
{ \small JSON Feed Example Document                              } & { \small JSON BinPack (46.5\%) }  & { \small ASN.1 (30.5\%) } \\ \hline
{ \small GitHub Workflow Definition                              } & { \small JSON BinPack and ASN.1 (53.6\%) } & { \small Apache Avro (53\%) } \\ \hline
{ \small GitHub FUNDING Sponsorship Definition (Empty)           } & { \small JSON BinPack, ASN.1 and Apache Avro (91.2\%) } & { \small Protocol Buffers (90.7\%) } \\ \hline
{ \small ECMAScript Module Loader Definition                     } & { \small JSON BinPack and ASN.1 (88.2\%) } & { \small Apache Avro (84.3\%) } \\ \hline
{ \small ESLint Configuration Document                           } & { \small JSON BinPack (94.39\%) }  & { \small ASN.1 (94.3\%) } \\ \hline
{ \small NPM Package.json Linter Configuration Manifest          } & { \small JSON BinPack (92.2\%) }  & { \small Apache Avro (82.6\%) } \\ \hline
{ \small .NET Core Project                                       } & { \small JSON BinPack (87.4\%) }  & { \small ASN.1 and Apache Avro (76.9\%) } \\ \hline
{ \small NPM Package.json Example Manifest                       } & { \small JSON BinPack (58\%) }    & { \small ASN.1 (33.6\%) } \\ \hline
{ \small JSON Resume Example                                     } & { \small JSON BinPack (51.8\%) }  & { \small ASN.1 (29.6\%) } \\ \hline
{ \small Nightwatch.js Test Framework Configuration              } & { \small JSON BinPack (95.1\%) }  & { \small ASN.1 (94\%) } \\

\bottomrule
\end{tabularx}
\end{table}

\subsection{Q2: How does JSON BinPack in schema-less mode compare to JSON in terms of space-efficiency?}
\label{sec:jsonbinpack-conclusions-schemaless}

As explained in \autoref{sec:methodology}, we benchmark JSON BinPack in
schema-less mode by serializing every input JSON \cite{ECMA-404} document with
a \emph{loose} JSON Schema \cite{jsonschema-core-2020} definition that matches
every instance.  As shown in \autoref{sec:benchmark} and
\autoref{table:jsonbinpack-summary-schema-less}, the schema-less mode of the
JSON BinPack binary serialization specification, denoted as \emph{JSON BinPack
(Schema-less)}, is as space-efficient or more space-efficient than every other
schema-less serialization specification considered by
\cite{viotti2022benchmark} for the \textbf{27} proposed input data.  Like CBOR
\cite{RFC7049} and MessagePack \cite{messagepack}, JSON BinPack in schema-less
mode is strictly space-efficient in comparison to JSON \cite{ECMA-404}.
However, JSON BinPack in schema-driven mode is strictly space-efficient in
comparison to JSON BinPack in schema-less mode.

In \cite{viotti2022benchmark}, we find that the most space-efficient
JSON-compatible binary schema-less serialization specifications are CBOR
\cite{RFC7049} and MessagePack \cite{messagepack}. CBOR \cite{RFC7049} results
in \textbf{22.5\%} and \textbf{22.4\%} median and average size reductions with
a maximum of \textbf{43.2\%} and a minimum of \textbf{6.8\%}.  Similarly,
MessagePack \cite{messagepack} results in \textbf{22.7\%} and \textbf{22.8\%}
median and average size reductions with a maximum of \textbf{43.2\%} and a
minimum of \textbf{6.8\%}. In comparison, JSON BinPack in schema-less mode
produces strictly space-efficient results: \textbf{30.6\%} and \textbf{30.5\%}
median and average size reductions with a maximum of \textbf{72.5\%} and a
minimum of \textbf{10.2\%} as shown in
\autoref{table:jsonbinpack-summary-results-uncompressed}.  Additionally, JSON
BinPack in schema-less mode provides significant improvements in terms of
space-efficiency in comparison to the best-performing schema-less binary
serialization specifications for documents that have a high-degree of nesting
according to the JSON \cite{ECMA-404} taxonomy defined in
\cite{viotti2022benchmark}.  For example, JSON BinPack produces bit-strings
that are \textbf{27.7\%}, \textbf{22\%} and \textbf{18.9\%} smaller than the
best-performing schema-less alternatives for \emph{GeoJSON Example Document}
(see \autoref{sec:benchmark-geojson}), \emph{OpenWeather Road Risk API Example}
(see \autoref{sec:benchmark-openweatherroadrisk}) and \emph{CommitLint
Configuration} (see \autoref{sec:benchmark-commitlint}), respectively.
Furthermore, JSON BinPack in schema-less mode produces space-efficient results
in comparison to every schema-driven binary serialization specification for
\emph{GeoJSON Example Document} (see \autoref{sec:benchmark-geojson}) and
\emph{TravisCI Notifications Configuration} (see
\autoref{sec:benchmark-travisnotifications}), only second to JSON BinPack
executed in schema-driven mode.

JSON BinPack in schema-less mode matches but not increases the size reduction
characteristics provided by the most space-efficient schema-less serialization
specifications in \textbf{11} out of \textbf{27} cases.  In terms of the
taxonomy for JSON \cite{ECMA-404} documents defined in
\cite{viotti2022benchmark}, this list includes \textbf{9} out of the
\textbf{12} documents considered \textbf{Tier 1 Minified $<$ 100 bytes} and
\textbf{2} out of the \textbf{2} documents considered \textbf{Tier 2 Minified
$\geq$ 100 $<$ 1000 bytes and boolean}.  These \textbf{11} documents represent
cases where both CBOR \cite{RFC7049} and MessagePack \cite{messagepack} perform
close to the optimal level in terms of space-efficiency for a schema-less
serialization specification.

\begin{table}[ht!] \caption{A summary of the top-performing and second
  top-performing schema-less serialization specifications for every document
  considered in \cite{viotti2022benchmark}. Percentages represent the size reduction achieved in comparison to JSON. More is better.}

\label{table:jsonbinpack-summary-schema-less}
\begin{tabularx}{\linewidth}{X|X|X}
\toprule
\textbf{Document Name} & \textbf{Best Performing Specifications} & \textbf{Second Best Performing Specifications} \\
\midrule

{ \small JSON-e Templating Engine Sort Example                   } & { \small JSON BinPack, CBOR and MessagePack (38.2\%) } & { \small Smile (20.5\%) } \\ \hline
{ \small JSON-e Templating Engine Reverse Sort Example           } & { \small JSON BinPack and MessagePack (39.5\%) } & { \small CBOR (38.3\%) } \\ \hline
{ \small CircleCI Definition (Blank)                             } & { \small JSON BinPack, CBOR and MessagePack (28.5\%) } & { \small UBJSON (7.1\%) } \\ \hline
{ \small CircleCI Matrix Definition                              } & { \small JSON BinPack (30.5\%) } & { \small CBOR and MessagePack (24.2\%) } \\ \hline
{ \small Grunt.js Clean Task Definition                          } & { \small JSON BinPack (38.7\%) } & { \small CBOR and MessagePack (35.4\%) } \\ \hline
{ \small CommitLint Configuration                                } & { \small JSON BinPack (37.5\%) } & { \small CBOR and MessagePack (22.9\%) } \\ \hline
{ \small TSLint Linter Definition (Extends Only)                 } & { \small JSON BinPack, CBOR and MessagePack (12.6\%) }   & { \small Smile (3.1\%) } \\ \hline
{ \small ImageOptimizer Azure Webjob Configuration               } & { \small JSON BinPack, CBOR and MessagePack (25.6\%) }   & { \small Smile (14.6\%) } \\ \hline
{ \small SAP Cloud SDK Continuous Delivery Toolkit Configuration } & { \small JSON BinPack, CBOR and MessagePack (43.1\%) }   & { \small BSON and UBJSON (34\%) } \\ \hline
{ \small TSLint Linter Definition (Multi-rule)                   } & { \small JSON BinPack, CBOR and MessagePack (30.6\%) }   & { \small Smile (20.4\%) } \\ \hline
{ \small CommitLint Configuration (Basic)                        } & { \small JSON BinPack, CBOR and MessagePack (32\%) }     & { \small UBJSON (24\%) } \\ \hline
{ \small TSLint Linter Definition (Basic)                        } & { \small JSON BinPack, CBOR and MessagePack (23.8\%) }   & { \small Smile and UBJSON (11.9\%) } \\ \hline
{ \small GeoJSON Example Document                                } & { \small JSON BinPack (38.4\%) }       & { \small MessagePack (14.7\%) }  \\ \hline
{ \small OpenWeatherMap API Example Document                     } & { \small JSON BinPack (29.3\%) }       & { \small MessagePack (22.6\%) }  \\ \hline
{ \small OpenWeather Road Risk API Example                       } & { \small JSON BinPack (32.2\%) }       & { \small Smile (13\%) }          \\ \hline
{ \small TravisCI Notifications Configuration                    } & { \small JSON BinPack (72.5\%) }       & { \small FlexBuffers (66.1\%) }  \\ \hline
{ \small Entry Point Regulation Manifest                         } & { \small JSON BinPack (38.2\%) }       & { \small Smile (31.5\%) }        \\ \hline
{ \small JSON Feed Example Document                              } & { \small JSON BinPack (10.2\%) }       & { \small MessagePack (9.7\%) }   \\ \hline
{ \small GitHub Workflow Definition                              } & { \small JSON BinPack (22.1\%) }       & { \small MessagePack and Smile (19.3\%) } \\ \hline
{ \small GitHub FUNDING Sponsorship Definition (Empty)           } & { \small JSON BinPack, CBOR and MessagePack (32.2\%) } & { \small Smile (29.5\%) } \\ \hline
{ \small ECMAScript Module Loader Definition                     } & { \small JSON BinPack, CBOR and MessagePack (37.2\%) } & { \small Smile (31.3\%) } \\ \hline
{ \small ESLint Configuration Document                           } & { \small JSON BinPack (15\%) }        & { \small MessagePack (14.8\%) }   \\ \hline
{ \small NPM Package.json Linter Configuration Manifest          } & { \small JSON BinPack (31.7\%) }      & { \small FlexBuffers (18.7\%) }   \\ \hline
{ \small .NET Core Project                                       } & { \small JSON BinPack (28.6\%) }      & { \small Smile (17\%) }           \\ \hline
{ \small NPM Package.json Example Manifest                       } & { \small JSON BinPack (13.3\%) }      & { \small Smile (12.2\%) }         \\ \hline
{ \small JSON Resume Example                                     } & { \small JSON BinPack (14.07\%) }     & { \small Smile (14.04\%) }        \\ \hline
{ \small Nightwatch.js Test Framework Configuration              } & { \small JSON BinPack (28\%) }        & { \small Smile (27.6\%) }         \\

\bottomrule
\end{tabularx}
\end{table}

\subsection{Q3: How does JSON BinPack in schema-driven and schema-less mode compare to compressed JSON?}

\begin{table}[ht!] \caption{A summary of the size reduction provided by the
  JSON BinPack binary serialization specification pre-production implementation
  in both schema-driven and schema-less mode in comparison to the best case
  scenarios of compressed JSON \cite{ECMA-404} given the compression formats
  and input data listed in \cite{viotti2022benchmark}.}

\label{table:jsonbinpack-summary-results-compressed}
\begin{tabularx}{\linewidth}{X|l|l|l|l|l|l}
\toprule
\multirow{2}{*}{\textbf{Mode}} & \multicolumn{5}{c|}{\textbf{Size Reductions in Comparison To Compressed JSON}} & \multirow{2}{*}{\textbf{Negative Cases}} \\ \cline{2-6}
& \textbf{Maximum} & \textbf{Minimum} & \textbf{Range} & \textbf{Median} & \textbf{Average} & \\
\midrule
JSON BinPack (Schema-driven) & 100\% & 5.65\% & 94.3 & 76.1\% & 66.8\% & 0 / 27 (0\%) \\ \hline
JSON BinPack (Schema-less)   & 69.6\% &    -146.4\% &  216.1 & 7.4\% &  -5.27\% & 13 / 27 (48.1\%) \\ \hline
\textbf{Averages} & \textbf{84.4\%} & \textbf{-70.3\%} & \textbf{155.2} & \textbf{41.7\%} & \textbf{30.7\%} & \textbf{24\%} \\
\bottomrule
\end{tabularx}
\end{table}

In \cite{viotti2022benchmark}, we conclude that general-purpose data
compression tends to yield negative results for JSON \cite{ECMA-404} documents
that are \textbf{Tier 1 Minified $<$ 100 bytes} according to the proposed
taxonomy given that the auxiliary data structures encoded by dictionary-based
compressors may exceed the size of such small input documents.  However,
leaving \textbf{Tier 1 Minified $<$ 100 bytes} documents aside, best-case
compressed JSON \cite{ECMA-404} is space-efficient in comparison to the
considered schema-less binary serialization specifications in 86.6\% of the
cases. Leaving \textbf{Tier 1 Minified $<$ 100 bytes} documents aside,
best-case compressed JSON \cite{ECMA-404} is strictly space-efficiency in
comparison to the considered schema-driven binary serialization specifications
in \textbf{33.3\%} of the cases.

While JSON BinPack in schema-less mode matches or outperforms the alternative
schema-less binary serialization specifications considered in
\cite{viotti2022benchmark} as shown in
\autoref{sec:jsonbinpack-conclusions-schemaless}, best-case compressed JSON
\cite{ECMA-404} is space-efficient in comparison to JSON BinPack in schema-less
mode in \textbf{13} out of the \textbf{27} considered cases as shown in
\autoref{table:jsonbinpack-summary-results-compressed}. Of these \textbf{13}
negative cases, \textbf{8} documents are considered textual according to the
taxonomy defined in \cite{viotti2022benchmark}. However, unlike the other
considered schema-driven binary serialization specifications, JSON BinPack in
schema-driven mode is space-efficient in comparison to best-case compressed
JSON \cite{ECMA-404} as shown in
\autoref{table:jsonbinpack-summary-results-compressed} in terms of the median
and average with size reductions of \textbf{76.1\%} and \textbf{66.8\%},
respectively.  Existing literature \cite{10.1145/2016716.2016718}
\cite{AnjosEdman2016SAsr} show that compressed textual schema-less
serialization specifications such as JSON \cite{ECMA-404} can outperform
compressed and uncompressed schema-driven binary serialization specifications
in terms of space-efficiency.  However, we conclude that a space-efficient
schema-driven serialization specification such as JSON BinPack can outperform
general-purpose data compression.

\section{Future Work}
\label{sec:future-work}
\textbf{On-going JSON Schema Support in JSON BinPack.} The pre-production JSON
BinPack implementation benchmark in this paper does not support every keyword
defined by the JSON Schema Core \cite{jsonschema-core-2020} and Validation
\cite{jsonschema-validation-2020} specifications. Every JSON Schema document is
supported by JSON BinPack by definition, as JSON BinPack is designed to
gracefully fallback to an schema-less encoding when encountering unrecognized
keywords. However, explicit support for JSON Schema keywords increases
space-efficiency when such keywords are in use. We intend to continue writing
encodings, canonicalization and mapping rules to cover every official JSON
Schema vocabulary.

\textbf{Production-ready JSON BinPack Implementation.} This study considers a
pre-production implementation of JSON BinPack written using the TypeScript
programming language to demonstrate the space-efficiency potential of the
proposed serialization specification. To make JSON BinPack suitable for
production usage, we intend to produce a new implementation using a systems
programming language.

\textbf{Runtime Efficiency Benchmark.} Following the development of a
production-ready implementation of JSON BinPack written using a systems
programming language, we hope to pursue a runtime-efficiency benchmark that
looks the characteristics such as serialization and deserialization speed, and
memory consumption.

\bibliographystyle{ACM-Reference-Format}
\bibliography{benchmark}

\end{document}